\documentclass[fleqn,usenatbib,useAMS]{mnras}

\usepackage{graphicx}	
\usepackage{amsmath}	
\usepackage{amssymb}	
\usepackage{multicol}   
\usepackage{bm}		    
\usepackage{lscape}	    

\newcommand{\kms}{\,km\,s$^{-1}$\,} 
\newcommand{\ms}{\,m\,s$^{-1}$\,} 
\newcommand{\Rnom}{\hbox{$\mathcal{R}^{\rm N}_{\odot}$\,}}

\usepackage[T1]{fontenc}
\usepackage{ae,aecompl}

\usepackage{newtxtext,newtxmath}
\usepackage{longtable}
\title[SPIRou Input Catalogue]{\textit{SPIRou Input Catalogue: Global properties of 440 M dwarfs observed with ESPaDOnS at CFHT}}

\author[P. Fouqu\'e et al.]{
\parbox[t]{\textwidth}{
Pascal Fouqu\'e,$^{1,2,5}$\thanks{Contact e-mail: \href{mailto:fouque@cfht.hawaii.edu}{fouque@cfht.hawaii.edu}}\thanks{Present address: 65-1238 Mamalahoa Hwy Kamuela, HI 96743 U.S.A.}\thanks{Based on observations obtained at the Canada-France-Hawaii Telescope (CFHT) which is operated by the National Research Council of Canada, the Institut National des Sciences de l'Univers of the Centre National de la Recherche Scientifique of France, and the University of Hawaii.}
Claire Moutou,$^{1,3}$
Lison Malo,$^{1,4}$
Eder Martioli,$^{7}$
Olivia Lim,$^{1,4}$
Arvind Rajpurohit,$^{11}$
Etienne Artigau,$^{4}$
Xavier Delfosse,$^{6}$
Jean-Fran\c{c}ois Donati,$^{2,5}$
Thierry Forveille,$^{6}$
Julien Morin,$^{9}$
France Allard,$^{12}$
Rapha\"el Delage,$^{1}$
Ren\'e Doyon,$^{4}$
Elodie H\'ebrard,$^{10}$
Vasco Neves$^{8,13}$
}
\vspace{0.2cm} \\
$^{1}$CFHT Corporation; 65-1238 Mamalahoa Hwy; Kamuela, Hawaii 96743; USA\\
$^{2}$Universit\'e de Toulouse; UPS-OMP; IRAP; Toulouse, France\\
$^{3}$Aix Marseille Universit\'e, CNRS, LAM, Laboratoire d'Astrophysique de Marseille, Marseille, France\\
$^{4}$Institute for Research on Exoplanets, D\'epartement de physique, Universit\'e de Montr\'eal, CP 6128, Succursale Centre-Ville, Montr\'eal, Quebec H3C 3J7, Canada \\
$^{5}$CNRS; IRAP; 14, avenue Edouard Belin, F-31400 Toulouse, France\\
$^{6}$Universit\'e Grenoble Alpes, CNRS, IPAG, 38000 Grenoble, France\\
$^{7}$Laboratorio Nacional de Astrofisica (LNA/MCTI), Rua Estados Unidos, 154, Itajuba, MG, Brazil\\
$^{8}$Instituto Federal do Paran\'a, 85860000, Campus Foz do Igua\c{c}u, Foz do Igua\c{c}u-PR, Brazil\\
$^{9}$LUPM, Universit\'e de Montpellier, CNRS, Place Eug\`ene Bataillon, F-34095 Montpellier, France\\
$^{10}$Department of Physics and Astronomy, York University, 4700 Keele St., Toronto, Ontario, M3J 1P3 Canada\\
$^{11}$Astronomy and Astrophysics Division, Physical Research Laboratory, 380009 Ahmedabad, India\\
$^{12}$Centre de Recherche Astrophysique de Lyon, UMR 5574, Universit\'{e} de Lyon, ENS de Lyon, Universit\'{e} Lyon 1, CNRS, F-69007, Lyon, France\\
$^{13}$Casimiro Montenegro Filho Astronomy Center, Itaipu Technological Park, 85867-900, Foz do Igua\c{c}u-PR, Brazil\\
}

\date{Accepted 2017 December 11. Received 2017 December 7; in original form 2017 November 16.}

\pubyear{2017}

\begin{document}
\label{firstpage}
\pagerange{\pageref{firstpage}--\pageref{lastpage}}
\maketitle

\begin{abstract}
Present and future high-precision radial-velocity spectrometers dedicated to the discovery of low-mass planets orbiting low-mass dwarfs need to focus on the best selected stars to make an efficient use of telescope time. In the framework of the preparation of the SPIRou Input Catalog, the CoolSnap program aims at screening M dwarfs in the solar neighborhood against binarity, rapid rotation, activity, ... To optimize the selection, the present paper describes the methods used to compute effective temperature, metallicity, projected rotation velocity of  a large sample of 440 M dwarfs observed in the visible with the high-resolution spectro-polarimeter ESPaDOnS at CFHT. It also summarizes known and newly-discovered spectroscopic binaries, and stars known to belong to visual multiple systems. A calibration of the projected rotation velocity versus measured line widths for M dwarfs observed by the ESPaDOnS spectro-polarimeter is derived, and the resulting values are compared to equatorial rotation velocities deduced from rotation periods and radii. A comparison of the derived effective temperatures and metallicities with literature values is also conducted. Finally, the radial velocity uncertainty of each star in the sample is estimated, to narrow down the selection of stars to be included into the SPIRou Input Catalogue (SPIC). 
\end{abstract}

\begin{keywords}
low-mass stars, radial velocity, effective temperature, metallicity, binarity -- planet search
\end{keywords}
\thanks{}




\section{Introduction}
Dwarf stars of spectral type M were the coolest stellar objects known until the discovery of field brown dwarfs \citep{becklin88,rebolo95,nakajima95} and the creation of new spectral types L, T and Y \citep{martin97,kirkpatrick99,martin99,kirkpatrick00}. M dwarfs are the most numerous stars in our Galaxy, amounting to about two-thirds in number and about 40\% in stellar mass \citep{kirkpatrick12}. They were not known from ancient astronomers, as none of them is visible to the naked eye: the brightest one, Gl~825, has a V magnitude of 6.7 and is an M0V, sometimes classified as K7V. Therefore, M stars were a good example of "invisible" matter, later recognised as a major contributor to the stellar mass of our Galaxy.

Although they share a common spectral class, they display a wide range in properties: their masses span a range of about a factor 9, similar to the range spanned by B, A, F, G and K stars altogether. Similarly, their bolometric luminosities span a range of 200. Their global properties vary a lot along the sub-classes from M0 to M9, crossing the limit between stars and brown dwarfs, and as other spectral types, they display a large variety of ages, from pre-main sequence stars of a few Myr to very old stars, with a corresponding range of radius and therefore gravity for a given mass. They also belong to different star populations (Galactic disk and halo), being classified as dwarfs, subdwarfs, extreme subdwarfs and even ultra subdwarfs according to their metallicity \citep{lepine07a}.

\citet{lepine11} estimate that there are about 11,900 M dwarfs brighter than $J=10$ in the whole sky. But given their wide range in absolute magnitudes, it is difficult to translate this figure to a given number of M dwarfs within a given distance limit, for instance 25~pc. All the early M dwarfs (up to M3.5V) will then be counted, but not the later spectral type ones. There is no current complete catalog of late M dwarfs up to a given distance.

In addition, it is well known that M dwarfs display a range of activity, rotational velocity and magnetic properties (\citet{west04,reiners07a,kiraga07,donati08,morin08a,morin08b,morin10,morin11,irwin11,reiners12,west15,newton17} among others), that is further investigated in this study and companion papers (\citet{moutou17} and Malo et al., in prep.). Although this class of stars was somehow neglected in the past due to their faintness at optical wavelengths, it started to emerge with the advent of near-infrared sky surveys, DENIS \citep{epchtein99} and 2MASS \citep{skrutskie06}, which opened the way to near-infrared spectrometers. As the small mass and radius of M dwarfs were favourable to reveal their planetary companions, and with the additional benefit that their habitable zones lie close enough to the star to allow discoveries of habitable planets, large surveys of these stars began (e.g. \citet{bonfils13,delfosse13}).

In the framework of the preparation of the new near-infrared high-resolution spectro-polarimeter SPIRou (Donati et al., in \citet{deeg18}), to be installed at CFHT in 2018, members of the SPIRou team decided in 2014 to embark upon an observational snapshot program of M dwarfs, nicknamed CoolSnap, using the ESPaDOnS visible high-resolution spectro-polarimeter at CFHT \citep{donati97}. The goal of this survey is a better knowledge of M dwarfs selected as prime targets to search for planetary-mass objects in the habitable zone before their inclusion into the SPIRou Input Catalogue (SPIC). The selection criteria used to build the CoolSnap sample are described in Malo et al., in prep.. Their activity and magnetic properties are described in \citet{moutou17}. Here, we concentrate on the global properties of the observed stars, such as effective temperature, metallicity, rotational velocity, binarity. These properties are important for our selection, as we want to avoid stars that are too active, fast-rotating objects, close multiple systems, which will all prevent us from detecting low-mass planets orbiting these stars.

Other near-infrared spectrographs are currently under development, such as HPF \citep{mahadevan12}, CARMENES \citep{quirrenbach14}, or GIARPS \citep{claudi16}. These projects can benefit from our study, as we benefited for instance from the CARMENCITA catalogue \citep{alonso15,cortes17}.

This paper is organized as follows: Section~\ref{sec:obs} describes how the stars were selected to build a sample of 440 M dwarfs. Section~\ref{sec:multiple} describes spectroscopic binaries either discovered during these observations or already known, and more generally the multiplicity of systems to which stars in our sample belong. Section~\ref{sec:st-teff-feh} explains how spectral type, effective temperature and metallicities are derived for our sample, and the limitation of the \textsc{\small{mcal}} method introduced by \citet{neves14}, and used to measure these properties. Section~\ref{sec:vsini} describes how projected rotation velocities are derived from the width of the LSD profile obtained from the observed spectra with an M2 template. Section~\ref{sec:discussion} concludes about stars which are good candidates for $RV$ search of low-mass planets using the SPIRou near-infrared spectro-polarimeter, from the point-of-view of the parameters measured in this study. Finally, Section~\ref{sec:conclusion} summarizes this work and link it to the other two papers in this series, namely \citet{moutou17} and Malo et al., in prep..

\section{Sample and observations}\label{sec:obs}
We performed our initial compilation of M dwarfs based on the following studies (see Malo et al., in prep. for more details):   
\begin{itemize}
\item An all-sky catalogue of bright M dwarfs \citep{lepine11}, which consists of 8889 K7-M4 dwarfs with $J<10$. This sample is based on the ongoing SUPERBLINK proper-motion survey. Spectral types are estimated from the $V-J$ colour index.
\item A catalogue of bright ($K<9$) M dwarfs \citep{frith13}, which consists of 8479 K7-M4 dwarfs. This catalogue rests on the PPMXL proper-motion survey.
\item An all-sky catalogue of nearby cool stars CONCH-SHELL \citep{gaidos14}, which consists of 2970 nearby ($d<50$~pc), bright ($J<9$) M- or late K-type dwarf stars, 86\% of which have been confirmed by spectroscopy. This sample is also selected from the SUPERBLINK proper-motion survey combined with spectra and photometric colour criteria.
\item A sample of spectroscopically confirmed nearby M dwarfs \citep{newton14}, which consists of 447 M dwarfs with measured metallicities, radial velocities and spectral types from moderate resolution ($R\sim2000$) near-infrared spectroscopy. This sample is drawn from the MEarth survey \citep{irwin11}.
\item A southern sample of M dwarfs within 25~pc \citep{winters15}, which consists of 1404 M0-M9.5 dwarfs with $6.7<V<21.4$. This sample is based on the RECONS program and supplemented by observations at the CTIO/SMARTS 0.9m telescope.
\item The CARMENES input catalogue of M dwarfs \citep{alonso15}, which consists of 753 spectroscopically confirmed K-M stars. 
\item A northern sample of mid-to-late M dwarfs from the MEarth project \citep{newton16b}, which consists of 387 nearby dwarfs with measured rotation periods.  
\end{itemize}

This compilation leads to an all-sky sample of about 14,000 K5-M9 stars. Since SPIRou will be installed at CFHT (latitude $20\degr$), we restrict our sample to stars observable with declination north of $-30\degr$, which gives a final sample of 10,142 stars.

We applied to this initial sample a merit function computed from the star flux in $H$ band and the expected radial velocity amplitude produced by a 3 Earth mass planet orbiting it in the Habitable Zone, which in turn depends upon mass, radius and temperature of the star, to select the 150 highest merit stars to be observed. Details about this merit function are given in Malo et al., in prep..

Observations were conducted with the ESPaDOnS spectro-polarimeter \citep{donati97} at the CFHT 3.6~m telescope on top of Maunakea (Hawaii), which provides a wide optical range from 367~nm to 1050~nm in a single shot at a resolving power of 65,000 (polarimetry) or 68,000 (pure spectroscopy in the so-called "star plus sky" mode, with one fiber on the target and one on the sky: we call it "S+S" hereafter). Data are reduced using the \textsc{\small{Libre-Esprit}} software \citep{donati97}. Least-Squares Deconvolution (LSD, \citealt{donati97}) is then applied to all the observations, to take advantage of the large number of lines in the spectrum and increase the signal-to-noise ratio (SNR per 2.6\kms pixel) by a multiplex gain of the order of 10. We used a mask of atomic lines computed with an \textsc{\small{Atlas}} local thermodynamic equilibrium (LTE) model of the stellar atmosphere \citep{kurucz93a}. The final mask contains about 4000 moderate to strong atomic lines with a known Land\'e factor. This set of lines spans a wavelength range from 350~nm to 1082~nm. The use of atomic lines only for the LSD masks relies on former studies of early and mid M dwarfs \citep{donati06}.

More details about the CoolSnap observations\footnote{Program IDs 14BF13/B07/C27, 15AF04/B02, 15BB07/C21/F13, 16AF25, 16BC27/F27 and 17AC30, P.I. E.~Martioli, L.~Malo and P.~Fouqu\'e} and the data reduction are given in \citet{moutou17} and Malo et al., in prep.. For the purpose of this paper, let us just state that two high signal-to-noise spectra (SNR $\sim 100$ at 800~nm) are taken for each M star of the sample (typically M0 to M6), separated by several days or weeks, in order to assess possible changes in the magnetic activity or in the heliocentric radial velocity (HRV). We observed 280 spectra in polarimetric mode for 118 stars. Removing four stars initially selected for the CoolSnap sample and observed, but for which classification issues (they most certainly are not M dwarfs) were later discovered, leads to 114 genuine M dwarfs in the CoolSnap sample. The 4 rejected stars are listed in Table~\ref{tab:rejected} for completeness.

\begin{table*}
	\centering
	\caption{List of 4 rejected stars.}
	\label{tab:rejected}
	\begin{tabular}{lll}
		\hline
		2MASS name & Common name & Reason for rejection \\
		\hline
J07100298-0133146 & & V=12.196 rather than 13.34 originally used: V-J=2.23 therefore corresponds to a K5V-K6V spectral type \\
J16275072-1926069 & TYC~6211-472-1 & $J-K_{\rm s}=1.3$ should have been removed from \citet{gaidos14} \\
J17294104-1748323 & TYC~6239-2457-1 & wrong PM, not a dwarf; SB1? (us: 3.2\kms in 35 days) \\
J18302580-0006226 & & $J-K_{\rm s}=2.0$ should have been removed from \citet{gaidos14} \\
idem & & $V-K_{\rm s}=7.3$ may be explained by a K giant with circumstellar material: it is an IRAS star \\
		\hline
	\end{tabular}
\end{table*}

In addition to our own measurements, we searched the ESPaDOnS archive in polarisation mode at the Canadian Astronomy Data Center (CADC\footnote{http://www.cadc-ccda.hia-iha.nrc-cnrc.gc.ca/en/cfht/}) from 2005 to 2015 (inclusively) and found 839 spectra for 71 additional M dwarfs (and 10 spectra for 2 stars in the CoolSnap sample, namely Gl~411 and Gl~905). The two samples have different characteristics, the stars from the archive often being active and rapid rotators and generally having a large number of spectra, while the CoolSnap sample is limited to 2 spectra taken at different epochs for each star.

Finally, we also searched the ESPaDOnS archives for stars observed in the purely spectroscopic S+S mode. We found 785 spectra for 255 additional stars, raising the total sample of M dwarfs observed with ESPaDOnS to 440.

Spectra of stars belonging to the complementary samples (polarimetric and spectroscopic) have generally been published, but we reanalyze them to derive their effective temperature, metallicity, and projected rotation velocity in a consistent way.

\section{Multiple systems}\label{sec:multiple}
Binarity (and higher multiplicity) is common among stars. Many techniques have been devised to disentangle physical association from apparent projection on the sky. A good historical review is given by \citet{dommanget00a}. For our purpose, multiplicity may be important for the following reasons: 
\begin{itemize}
    \item we may discover that an object initially identified as a single star is in fact a close binary. The selection criterion may then be invalidated when the magnitude or color encompasses both stars;
    \item if the components are too close to be separated in the fiber entrance of the spectrograph, both spectra are recorded and the object may then reveal as a single-line or double-line spectroscopic binary;
    \item even when the separation is large enough, and assuming that the system is physical, the planet formation mechanism may have been affected by the binarity;
    \item wide multiple physical systems composed of a FGK primary and an M secondary allow a calibration of the metallicity of the M dwarf, assuming that it shares the same metallicity as the primary component of the system (see e.g. \citet{bonfils05}).
\end{itemize}

The release of GAIA data (DR1 and soon DR2) will allow us to confirm the status of the binaries in our sample, and discard the optical systems which are not physical. GAIA will certainly also discover new astrometric binaries in this sample. However, it is still important for future observations to know whether a star has a close companion, since the light from the companion may contribute significantly to the measured flux, which may affect the measured parameters (magnitudes, colors, ...).

It is obvious that only a fraction of these systems may affect our observations or the future detection of planetary systems orbiting the stars of our sample. As the fiber diameter is 1.58" for ESPaDOnS and 1.33" for SPIRou, binaries separated by less than 1" will contaminate the observed spectrum. Components separated by more than 2" should be easy to separate under reasonable seeing. However, at this separation, some parameters may still be affected, such as visual or near-infrared magnitudes. 

On another hand, physical separations matter in the rate of formation of planetary systems. Therefore, close physical multiplicity of the stellar system may affect the formation of planets. More details are given in \citet{thebault14}.

In order to identify the physical systems (visual or spectroscopic) in our catalogue, we started to build a catalogue of multiple systems involving M dwarfs. We defer to a future publication details and statistics about this catalogue, for instance a confirmation of physical systems based on future released data from GAIA (DR2 and following), and an evaluation of the multiplicity rate among M stars, compared to earlier spectral types, based on a complete distance-limited sample.

\subsection{Spectroscopic binaries}
Spectroscopic binaries are easily identified when two peaks appear in the LSD profile (SB2). Sometimes, only one component is visible in the spectrum (generally because the other component is much fainter), and we have an SB1. Given that the accuracy of the heliocentric radial velocity (HRV) measured by ESPaDOnS and reduced with \textsc{\small{Libre-Esprit}} \citep{donati97} is about 20-30\ms \citep{moutou07}, SB1 are revealed when the radial velocity corrected to the heliocentric reference frame $HRV$ significantly differs between the two spectra. In Table~\ref{tab:spectro-bin2}, we list the stars in our sample which have been observed and revealed themselves as SB2 (21 stars including uncertain ones), or even SB3 (2 stars), together with already known spectroscopic binaries (28 SB2, 4 SB3, two quadruple systems SB1+SB2 and SB2+SB2), which should have been excluded when assembling the observational sample.

Among the 57 SB listed in this Table, about one half also appears in Table~\ref{tab:vis-bin}, because they belong to multiple systems with both visual and spectroscopic components.

In Table~\ref{tab:spectro-bin1}, we list the stars in our sample which have been observed and revealed themselves as SB1 (2 stars), together with already known single-line spectroscopic binaries (8 stars), also missed when assembling the sample or discovered by others during our survey. Radial velocity variations may also be due to activity-induced rotational modulation for stars with strong magnetic fields, rather than binarity. A few special cases with discrepant or anomalous results are listed in Table~\ref{tab:special}.

\begin{table*}
	\centering
	\caption{Single-line spectroscopic binaries detected in the observations of the CoolSnap sample or listed in the literature and recovered from the ESPaDOnS "S+S" archive. The heliocentric radial velocities (in \kms) and corresponding heliocentric Julian dates ($-2450000$, at mid-exposure, TT) are given, as measured in our observations.}
	\label{tab:spectro-bin1}
	\begin{tabular}{llllccl}
		\hline
		2MASS name & Common name & SB type & Reference & $HRV$ & $HJD$ & Comment \\
		\hline
J00582789-2751251 & Gl~46 & SB1 & this work & 23.0, 20.2 & 7262.029, 7284.933 & \\
J08313744+1923494 & GJ~2069B & SB1 & \citet{delfosse99b} & 7.5, 7.5 & 6813.735, 6814.733 & also in Table~\ref{tab:vis-bin} \\
J09142298+5241125 & Gl~338A & SB1 & \citet{cortes17} & 12.3, 11.2 & 4275.764, 6813.758 & also in Table~\ref{tab:vis-bin} \\
J10141918+2104297 & GJ~2079 & SB1? & \citet{shkolnik12} : Table~2 & 13.0 & 6771.836 & also in Table~\ref{tab:vis-bin} \\
J11032125+1337571 & LP~491-51 & SB1 & \citet{cortes17} & $-24.0$ & 4546.875 & \\
J16240913+4821112 & Gl~623 & SB1 & \citet{nidever02} & $-26.8$, $-27.2$ & 7226.775, 7402.173 & also in Table~\ref{tab:vis-bin} \\
J17093153+4340531 & GJ~3991 & SB1 & \citet{reid97,delfosse99b} & 1.0, $-54.6$ & 7085.137, 7121.138 & \\
J18495543-0134087 & & SB1 & \citet{malo14a} & 118.6, 115.3 & 5747.749, 5747.757 & \\
J22384530-2036519 & Gl~867B & SB1 & \citet{davison14} & 4652.100 & $-2.66$ & also in Table~\ref{tab:vis-bin} \\
J22524980+6629578 & & SB1? & this work & $-8.3$, $-7.6$ & 7611.944, 7680.825 & \\
		\hline
	\end{tabular}
\end{table*}

\begin{table*}
	\centering
	\caption{Some special cases of spectroscopic binaries, with discrepant or anomalous results.}
	\label{tab:special}
	\begin{tabular}{lll}
		\hline
		2MASS name & Common name & Comment \\
		\hline
J03373331+1751145 & GJ~3239 & SB2, but we measured $v \sin i$ for the primary component (see Table~\ref{tab:results}). \\
J08313744+1923494 & GJ~2069B & close VB with an $RV$ drift of 600\ms over 850~days \citep{delfosse99b}, \\
& & strong magnetic field \citep{reiners09}. \\
J10182870-3150029 & TWA~6 & non-gaussian large LSD profile: see \citet{skelly08}. \\
J11250052+4319393 & LHS~2403 & 3 low SNR (25) spectra possibly contaminated by the Moon. \\
J14170294+3142472 & GJ~3839 & close visual binary and SB2 \citep{delfosse99b}, \\
& & this work: not clearly SB2, but asymmetrical LSD profile. \\
J12141654+0037263 & GJ~1154 & unresolved SB2 (variable spectral line-width) \citep{bonfils13}, \\
& & strong large-scale magnetic field \citep{reiners09,morin10}. \\
J14493338-2606205 & Gl~563.2A & SB2, but we measured $v \sin i$ for the primary component (see Table~\ref{tab:results}). \\
J23315208+1956142 & Gl~896A & SB1 \citep{delfosse99b}, \\
& & magnetic activity \citep{morin08b}. \\
J23315244+1956138 & Gl~896B & SB1 \citep{delfosse99b}, \\
& & magnetic activity \citep{morin08b}. \\
		\hline
	\end{tabular}
\end{table*}

\subsection{Visual multiple systems detected by imagery}
As stated above, it is important to know whether a star in our survey belongs to a physical multiple system. Unfortunately, there is no recent compilation of such systems. Rather than just checking for the multiplicity status of the stars in our sample, we embarked into a parallel project of listing all multiple systems involving an M dwarf, in order to get better statistics, not biased by the selection process which led to our sample. For this purpose, we surveyed the literature for physical systems detected by imagery, including adaptive optics, coronagraphy or lucky imaging observations of M dwarfs.

We started by checking the information provided by the Washington Double Star Catalogue \citep{mason01}, in its constantly updated on-line version at CDS (hereafter WDS), the Catalogue of Visual Double Stars observed by the Hipparcos satellite \citep{dommanget00a,dommanget00b}, the MSC Catalog of Physical Multiple Stars \citep{tokovinin97}, the Catalog of Common Proper-Motion Companions (hereafter CPM) to Hipparcos stars \citep{gould04} and the Catalogue of Faint Companions to Hipparcos stars \citep{lepine07b}. We then surveyed the literature for additional binary stars or additional information on the systems described in the above references. Finally, some optical binaries were discovered by us at the telescope, using images from the guider. 

The compilation used in this paper is not complete, as we preferred waiting for the second release of GAIA in April 2018, to discard unphysical multiple systems or components when GAIA measures discrepant parallaxes or proper motions. In its present version, it contains 671 multiple systems, among which 393 have an M dwarf primary. We used this limited version for investigating the multiplicity of stars in our sample of 440 M dwarfs. The resulting Table is given in Appendix.

\section{Measure of spectral type, effective temperature and metallicity}\label{sec:st-teff-feh}

\subsection{Spectral type}

\begin{figure}
	\includegraphics[width=\columnwidth]{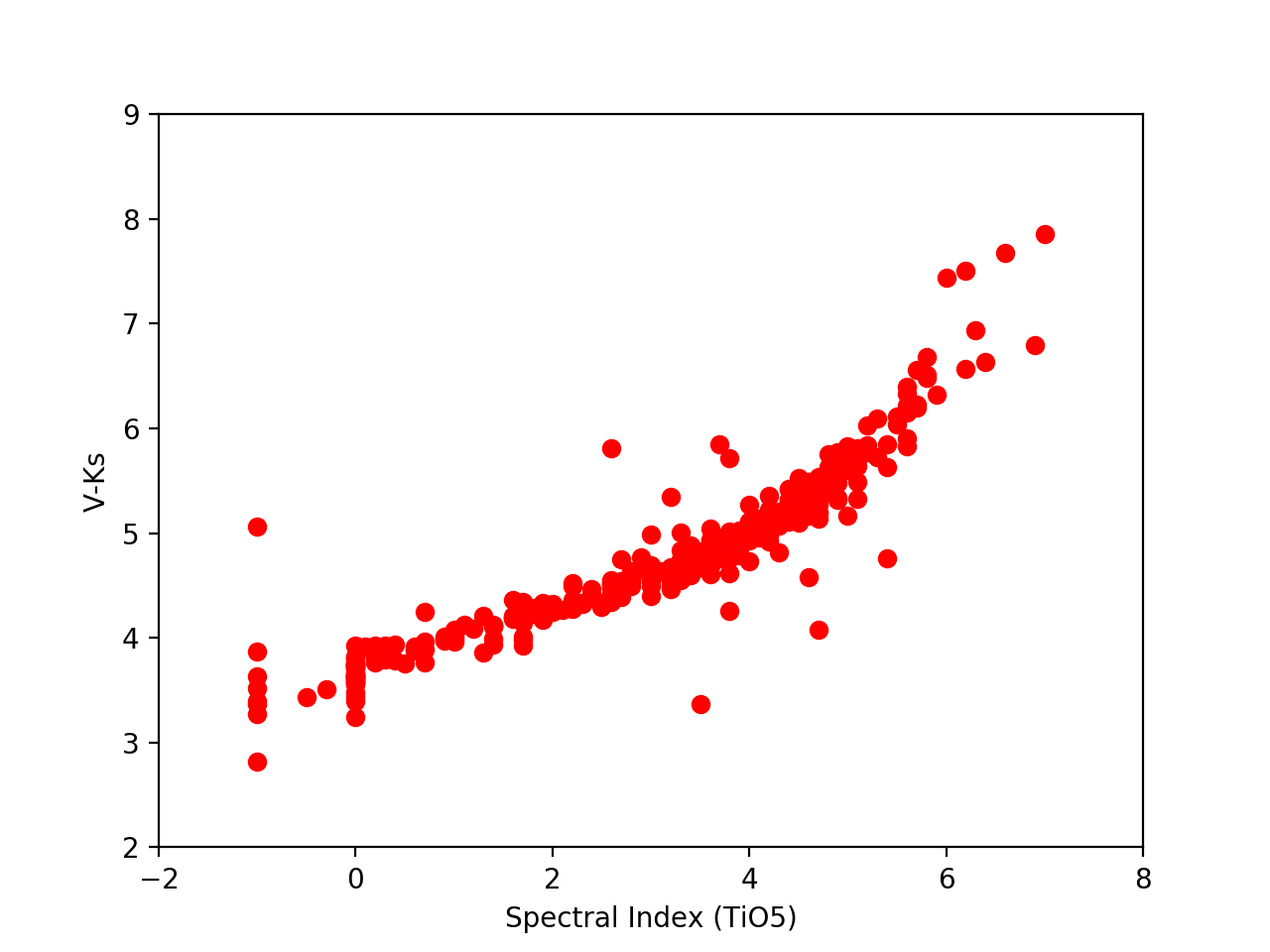}
    \caption{Correlation between the spectral type measured from the TiO~5 spectral index, with the $V-K_{\rm s}$ color.}
    \label{fig:ST_VmK}
\end{figure}

We estimate the spectral type of our stars from a measurement of the TiO~5 spectral index at 713~nm, as defined and calibrated in \citet{reid95}. It is well-adapted to the range of spectral types of our sample, at least up to M6.5V. Standard numerical values are adopted, from -1 for K7V, 0 for M0V to 6 for M6V. The correlation with the $V-K_{\rm s}$ color is clear, as displayed in Figure~\ref{fig:ST_VmK}. Some stars with an earlier spectral type than our M0 limit (negative spectral indices) or for which we could not measure the spectral type are listed in Table~\ref{tab:k7m0}. As the limit between spectral classes K7V and M0V is somewhat fuzzy, we prefer not to exclude those stars a priori, without a clear confirmation of a K spectral type. The value of $V-K_{\rm s}$ may help, as the average value of 25 M0 stars in our sample is $3.65\pm0.02$. Other outliers are generally close visual binaries, where the photometry may be contaminated. They are listed in Table~\ref{tab:outliers}.

\begin{table*}
	\centering
	\caption{List of 13 stars with undetermined or negative spectral types.}
	\label{tab:k7m0}
	\begin{tabular}{llclcl}
		\hline
		2MASS name & Common name & Spectral type (TiO~5) & Spectral type & $V-K_{\rm s}$ & Possible explanation \\
		\hline
J00161455+1951385 & GJ~1006A & $-1.0$ & M4V & 5.058 & \\
J00233468+2014282 & FK~Psc & $-0.3$ & K7.5V & 3.505 & \\
J00340843+2523498 & V493~And & $-0.5$ & M0V & 3.436 & \\
J01373940+1835332 & TYC~1208-468-1 & $-1.0$ & K3V+K5V & 3.868 & \\
J02272804+3058405 & BD+30~397B & $-1.0$ & M2V & & young M dwarf in $\beta$Pic \citep{shkolnik09} \\
J02272924+3058246 & AG~Tri & none & K7V & 3.205 & young M dwarf in $\beta$Pic \citep{shkolnik09} \\
J08081317+2106182 & LHS~5133 & $-1.0$ & K7V & 3.392 & \\
J10112218+4927153 & Gl~380 & $-1.0$ & K7V & 3.636 & \\
J11220530-2446393 & TWA~4 & $-1.0$ & K5V & 3.519 & \\
J12245243-1814303 & Gl~465 & none & M3V & 4.300 & large rotation ($v \sin i = 63$\kms) \\
J16575357+4722016 & Gl~649.1B & none & M1.5Ve & & maybe Gl~649.1A (K3V) at 5.1" was observed \\
J20560274-1710538 & TYC~6349-200-1 & $-1.0$ & & 3.370 & \\
J22465311-0707272 & & $-1.0$ & & 2.822 & photometry may be contaminated by a star at 4.3" \\
		\hline
	\end{tabular}
\end{table*}

\begin{table*}
	\centering
	\caption{Stars with discrepant $V-K_{\rm s}$ colors for their spectral index.}
	\label{tab:outliers}
	\begin{tabular}{llccl}
		\hline
		2MASS name & Common name & Spectral type (TiO~5) & $V-K_{\rm s}$ & Comment \\
		\hline
J01034013+4051288 & NLTT~3478 & 3.5 & 3.370 & visual binary 0.3" \\
J01034210+4051158 & NLTT~3481 & 4.6 & 4.584 & visual binary 2.5" \\
J01591260+0331113 & NLTT~6638 & 3.2 & 5.351 & SB2 and visual binary \\
J08313744+1923494 & GJ~2069B & 4.7 & 4.081 & visual binary 1.0" \\
J08524466+2230523 & NLTT~20426 & 3.8 & 4.260 & visual binary 4.6" \\
J11314655-4102473 & Gl~431 & 3.8 & 5.719 & \\
J17462507+2743014 & Gl~695BC & 3.7 & 5.847 & visual binary 0.8" \\
J18450905-0926438 & TYC~5696-202-2 & 2.6 & 5.815 & \\
J22171870-0848186 & Gl~852B & 5.4 & 4.759 & visual binary 1.0" \\
		\hline
	\end{tabular}
\end{table*}

\subsection{The \textsc{\small{mcal}} method}

Three important parameters used to characterize stars are effective temperature, metallicity and gravity. In the case of M dwarfs, they are notoriously difficult to measure, especially because no continuum exists in the optical spectrum. There is a long list of publications dealing with several methods to measure mainly the two first, without reaching definite conclusions, for instance \citet{bonfils05,woolf05,casagrande08,onehag12,rojas-ayala12,rajpurohit13}.

In this work, we chose to use the \textsc{\small{mcal}} method of measurement described in \citet{neves14}. In short, it is based on measurements of  pseudo-equivalent widths of lines in high-resolution optical spectra obtained by \citet{bonfils13} using the HARPS spectrometer, which are then correlated to known values of $T_{\rm eff}$ and [Fe/H] from \citet{casagrande08} and \citet{neves12}, respectively. A caveat is that gravity is not used in this correlation, so young stars with low gravity probably get assigned a biased temperature and metallicity.

For this study, we started by using the \citet{neves14} calibration: the calibrating $T_{\rm eff}$ values come from \citet{casagrande08} $T_{\rm eff}$ versus color relations. The authors enhanced the infrared flux method (IRFM, \citet{blackwell90}), to apply it to M dwarfs by adding information from the optical range. Their new method is called MOITE (Multiple Optical-Infrared TEchnique). In this method, the bolometric flux comes from optical and infrared photometry for about 80\%, and the rest comes from Phoenix models\footnote{ftp.hs.uni-hamburg.de/pub/outgoing/phoenix/GAIA/} described for instance in \citet{hauschildt99}. 

Although this method allows one to derive metallicities, \citet{neves14} prefer to use their own metallicity values. These are based on a technique pioneered by \citet{bonfils05}. It starts with binary stars where the primary component is a star of spectral type F, G or K which has a spectroscopically measured metallicity, and the secondary is an M dwarf assumed to share the metallicity of the primary. These binary M dwarfs serve in turn to calibrate an $M_{\rm k}$ versus $V-K_{\rm s}$ color-magnitude diagram: the main-sequence locus at an average metallicity is identified, and the color or absolute magnitude shift from this locus gives a measure of the metallicity of new M stars. Subsequently, \citet{johnson09} corrected the calibration for metal-rich M stars, and \citet{schlaufman10} refined that latter calibration. \citet{mann13b} compiled and measured metallicities of solar-type primaries in 112 wide binary systems involving an M dwarf secondary. \citet{maldonado15} used a similar method to \textsc{\small{mcal}} to calibrate stellar parameters of 53 M dwarfs observed with HARPS.

\citet{neves12} refined once more over \citet{schlaufman10}. Using this calibration, \citet{neves13} computed the metallicity of all the M dwarfs in the \citet{bonfils13} sample and \citet{neves14} selected the more suitable for their calibration of the pseudo-equivalent widths versus metallicity and effective temperatures. Their Table~2 contains 65 stars, and the calibrating values are given in the columns labelled [Fe/H]\_N12 and Teff\_C08. It should be noted that some of these values differ from the similar previous Table~A.1 in \citet{neves13}, probably because of a change in the adopted $V$ magnitude of the star, which in turn produces a change in the distance to the main-sequence locus and therefore of its computed $T_{\rm eff}$ from colors.

\subsection{Limits of the method}

Not all spectra are usable when applying the \textsc{\small{mcal}} method. Some spectra have low SNR, giving an ill-defined peak in the LSD profile, or an inaccurate RV. Two stars (vB8 and vB10) have 9 spectra each in the Polar archive (published in \citet{morin10}, with SNR between 68 and 107, but they have very late spectral types (M7V and M8V, respectively) outside of the calibration range of the method. In polarimetric mode, we are therefore working on 1090 spectra taken with a large enough signal-to-noise ratio (typically, SNR per 2.6\kms pixel $>100$), for 182 stars, removing the two very-late dwarfs mentioned above and 2MASS J09002359+215054, which only has one spectrum with an SNR of 30 in the CoolSnap sample. Similarly, some S+S spectra have a low SNR which does not meet our original quality criterion for polarimetry (SNR $>100$). We only exploited S+S spectra of good quality (well-detected LSD profile, correct RV), reducing the number of useful spectra to 706 for 298 stars (including 45 with polarimetric spectra too), which added to the 182 stars with useful polarimetric spectra leads to a total of 435 stars which can a priori be used to measure global parameters.

But in fact, as explained by \citet{neves14}, some very active stars are not suitable to the measurement of $T_{\rm eff}$ and [Fe/H] by this method. As many stars in the ESPaDOnS archive are active, this can drastically reduce the sample of stars where those parameters can be measured. To identify very active stars, the method measures an H$\alpha$ index as defined in \citet{gomes11}. A small value of about 0.03 corresponds to inactive stars. The adopted cut-off is a value of 0.25, roughly corresponding to a luminosity ratio $\log L_{\rm H\alpha}/L_{\rm bol}$ of $-4.0$, above which $H\alpha$ and magnetic flux become independent of the rotation rate, as shown in \citet{reiners09}. According to this cut-off between saturated (or very active) and non-saturated stars, our CoolSnap sample contains 10/113, the polarimetric archive 40/69, and the spectroscopy archive 146/253 very active stars, for which metallicity and effective temperatures cannot be reliably measured by the \textsc{\small{mcal}} method. An additional 33 non-saturated stars are spectroscopic binaries, for which the method does not work properly either (see above). Finally, a few non-saturated rapid rotators are not well-suited either to this technique, as the measurement of pseudo-equivalent widths is affected by the broadening of the lines due to rotation, and the calibration therefore returns too low temperatures. We do not consider measured effective temperatures and metallicities for 20 non-saturated stars with a $v \sin i$ larger than 8\kms. We are left with 192 stars on which comparisons with other methods can be secured.

The main source of accurate $T_{\rm eff}$ comes from the work of \citet{boyajian12}, who measure M dwarf radii using the CHARA interferometer. They then compute the bolometric flux from multi-band photometry and derive a value of $T_{\rm eff}$. This seems to be a straightforward method, if the template spectra fitted to the photometry are reliable. \citet{mann13a} argue that when compared to their actual low-resolution spectra, there are systematic differences, leading to underestimated bolometric fluxes and temperatures. Finally, \citet{mann15} use the same method to measure the bolometric flux, but use the CIFIST suite of the BT-Settl version of the PHOENIX atmosphere models \citep{allard13}, to measure $T_{\rm eff}$ and derive the corresponding radii. 

Both \citet{mann15} and \citet{rajpurohit13}, who measured $T_{\rm eff}$ by fitting BT-Settl synthetic spectra, show that $T_{\rm eff}$ values from \citet{casagrande08} are too low due to the assumption that M dwarf can be treated as black bodies beyond 2$\mu$m. As the \citet{casagrande08} temperature scale is used in the original \textsc{\small{mcal}} method used by \citet{neves14}, it is important to confirm this result. For this purpose, we compared the original \citet{neves14} calibration to other sources of measurements, for instance \citet{woolf05, woolf06}, who use CaH2 and TiO5 molecular band strength indices, \citet{onehag12,lindgren16,lindgren17}, who fit synthetic spectra to high-resolution VLT-CRIRES spectra in the $J$-band, which are free from large molecular-band contributions, or \citet{rojas-ayala12}, who measure equivalent widths of CaI and NaI lines in the near-infrared and a spectral index quantifying the absorption due to H$_2$O opacity. We found a faire agreement for the metallicities (within 0.2\,dex), but the effective temperatures obtained using the original calibration are systematically low by about 200 K.

We therefore adopt the \citet{mann15} $T_{\rm eff}$ scale while retaining the \citet{neves14} metallicity scale. We modified the original \textsc{\small{mcal}} code to recompute the coefficients of the calibration relations using the more recent and accurate source of $T_{\rm eff}$. The code contained a revised table of 68 calibrators, adding 3 stars to \citet{neves14} Table~2 (Gl~388, Gl~551, Gl~729). Among these calibrators, only 29 have $T_{\rm eff}$ and [Fe/H] values in \citet{mann15}. We therefore use these 29 stars with \citet{mann15} $T_{\rm eff}$ (ranging from 3056 to 3848\,K) and \citet{neves14} [Fe/H] (ranging from $-0.51$ to 0.19\,dex) to re-calibrate the matrices given in  \citet{neves14}. The median differences between \citet{mann15}- and \citet{neves14}-based calibrations are: $\Delta T_{\rm eff}=180\pm80$\,K and $\Delta {\rm [Fe/H]}=0.04\pm0.12$\,dex. Similarly, median differences between the \citet{mann15}-based calibration and \citet{rojas-ayala12} for 21 stars in common are: $\Delta T_{\rm eff}=240\pm170$\,K and $\Delta {\rm [Fe/H]}=0.08\pm0.11$\,dex. This confirms the offset of about 200~K in temperature and the fair agreement in metallicity.

The list of stars used in this comparison is given in Table~\ref{tab:calibrators_teff_feh} in Appendix. Spectroscopic binaries have been removed from this comparison: SB2 have double lines which probably affect the determination of the continuum, and there is a risk to mix both components in the measurements of the lines. SB1 are a priori more immune, but the secondary may affect the line depth, which is used in the determination of both effective temperature and metallicity in the \textsc{\small{mcal}} method.

\begin{figure}
	\includegraphics[width=\columnwidth]{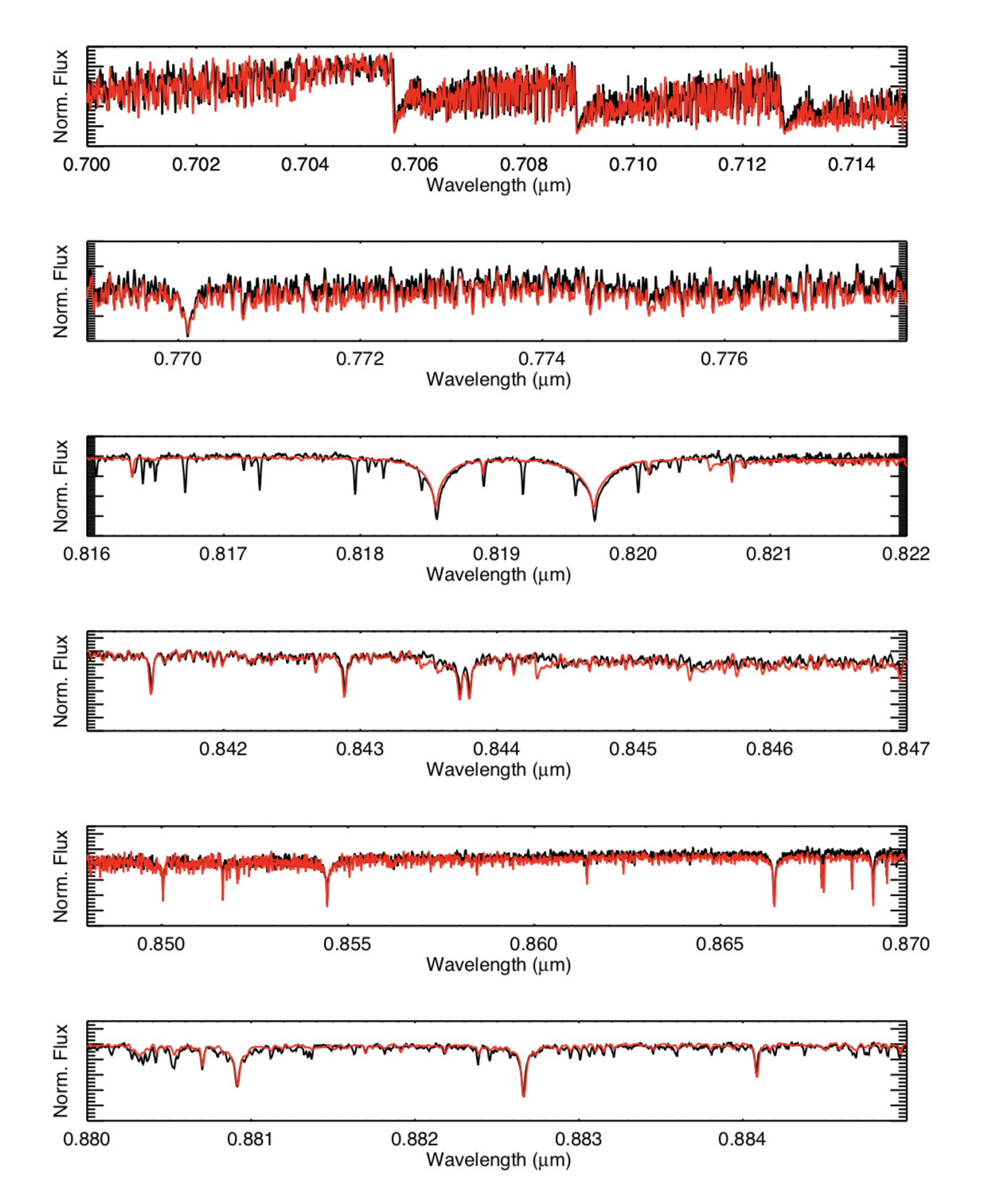}
    \caption{Comparison of an observed ESPaDOnS spectrum (in black) with the corresponding synthetic spectrum (in red) from a BT-Settl model for $T_{\rm eff}=3300$\,K, [Fe/H]=$-0.10$\,dex and $\log g=5.0$ between 700~nm and 885~nm.}
    \label{fig:1941096p}
\end{figure}

Promising new techniques to derive effective temperature, metallicity and gravity of M dwarfs have been pioneered by \citet{rajpurohit13}, using high-resolution stellar spectra and up-to-date model atmospheres. They look for the best combination of the 3 parameters used as an input to generate BT-Settl synthetic spectra which reproduce the observed spectra. We are in the process of applying this method described in \citet{rajpurohit17} to our spectra. Unfortunately, preliminary results show a good agreement only for effective temperatures, but no correlation for metallicities. An example of fitted spectrum is given in Figure~\ref{fig:1941096p}. A more thorough comparison of our results with BT-Settl synthetic spectra will be deferred to a future paper. 

Finally, a similar comparison using specific wavelength windows in which line parameters were corrected to provide an optimal fit to some standard stars with known parameters is also in progress (Kulenthirarajah et al., in prep.).

\subsection{Comparison of results}

\begin{figure}
	\includegraphics[width=\columnwidth]{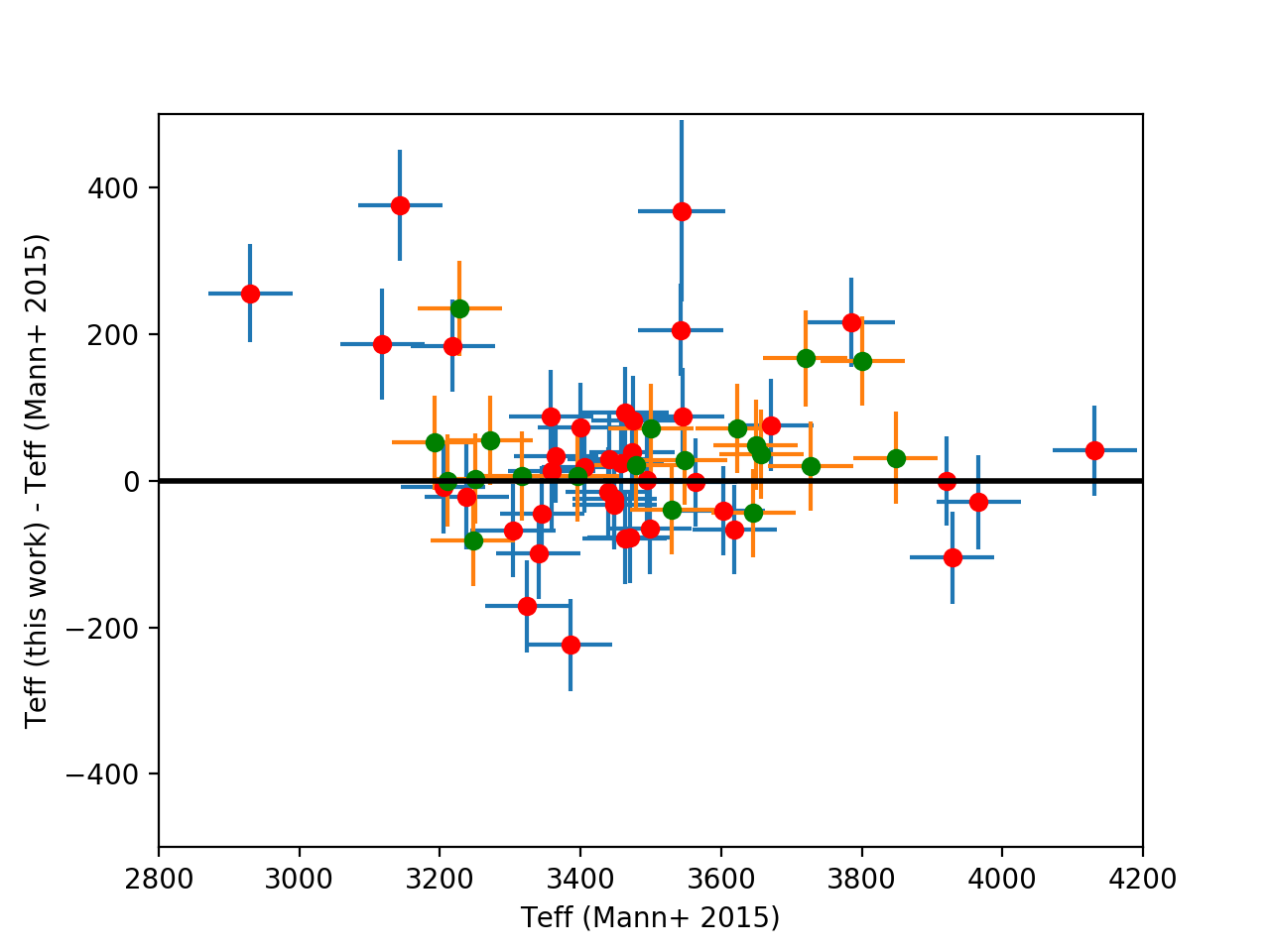}
    \caption{Difference between our effective temperatures and reference values from \citet{mann15}. Green points (with orange errorbars) correspond to stars used in the re-calibration of the \textsc{\small{mcal}} method, and red points (with blue errorbars) to additional stars.}
    \label{fig:diff_teff_Neves_Mann}
\end{figure}

Figure~\ref{fig:diff_teff_Neves_Mann} shows a comparison of our effective temperatures to corresponding values from \citet{mann15}. We adopt their uncertainty on $T_{\rm eff}$ as listed (typically 60\,K) and a quadratic sum of the uncertainty returned by \textsc{\small{mcal}} and a systematic uncertainty of 60\,K for our measurements, based on the observed dispersion between the two sets. The agreement is not surprising as our re-calibration of \textsc{\small{mcal}} method is based on 29 effective temperatures from \citet{mann15} (green points), but we have more measured stars (red points) and not all 29 calibrators have an ESPaDOnS spectrum. After rejecting 3 outliers from the sample (LHS~1723, Gl~297.2B and HH~And=Gl~905), the mean difference between the two systems computed from 57 stars is $T_{\rm eff}$ (this work) - $T_{\rm eff}$ (reference) = $20\pm12$\,K with an rms of 90\,K. Given that Mann's temperatures have a typical uncertainty of 60\,K, it shows that our effective temperatures should have a similar accuracy, and we therefore adopt a systematic uncertainty of 60\,K for our measurements. 

\begin{figure}
	\includegraphics[width=\columnwidth]{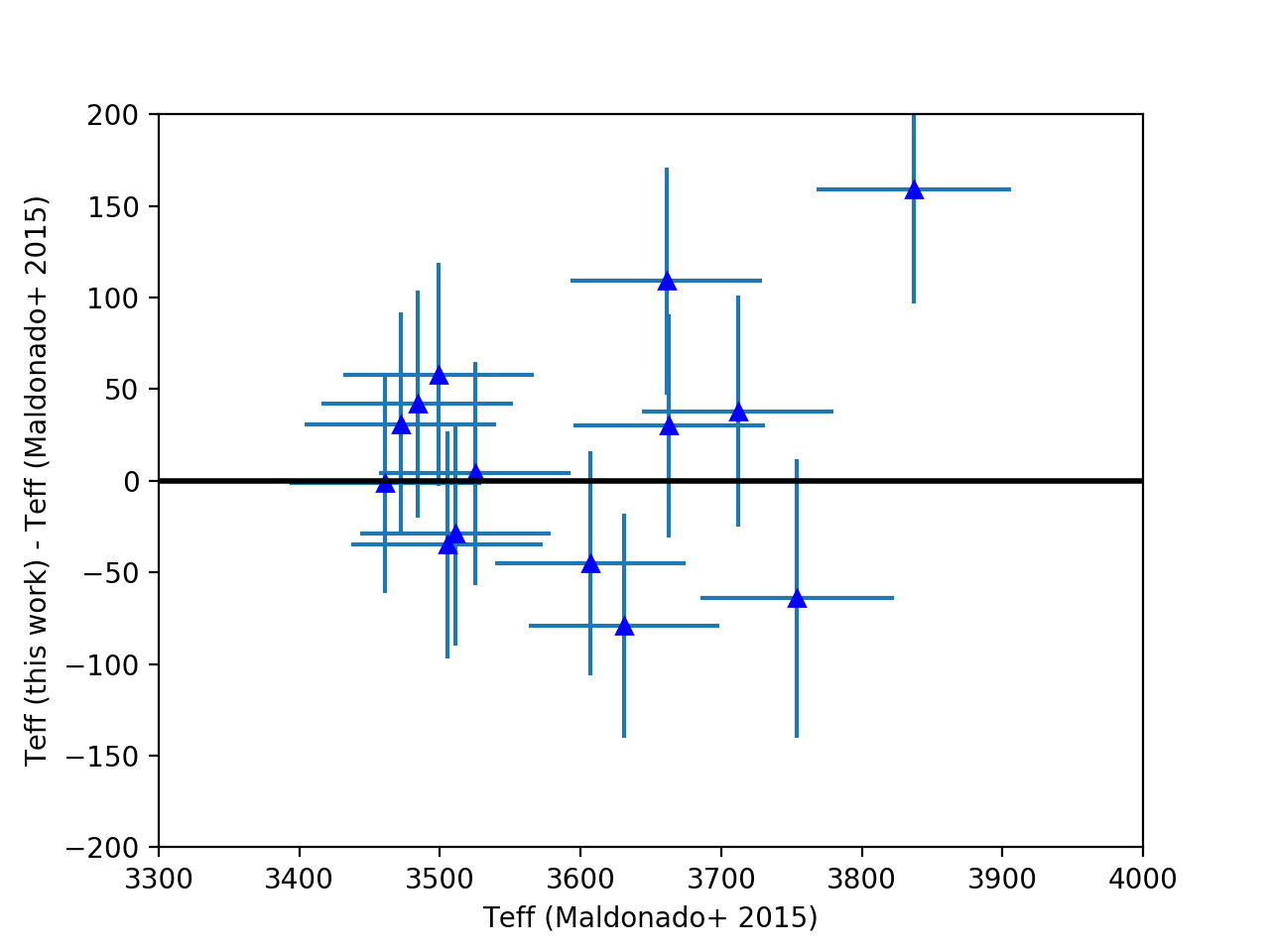}
    \caption{Difference between our effective temperatures and values from \citet{maldonado15}.}
    \label{fig:diff_teff_Neves_Maldonado}
\end{figure}

We also compare our results to the work of \citet{maldonado15}, who use a similar method to \textsc{\small{mcal}} to estimate effective temperatures and metallicities. We adopt their uncertainty on $T_{\rm eff}$ as listed (typically 68\,K) and a quadratic sum of the uncertainty returned by \textsc{\small{mcal}} and a systematic uncertainty of 60\,K for our measurements. Unfortunately, their sample is limited to early-type stars, but the agreement with our effective temperatures is also satisfactory, as can be seen in Figure~\ref{fig:diff_teff_Neves_Maldonado} (mean difference, this work minus \citet{maldonado15}: $+16\pm17$\,K, $\sigma=64$\,K).

\begin{figure}
	\includegraphics[width=\columnwidth]{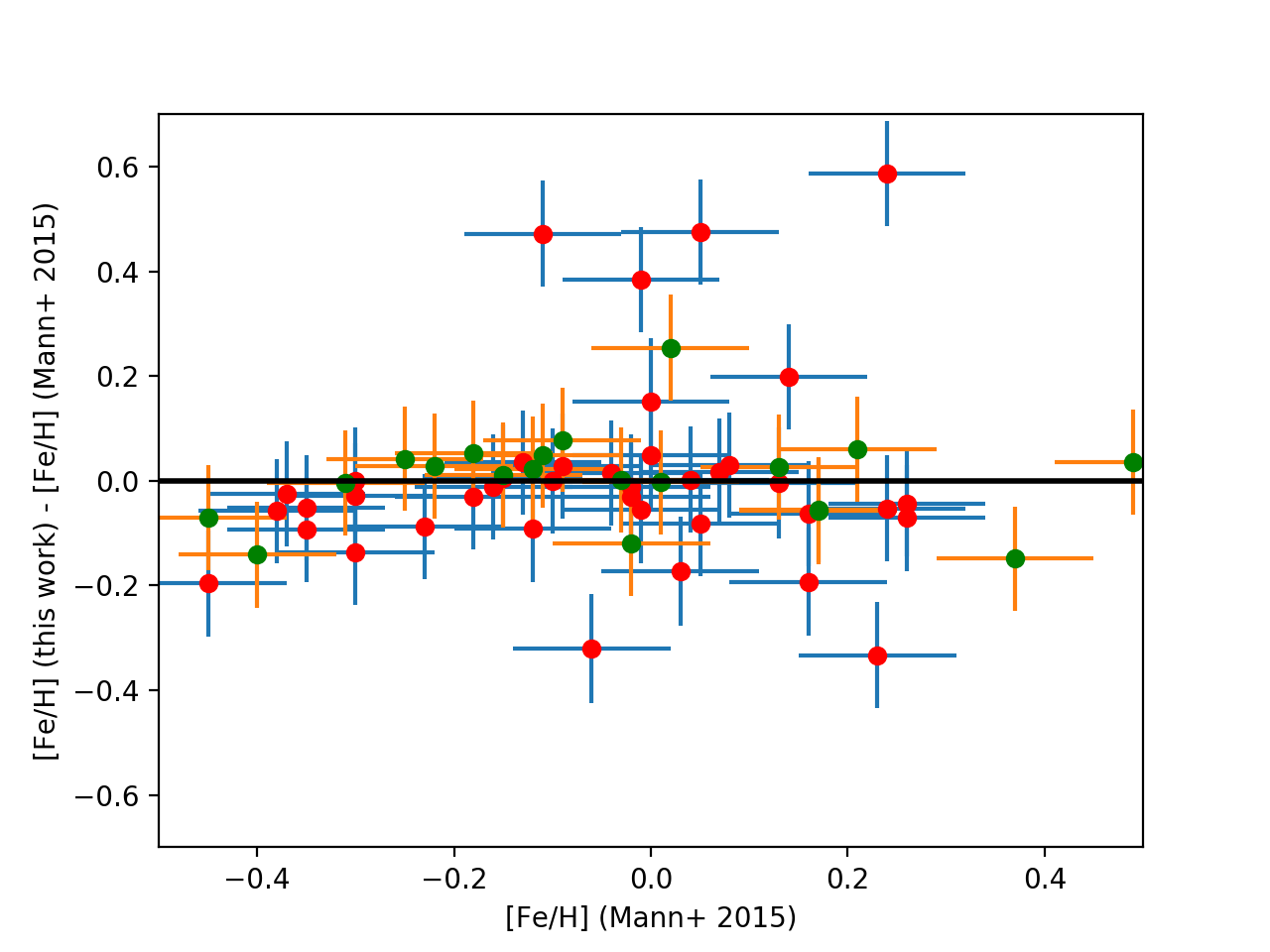}
    \caption{Difference between our metallicities and reference values from \citet{mann15}. Green points (with orange errorbars) correspond to stars used in the re-calibration of the \textsc{\small{mcal}} method, and red points (with blue errorbars) to additional stars.}
    \label{fig:diff_feh_Neves_Mann}
\end{figure}

For the metallicity comparison, Figure~\ref{fig:diff_feh_Neves_Mann} displays the results from \citet{mann15} compared to ours. We adopt an uncertainty on the [Fe/H] values from their paper (typically 0.08\,dex), and a quadratic sum of the uncertainty returned by \textsc{\small{mcal}} and a systematic uncertainty of 0.10\,dex, based on the observed dispersion between the two sets. This is a  more meaningful comparison than for effective temperatures, as Mann's metallicities have not been used in our re-calibration. It shows a generally good agreement, but some of our metallicities seem too high. These correspond to K7V-M0V stars, which have effective temperatures slightly out of our calibration domain. Rejecting the same 3 stars, the mean difference between the two systems is [Fe/H] (this work) - [Fe/H] (reference) = $0.014\pm0.020$\,dex with an rms of 0.15\,dex. Given that \citet{mann15} claim an accuracy of 0.08\,dex, our accuracy would be about 0.13\,dex. However, removing four K7V-M0V stars with discrepant metallicities still gives a negligible offset of $-0.021\pm0.011$\,dex, but with an rms of 0.08\,dex. We therefore adopt a systematic uncertainty of 0.10\,dex for our values of [Fe/H] when the effective temperatures fall within the limits of our calibration (3056 to 3848\,K), to be added quadratically to the generally negligible uncertainty returned by \textsc{\small{mcal}}.

\begin{figure}
	\includegraphics[width=\columnwidth]{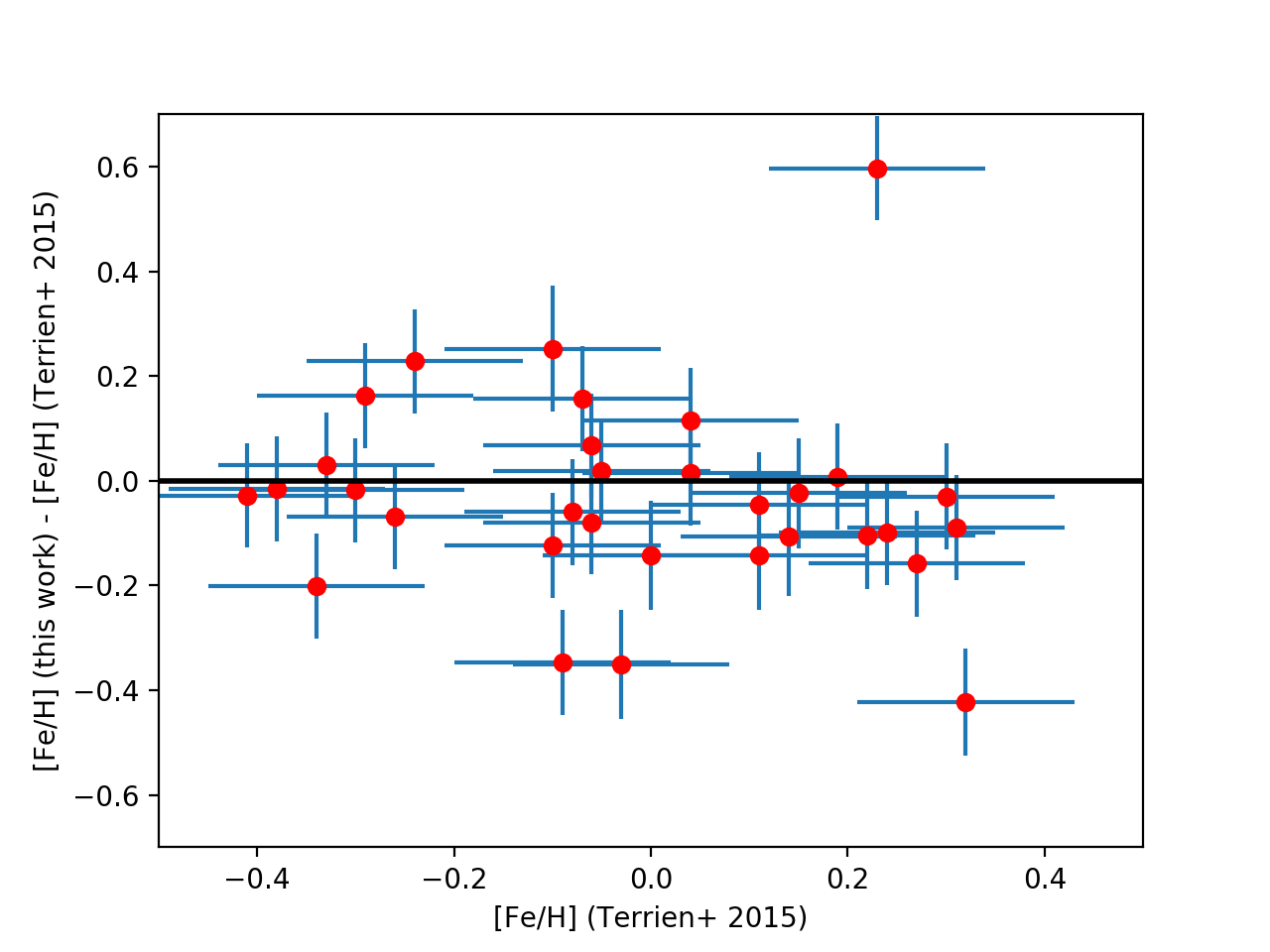}
    \caption{Difference between our metallicities and values from \citet{terrien15b}.}
    \label{fig:diff_feh_Neves_Terrien}
\end{figure}

A more independent comparison for metallicities has been made with the results obtained by \citet{terrien12}, who measure equivalent widths of Na, Ca and K lines in the near-infrared (H- and K-bands), and correct for effective temperature effects using $\rm H_2$O indices. Thirty-three non-active stars were found in common with \citet{terrien15b}, and the comparison is displayed in Figure~\ref{fig:diff_feh_Neves_Terrien}. We adopt a uniform uncertainty of 0.11\,dex on the [Fe/H] values from \citet{terrien15b}, as stated in their paper, and a quadratic sum of the uncertainty returned by \textsc{\small{mcal}} and a systematic uncertainty of 0.10\,dex. The agreement is satisfactory (mean difference, this work minus \citet{terrien15b}: $-0.055\pm0.026$\,dex, $\sigma=0.15$\,dex).

\section{Measure of the projected rotation velocity}\label{sec:vsini}
In order to measure the rotation of these stars from our polarimetric observations, we need a calibration of the rotational velocity of M dwarfs from the observed width of the LSD profile given by the Least-Squares Deconvolution (LSD) technique, described in \citet{donati97}.

We use M dwarfs of known $v \sin i$ from the literature for which high-resolution spectra have been obtained with ESPaDOnS, most of them from archival data and some from the CoolSnap program itself. We have combined both polarimetric and S+S spectra, assuming that the spectral resolution is the same (in fact 65,000 vs. 68,000).

\subsection{Sample and measurement techniques}
We based our compilation of $v \sin i$ values from the literature on the catalogue of 334 M dwarfs in \citet{reiners12}. We only retained stars with a measured value of $v \sin i$, not those with an upper limit. We then added a few stars from \citet{reiners07b}, \citet{donati08}, \citet{morin08b}, \citet{reiners09}, \citet{morin10} which were missing from the 2012 compilation. Very recently, \citet{reiners18} published a spectroscopic survey of 324 M dwarfs, where resolved values of $v \sin i$ are listed for 78 stars. This allowed us to revise old values of $v \sin i$ and add new calibrators.

Cross-matching the 440 M dwarfs observed with ESPaDOnS in our sample with the list of $v \sin i$ calibrators, we end up with 62 common stars with $v \sin i$ values ranging from 1.0 to 55.5\kms. Removing two stars which are SB2 (Gl~268 and Gl~735) gives 60 calibrators listed in Table~\ref{tab:calibrators_vsini}.

To calibrate our $v \sin i$ measurements, we used three approaches: 
a first approach uses a calibration of $v \sin i$ versus the observed width of the LSD profile, taking into account an intrinsic width which depends on the spectral type of the star. This is the approach adopted by \citet {delfosse98a} for the ELODIE spectrograph at OHP, \citet{melo01} for the FEROS spectrograph, \citet{boisse10} for SOPHIE at OHP, \citet{houdebine15} both for SOPHIE and HARPS at ESO, La Silla. We find that the intrinsic width, defined as the lower envelope of the observed width, slightly depends on the spectral type. However, it has to be recalled that we use a single template spectrum (mask) for all the stars that we correlate with the observed spectrum. So any mismatch between the actual spectral type of the star and the spectral type of the mask (M2) translates into a modification of the LSD profile.

A second approach uses an FeH line at 995.0334~nm to better estimate the intrinsic broadening of the line due to rotation. This line is  insensitive to gravity and magnetic field \citep{reiners07a} and should give a more direct comparison among stars of different spectral types than the LSD profiles. The measurement quality, however, is worse than when thousands of lines are used.

Finally, a third approach uses a few slow rotators for which the value of $v \sin i$ is known from the literature, and a high SNR polarimetric spectrum taken with ESPaDOnS exists. By broadening the LSD profile of a calibrator using different values of $v \sin i$ and comparing to the observed spectrum of a given star, we can then select the best calibrator and deduce the best value of $v \sin i$ reproducing the observed spectrum. This assumes that rotation is the main contributor to the width of the LSD profile, which means that we assume that convective turbulence and magnetic field broadening can be neglected. All methods better work for stars where the projected rotational velocity has a significant impact on the global line broadening.

\subsection{First approach: measure of the LSD profile}
The LSD software \citep{donati97} uses a line list built from an ATLAS9 local thermodynamic equilibrium model \citep{kurucz93a,kurucz93b} matching the properties of M2 stars, which contains about 5000 atomic lines weighed by their intensity. The multiplex gain is about 10 in signal-to-noise ratio.

\subsubsection{Variation of $\sigma_{\circ}$ with spectral type}
\begin{figure}
	\includegraphics[width=\columnwidth]{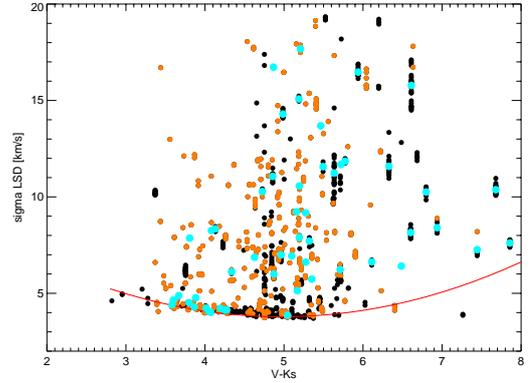}
    \caption{Variation of the LSD profile width with color, with the adopted lower envelope fit (red line). Black points correspond to polarimetric measurements and orange points to S+S spectra. Calibrators are marked with large cyan filled circles.}
    \label{fig:sigLSD_color}
\end{figure}

A necessary step in the calibration of $v \sin i$ from the width of the LSD profile is to estimate at each spectral type the minimum value of the width which can be measured. We measure the width of the LSD profile by fitting a gaussian profile and measuring the value of $\sigma$, and we use the $V-K_{\rm s}$ color as a quantitative estimate of the spectral type of the stars in our sample. We reject spectra with a SNR lower than 30. A diagram of $\sigma$ vs. $V-K_{\rm s}$ is displayed on Fig.~\ref{fig:sigLSD_color} and clearly shows an accumulation of points at small values of $\sigma$. The minimum value of $\sigma$ could be measured as the mode of the distribution in color bins. In practice, we fit a lower envelope by eye, and it can be seen that it fits both polarimetric measurements (black points) and S+S spectra (orange points). The minimum value of this lower envelope is about 4\kms, corresponding to a FWHM of the LSD profile of 9\kms. It is obtained at a $V-K_{\rm s}$ of about 5, corresponding roughly to an M4 spectral type. For earlier- or later- spectral types, the minimum values are higher.

Equation~\ref{eq:calsig0_LSD} describes the variation of $\sigma_{\circ}$ with the $V-K{\rm s}$ color for the LSD profile:

\begin{equation}
    \begin{alignedat}{3}
    \sigma_{\circ} = &11.39 &&-3.06\,(V-K{\rm s}) &&+0.308\,(V-K{\rm s})^2  \\
    &\pm0.32 &&\pm0.12 &&\pm0.011.
    \end{alignedat}
	\label{eq:calsig0_LSD}
\end{equation}

\subsubsection{Calibration of $v \sin i$ vs. $\sigma$}
Once we have an estimate of the intrinsic width $\sigma_{\circ}$ at a given color or spectral type, we need to subtract it quadratically from the measured width to get a measurement of the rotational broadening. As the intrinsic width is given by the lower envelope fitting the mode of the widths distribution, we are unable to measure the projected rotation velocity of slow rotators having a measured width similar or even smaller than the intrinsic width, due to measurement uncertainties. We discard these rotators in the calibration of $v \sin i$ vs. rotational broadening, and adopt an upper limit of 2\kms for their value of $v \sin i$. In summary, we define the rotational broadening as $\Delta$, given by Equation~\ref{eq:delta}:

\begin{equation}
    \Delta=\sqrt{\sigma^2-\sigma_{\circ}^2}.
	\label{eq:delta}
\end{equation}

Table~\ref{tab:calibrators_vsini} gives a list of the 60 stars used to calibrate these relations. Stars with an * have not been used in the calibration of the FeH relation (see below). When an uncertainty is not given in the reference of $v \sin i$, we adopt 10\% of $v \sin i$, with a minimum value of 1.5\kms. 

\begin{table*}
	\centering
	\caption{List of stars with known $v \sin i$ used to calibrate Equation~\ref{eq:calvsini_LSD} (all stars) and Equation~\ref{eq:calvsini_FeH} (except stars with *).}
	\label{tab:calibrators_vsini}
	\begin{tabular}{llccccccl}
	\hline
    \noalign{\vskip 0.1cm}
2MASS name & Common name & $V-K_{\rm s}$ &  $\sigma_{\circ}$ & $\langle\sigma\rangle$ & 
$\sqrt{\langle\sigma\rangle^2-\sigma_{\circ}^2}$ & 
Literature $v \sin i$ & Original error & Reference \\
    \hline
J01023895+6220422 & Gl~49 & 4.194 & 3.97 & 4.22 & 1.43 & 1 & & \citet{donati08} \\
J01031971+6221557 & Gl~51 & 5.635 & 3.92 & 11.24 & 10.53 & 12.0 & & \citet{morin10} \\
J01592349+5831162 & Gl~82 & 5.194 & 3.80 & 10.57 & 9.86 & 13.8 & & \citet{reiners12} \\
J02085359+4926565 & GJ~3136 & 4.867 & 3.79 & 16.73 & 16.29 & 24.1 & 2.4 & \citet{reiners18} \\
J02333717+2455392 & Gl~102 & 5.351 & 3.83 & 5.74 & 4.27 & 3.0 & 1.5 & \citet{reiners18} \\
J02515408+2227299 & & 5.208 & 3.80 & 17.68 & 17.26 & 27.2 & 2.7 & \citet{reiners18} \\
J03462011+2612560 & HD~23453 & 3.799 & 4.21 & 4.52 & 1.64 & 3.3 & 4.0 & \citet{reiners18} \\
J03472333-0158195 & G~80-21 & 4.626 & 3.82 & 6.87 & 5.71 & 5.2 & 1.5 & \citet{reiners18} \\
J04374092+5253372 & Gl~172 & 3.601 & 4.36 & 4.65 & 1.62 & 3.4 & 1.5 & \citet{reiners18} \\
J04593483+0147007 & Gl~182 & 3.807 & 4.20 & 7.86 & 6.65 & 10.4 & & \citet{reiners12} \\
J05082729-2101444 & & 5.832 & 4.02 & 23.43 & 23.09 & 25.2 & 2.5 & \citet{reiners18} \\
J05312734-0340356 & Gl~205 & 3.866 & 4.16 & 4.29 & 1.06 & 1.5 & & \citet{reiners07a} \\
J05363099+1119401 & Gl~208 & 3.669 & 4.30 & 4.88 & 2.31 & 3.8 & 1.5 & \citet{reiners18} \\
J06000351+0242236 & GJ~3379 & 5.274 & 3.81 & 6.62 & 5.41 & 4.9 & 1.5 & \citet{reiners18} \\
J06103462-2151521 & Gl~229 & 4.016 & 4.06 & 4.23 & 1.19 & 1.0 & & \citet{reiners07a} \\
J07444018+0333089 & Gl~285 & 5.321 & 3.82 & 7.71 & 6.70 & 4.0 & 1.5 & \citet{reiners18} \\
J08115757+0846220 & Gl~299 & 5.169 & 3.80 & 5.13 & 3.45 & 3.0 & 1.7 & \citet{delfosse98a} \\
J08294949+2646348 & GJ~1111 & 7.680 & 6.06 & 10.37 & 8.42 & 10.5 & 1.5 & \citet{reiners18} \\
J08313744+1923494 & GJ~2069B & 4.081 & 4.03 & 8.26 & 7.21 & 6.5 & 1.7 & \citet{delfosse98a} \\
J09002359+2150054 & LHS~2090 & 7.503 & 5.77 & & & 14.3 & 1.5 & \citet{reiners18} \\
J09142485+5241118 & Gl~338B & 3.584 & 4.38 & 4.41 & 0.52 & 2.3 & 1.5 & \citet{reiners18} \\
J09445422-1220544 & G~161-71 & 6.149 & 4.22 & 23.67 & 23.29 & 31.2 & 3.1 & \citet{reiners18} \\
J10121768-0344441 & Gl~382* & 4.250 & 3.94 & 4.13 & 1.23 & 1.8 & & \citet{reiners07a} \\
J10193634+1952122 & Gl~388 & 4.871 & 3.79 & 6.00 & 4.65 & 3.0 & & \citet{reiners07a} \\
J10285555+0050275 & Gl~393 & 4.276 & 3.93 & 4.14 & 1.28 & 1.5 & & \citet{reiners07a} \\
J10481258-1120082 & GJ~3622 & 7.858 & 6.36 & 7.61 & 4.17 & 2.1 & 1.5 & \citet{reiners18} \\
J10562886+0700527 & Gl~406 & 7.444 & 5.68 & 7.26 & 4.53 & 3.0 & & \citet{reiners07b} \\
J11023832+2158017 & Gl~410 & 3.884 & 4.15 & 4.77 & 2.36 & 2.6 & 1.5 & \citet{reiners18} \\
J11053133+4331170 & Gl~412B* & 6.611 & 4.62 & 15.78 & 15.09 & 8.2 & 2.7 & \citet{reiners18} \\
J11314655-4102473 & Gl~431 & 4.986 & 3.79 & 14.29 & 13.78 & 20.5 & & \citet{reiners12} \\
J12141654+0037263 & GJ~1154 & 6.110 & 4.19 & 6.63 & 5.14 & 6.0 & & \citet{reiners09} \\
J12185939+1107338 & GJ~1156 & 6.328 & 4.36 & 11.58 & 10.73 & 15.5 & 1.6 & \citet{reiners18} \\
J13003350+0541081 & Gl~493.1 & 5.774 & 3.99 & 11.85 & 11.16 & 16.4 & 1.6 & \citet{reiners18} \\
J13004666+1222325 & Gl~494 & 4.131 & 4.00 & 8.34 & 7.32 & 9.7 & & \citet{browning10} \\
J13295979+1022376 & Gl~514 & 4.049 & 4.05 &  4.06 &  0.32 & 1.5 & & \citet{reiners07a} \\
J13314666+2916368 & GJ~3789* & 5.273 & 3.81 & 47.89 & 47.74 & 55.5 & 8.4 & \citet{reiners18} \\
J13454354+1453317 & Gl~526 & 4.075 & 4.03 & 4.01 & & 2.0 & & \citet{reiners07a} \\
J14172209+4525461 & & 5.465 & 3.86 & 13.69 & 13.13 & 15.9 & 1.6 & \citet{reiners18} \\
J15215291+2058394 & GJ~9520 & 4.337 & 3.91 & 6.12 & 4.71 & 4.3 & 1.5 & \citet{reiners18} \\
J15303032+0926014 & NLTT~40406 & 6.485 & 4.50 & 6.41 & 4.57 & 16.3 & 1.6 & \citet{reiners18} \\
J15553178+3512028 & G~180-11 & 5.617 & 3.92 & & & 21.9 & & \citet{jenkins09} \\
J16301808-1239434 & Gl~628 & 5.043 & 3.79 & 3.87 & 0.77 & 1.5 & & \citet{reiners07a} \\
J16352740+3500577 & GJ~3966 & 5.163 & 3.80 & & & 15.8 & & \citet{reiners12} \\
J16553529-0823401 & Gl~644C & 6.798 & 4.82 & 10.26 & 9.05 & 5.4 & 1.5 & \citet{reiners18} \\
J16570570-0420559 & GJ~1207 & 5.159 & 3.80 & 9.23 & 8.41 & 10.7 & & \citet{reiners12} \\
J18021660+6415445 & G~227-22 & 5.721 & 3.96 & 11.68 & 10.99 & 11.3 & 1.5 & \citet{reiners18} \\
J18073292-1557464 & GJ~1224 & 5.711 & 3.96 & 6.23 & 4.81 & 2.2 & 1.5 & \citet{reiners18} \\
J18130657+2601519 & GJ~4044 & 5.193 & 3.80 & 7.89 & 6.92 & 5.9 & 1.5 & \citet{reiners18} \\
J18185725+6611332 & GJ~4053 & 5.495 & 3.87 & 11.57 & 10.90 & 15.3 & 1.5 & \citet{reiners18} \\
J19165762+0509021 & Gl~752B & 6.937 & 4.98 & 8.40 & 6.76 & 2.7 & 2.2 & \citet{reiners18} \\
J19510930+4628598 & GJ~1243 & 5.188 & 3.80 & 15.08 & 14.59 & 22.5 & 2.3 & \citet{reiners18} \\
J19535508+4424550 & GJ~1245B & 6.603 & 4.61 & 8.15 & 6.72 & 7.0 & & \citet{reiners07b} \\
J20294834+0941202 & Gl~791.2* & 5.757 & 3.98 & 21.17 & 20.80 & 32.0 & 2.0 & \citet{delfosse98a} \\
J22011310+2818248 & GJ~4247* & 5.228 & 3.81 & 21.42 & 21.08 & 35.4 & 3.5 & \citet{reiners18} \\
J22464980+4420030 & Gl~873 & 4.963 & 3.79 & 7.00 & 5.88 & 3.5 & 1.5 & \citet{reiners18} \\
J22515348+3145153 & Gl~875.1 & 4.726 & 3.80 & 10.28 & 9.55 & 13.4 & 1.5 & \citet{reiners18} \\
J23292258+4127522 & GJ~4338B* & 5.274 & 3.81 & 9.17 & 8.34 & 14.5 & & \citet{reiners12} \\
J23315208+1956142 & Gl~896A & 4.857 & 3.79 & 11.06 & 10.39 & 17.5 & & \citet{morin08b} \\
J23315244+1956138 & Gl~896B* & 5.938 & 4.08 & 16.47 & 15.96 & 24.2 & 1.4 & \citet{delfosse98a} \\
J23545147+3831363 & & 5.097 & 3.79 & 6.94 & 5.81 & 3.6 & 1.5 & \citet{reiners18} \\
    \hline
	\end{tabular}
\end{table*}

\begin{figure}
	\includegraphics[width=\columnwidth]{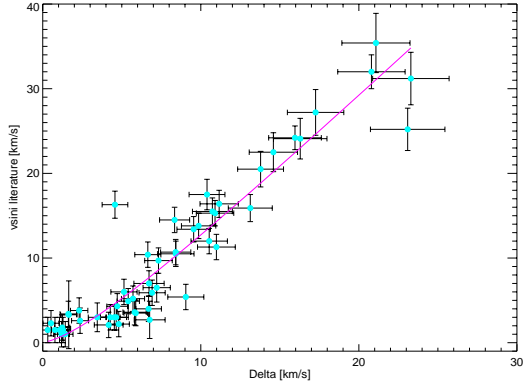}
    \caption{Literature value of $v \sin i$ with respect to the measured width attributed to rotation. A fit given by Equation~\ref{eq:calvsini_LSD} is overplotted.}
    \label{fig:vsiniLSD_calib}
\end{figure}

We then plot the literature measurements of $v \sin i$ vs $\Delta$ in Fig.~\ref{fig:vsiniLSD_calib}. The largest rotator (GJ~3789 at $v \sin i = 55.5$\kms) does not fit well the trend and is then rejected in order not to bias the calibration. Gl~412B is a clear outlier (strong magnetic slow rotator) and is removed too before the fit. Finally, we could not measure the value of $\sigma$ for 3 stars because their spectra have an SNR smaller than 30, and one star has a $\sigma$ value slightly smaller than the adopted $\sigma_{\circ}$ for its color. 

The functional shape of the fitting curve is not exactly linear: at large values of $\Delta$ we want $v \sin i$ proportional to $\Delta$ and at small values of $\Delta$ we want small $v \sin i$. We adopt the following function:

\begin{equation}
    v \sin i = \Delta \ \frac{a \ \Delta + b}{\Delta + c}.
	\label{eq:calvsini_LSD}
\end{equation}

Resulting values of $a$, $b$ and $c$ in Equation~\ref{eq:calvsini_LSD} over 54 calibrators are:

\begin{eqnarray} \nonumber
a &=& 1.75\pm0.06,\\ \nonumber
b &=& 2.10\pm0.68,\\ \nonumber
c &=& 5.41\pm0.63.
\end{eqnarray}

This gives a reasonable fit valid up to about 40\kms, displayed in Fig.~\ref{fig:vsiniLSD_calib}.

\subsection{Second approach: measure of the 995.0334~nm FeH line}
We have selected two FeH lines recommended by \citet{reiners07a}, because the continuum is well defined around 1$\mu$m and these two lines are insensitive to gravity and magnetic effects. However, \citet{reiners07a} used the Coud\'e Echelle Spectrograph (CES) at La Silla Observatory (Chile), which has a resolution of 200,000. ESPaDOnS in polarimetric mode has a typical resolution of 65,000 and one of the Reiners' line is blended in our spectra. We therefore only measure the FeH line at 995.0334~nm (air wavelength), which is very well defined in most of our spectra.

We fit a gaussian with a linear baseline to this line, and estimate the quality of the fit using various criteria. In some cases, the fit produces spurious results, for instance for spectroscopic binaries, low SNR spectra, K dwarfs where the FeH lines tend to disappear, ... The criteria are:
\begin{itemize}
    \item the wavelength shift with respect to the expected position must be smaller than 0.02~nm.
    \item the value of the $\chi^2$ per d.o.f. must be smaller than 0.7.
    \item the signal must be in absorption and its amplitude must be large enough compared to the noise: after fitting the gaussian profile, we subtract it from the spectrum and measure the residual noise: we accept a line if the ratio of its amplitude to the noise is larger than 3.
    \item finally, we reject the fit when the $\sigma$ is smaller than 1 pixel or much larger than the corresponding $\sigma$ of the LSD profile by a factor 3.
\end{itemize}

With these criteria, about 865 of our 1900 spectra provide a valuable fit of the 995.0334~nm FeH line.

\begin{figure}
	\includegraphics[width=\columnwidth]{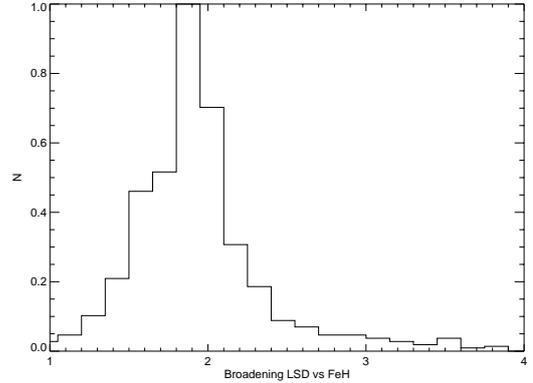}
    \caption{Histogram of the broadening of the LSD profile with respect to the 995.0334~nm FeH line, displayed as the ratio of the widths.}
    \label{fig:sigma_broad}
\end{figure}

A comparison of the LSD profile widths $\sigma$ to the corresponding values for the FeH line is shown as a histogram of the corresponding broadening in Fig.~\ref{fig:sigma_broad}, displayed as the ratio of the widths. It appears that in average the LSD profile is about twice larger than a single FeH line. We checked that this ratio does not significantly depend of the color of the star.

This result is confirmed by an analysis of a BT-Settl synthetic spectrum at $T_{\rm eff} = 3500$~K, [Fe/H]=0.0 and $\log g = 5.0$, where we measure an average line width of 0.24~nm for 3 Ti~I lines around 974~nm, and 0.11~nm for 2 FeH lines around 993~nm. A possible interpretation of this difference in line widths between atomic lines and molecular FeH lines comes from the low dissociation energy of FeH, namely 1.63~eV. So the molecule will be dissociated in regions where the turbulence is strong. A quick calculation gives a corresponding collision velocity of 2.4\kms. Higher velocity collisions would destroy the molecule and reduce the pressure broadening accordingly.

A similar diagram to Fig.~\ref{fig:sigLSD_color} for the FeH line is displayed in Fig.~\ref{fig:sigFeH_color} and shows a lower envelope which is flatter than for the LSD profile width and not defined very accurately, as a single line measurement is noisier than the LSD profile. This envelope is fit by Equation~\ref{eq:calsig0_FeH}:

\begin{figure}
	\includegraphics[width=\columnwidth]{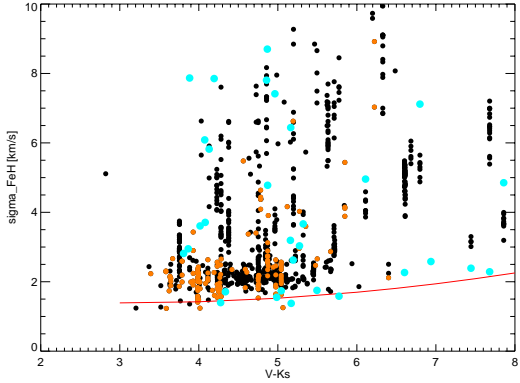}
    \caption{Variation of the 995.0334~nm FeH line width with color, with the adopted lower envelope fit (red line). Symbols are the same as in Fig.~\ref{fig:sigLSD_color}.}
    \label{fig:sigFeH_color}
\end{figure}

\begin{equation}
    \begin{alignedat}{3}
    \sigma_{\circ} &= 1.72 &&-0.215\,(V-K{\rm s}) &&+0.0352\,(V-K{\rm s})^2 \\
    &\pm0.48 &&\pm0.187 &&\pm0.0168.
	\end{alignedat}
	\label{eq:calsig0_FeH}
\end{equation}

\begin{figure}
	\includegraphics[width=\columnwidth]{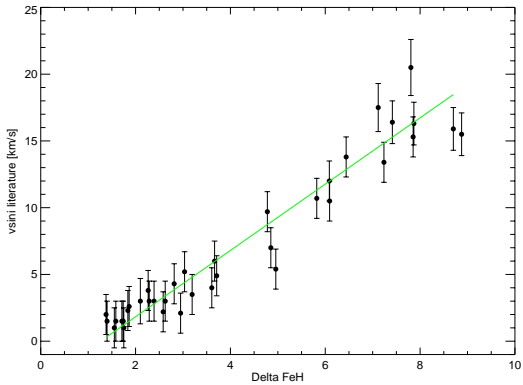}
    \caption{Literature value of $v \sin i$ with respect to the measured width of the 995.0394~nm FeH line, attributed to rotation. A linear fit is used for the calibration.}
    \label{fig:vsiniFeH_calib}
\end{figure}

The corresponding calibration of $v \sin i$ versus $\Delta$ as defined in Equation~\ref{eq:delta} can be fit by a similar formula to Equation~\ref{eq:calvsini_LSD}, but we find in practice that a linear fit is accurate enough. It is given by Equation~\ref{eq:calvsini_FeH} and shown in Fig.~\ref{fig:vsiniFeH_calib}:

\begin{equation}
    \begin{alignedat}{2}
    v \sin i = &-3.14 &&+2.48\,\Delta \\
    &\pm0.59 &&\pm0.13.
	\end{alignedat}
	\label{eq:calvsini_FeH}
\end{equation}

\subsection{Third approach: convolution with slow rotator templates}

A different technique consists in using a few slow rotators with high SNR spectra obtained with ESPaDOnS and for which the value of $v \sin i$ is well measured by high SNR spectra at higher resolution. The method is described in details in \citet{malo14b} and uses 6 calibrators, ranging in spectral type from M1.0 to M3.5, listed in Table~\ref{tab:template_vsini}. The reference values of $v \sin i$ all come from \citet{reiners07a}, who used very high resolution spectra (200,000) from the CES spectrograph at La Silla Observatory, which ensures reliability and homogeneity. The SNR of the ESPaDOnS spectrum used as template is given in the last column of Table~\ref{tab:template_vsini}, and is measured per CCD pixel at 810~nm on the intensity spectrum.

For each calibrator, we artificially broaden its spectrum using different values of $v \sin i$, and for each star in our sample, we look for the best fit of its spectrum among the library of broadened spectra of the calibrators. We then adopt as the value of $v \sin i$ for this spectrum the best match.

\begin{table}
	\centering
	\caption{List of slow rotators used as templates in the alternative technique, with their adopted values of $v \sin i$ in \kms, and the signal-to-noise ratio of the spectrum.}
	\label{tab:template_vsini}
	\begin{tabular}{cccc}
		\hline
	    Common name & Spectral type & $v \sin i$ & SNR\\
		\hline
		Gl~273 & M3.5 & 1.0 & 499\\
		Gl~382 & M1.5 & 1.8 & 297\\
		Gl~393 & M2.0 & 1.5 & 356\\
		Gl~514 & M1.0 & 1.5 & 293\\
	    Gl~526 & M1.5 & 2.0 & 433\\
	    Gl~628 & M3.5 & 1.5 & 219\\
		\hline
	\end{tabular}
\end{table}

A comparison of the results of this technique with the value of $v \sin i$ calibrated from the measure of the width of the LSD profile gives a good agreement at intermediate projected rotation velocity (typically from 4 to 30\kms). For slower rotators, there are differences due both to the calibration of $\sigma_{\circ}$ for the LSD profile method, and the adopted template $RV$ for the template method. For rapid rotators (and a few specific stars such as Gl~412B), non gaussian LSD profiles affect both methods and lead to differences between the two approaches too.

\subsection{Adopted projected rotation velocity}

From the 3 methods exposed above, we adopt a value of $v \sin i$ which is defined as follows, where $v \sin i_{\rm LSD}$ is obtained from the calibrated LSD intensity profile, $v \sin i_{\rm FeH}$ from the FeH line and $v \sin i_{\rm c}$ from the template convolution:
\begin{itemize}
\item All three methods are used and compared for each star, with the goal of obtaining a single value per star with an error bar representative of data quality,  measurement dispersion and calibration uncertainties. 
\item The median value of the three measurement is adopted, when $v \sin i_{\rm LSD}$ is larger than 3\kms (resolved profiles) and $v \sin i_{\rm FeH}$ is measured.
\item When $v \sin i_{\rm LSD}$ is smaller than 3\kms (unresolved profiles),  $v \sin i_{\rm c}$ is not included in the adopted value calculation.
\item When $v \sin i_{\rm LSD}$ was found smaller than 2\kms, we estimate that the rotation profile is unresolved in ESPaDOnS spectra and such values are reported as "<2".
 \end{itemize} 

For stars with a strong magnetic field, $v \sin i_{\rm FeH}$ from the FeH line should be preferred over the other two methods, as it is insensitive to the magnetic field. However, as the measurement is based on a single line it is more noisy, and in addition these stars are generally rapid rotators, which makes the line blended with nearby lines.

\section{Discussion}\label{sec:discussion}

\subsection{Comparison between projected and equatorial rotation velocities}
We found about 150 stars in our sample with a known rotation period, either measured from time series photometry or spectroscopy of chromospheric indicators \citep{suarez15}. We did not use rotation periods deduced from spectroscopic measurements when they are converted from chromospheric indicators such as $R'_{\rm HK}$ or projected rotation velocities $v \sin i$. Uncertain values are given in parentheses. From this period and the adopted radius of the star we compute the equatorial rotation velocity, using $v_{\rm eq} = 50.59 \ R/P_{\rm rot}$, where $v_{\rm eq}$ is in \kms, $R$ in \Rnom (assumed to be 695,700~km from \citet{prsa16}), and $P_{\rm rot}$ in days. An alternative approach pioneered e.g. by \citet{donati08} consists in comparing $R \sin i$ to the adopted radius, under the hypothesis that $v \sin i$ is measured more accurately than $R$, at least for rapid rotators. In our case, $v \sin i$ depends on the adopted calibrations and averaging process, so it is probably not more accurate than the star radius, which is estimated from the star color $V-J$ by a relation that we calibrated on interferometrically measured radii from \cite{boyajian12}.

We divided our sample into two parts; the slow rotators ($v_{\rm eq}<3$\kms), for which we want to check that small equatorial velocities are confirmed by a small value of $v \sin i$ from our measurement, and the resolved rotators, for which our measurement of $v \sin i$ should be smaller than the computed $v_{\rm eq}$. Both tables are given in Appendix.

We confirm that the calculus of $v_{\rm eq}$ from the estimated radius and measured $P_{\rm rot}$ agrees with our measured value of $v \sin i$ for average inclinations: about two-thirds of the expected slow rotators are not resolved with our spectrograph ($v \sin i < 2$\kms). Those having a measured value of $v \sin i$ may indicate that our calibration is slightly inaccurate (supposedly resolved projected rotation velocities are in fact upper limits). In a few cases, it may be due to a metallicity effect in the calibration of Equation~\ref{eq:calsig0_LSD} and Equation~\ref{eq:calsig0_FeH}, which has not been taken into account and may affect metal-poor and metal-rich stars (see \citet{melo01} for an explanation of this expected effect).

However, in about half the cases of resolved rotators, $v \sin i$ taken at face value is larger than  $v_{\rm eq}$. It is not unexpected that the distribution of $\sin i$ is biased toward larger values, as there is an observational bias against low inclination systems where photometric variations are more difficult to detect. However, the magnitude of the effect is too large to be attributed to this bias. This surprising effect has already been evidenced by e.g. \citet{reiners12} (see their Fig.~10), who attribute it to possibly inaccurate photometric rotation periods. We can also add inaccurate radii, for instance for young stars, as we use a mean relation only valid for old stars. But these inaccuracies can probably only explain a few cases, not the majority.

\subsection{From fundamental properties to radial velocity uncertainty}
\begin{figure}
	\includegraphics[width=\columnwidth]{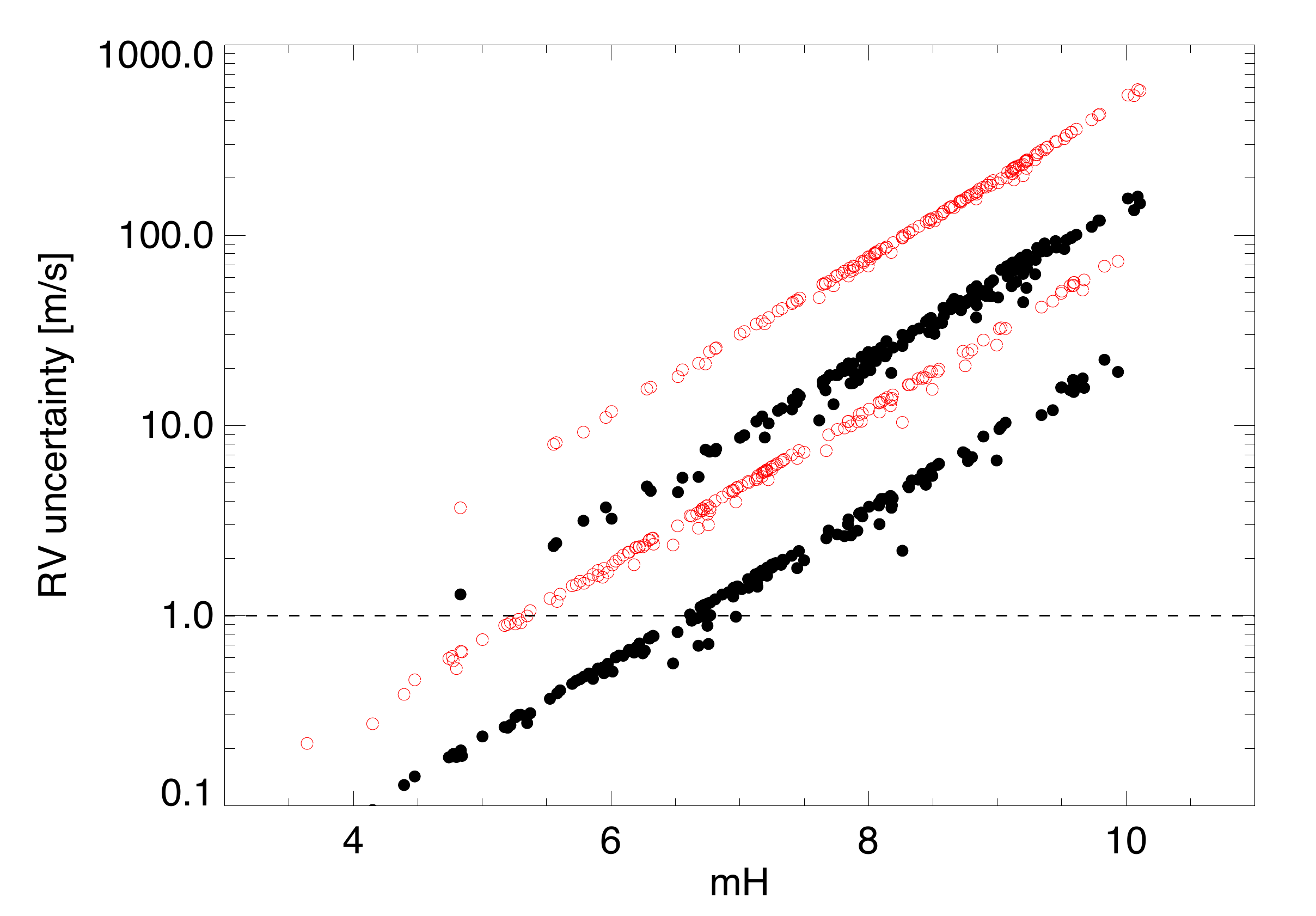}
    \caption{The expected radial-velocity uncertainty that would be achieved with SPIRou in 600s exposures, as a function of the stellar magnitude. A range of values for a given magnitude is obtained, depending on the performance of telluric corrections (from black: optimistic to red: conservative). The two different black and red sequences roughly mimic rapid- (10\kms, upper sequence) and slow- (1\kms, lower sequence) rotator cases. The horizontal line shows a realistic noise floor for such observations.}
    \label{fig:rv}
\end{figure}

Using the measured effective temperatures and collected apparent magnitudes in the $H$ band, it was then possible to estimate the potential of SPIRou observations for this sample of stars. When effective temperatures were not available, we used firstly the values from \citet{mann15}, then their Equation 7 deriving $T_{\rm eff}$ from the $V-J$ colour index and a correction for unknown metallicity based on the $J-H$ colour index, with coefficients given in their Table 2. We then use an exposure simulator for SPIRou to estimate the signal-to-noise ratio obtained in a typical visit of 600s integration time, with median seeing conditions of Maunakea (0.6" in the $H$ band). Then, from the SNR estimates, we used the $RV$ content as calculated in \citet{figueira16} to estimate the range of $RV$ uncertainties per visit. The quantity depends upon the rotational velocity, the effective temperature, and the performance of telluric corrections, in addition to the SNR. In Figure \ref{fig:rv}, we show two extreme conditions for each star where an effective temperature and rotational velocity are available: the conservative configuration where all regions contaminated by telluric lines more than 2\% depth are masked, and the optimistic configuration where these telluric lines are almost completely corrected for (see details in \citet{figueira16}, their cases 2 and 3). It is difficult, at this point, to predict where telluric corrections with SPIRou will stand: the proposed method is a PCA-based approach using a library of observed telluric spectra in varying conditions \citep{artigau14}; its performance in real conditions still needs to be assessed.  As a first estimate, we used the $RV$ uncertainty calculated for a rotational velocity of 1 (resp., 10)\kms for all stars having a $v \sin i$ less than (resp., greater than) 5\kms, which explains why data points are not covering the parameter space randomly.

Finally, as the $RV$ uncertainty is derived by photometric band, we applied the correction factor found for Barnard's star between models and observations of this M4 star (Artigau et al, subm.). These correction factors enhance the contribution of the $H$ and $K$ bands with respect to the bluer part of the spectrum; it is not yet known how they vary across the spectral type of M stars and with their metallicity.

Figure \ref{fig:rv} shows that an $RV$ uncertainty of 1 m/s is achieved for all slowly rotating stars brighter than an $H$ magnitude of 7 in 600s. Fainter stars, or faster rotators, would need a longer exposure time to achieve this precision. When the conservative approach of telluric masking is used, the limit drops by almost two magnitudes, showing the importance of devoting telescope time and pipeline development efforts to recover the stellar signal in these contaminated area. Finally, it seems that stars rotating at more than 10\kms will never achieve the 1\ms level, even when perfect telluric corrections are applied, down to an $H$ magnitude of 4.5. This must be taken into account when considering the targets for planet searches. 

\subsection{Multiplicity and planet formation}
Among the 153 systems listed in Table~\ref{tab:vis-bin}, more than half (88) have an apparent separation smaller than 2.0", preventing in most cases a clear separation of the two components with our instruments, the fiber of which have diameters of 1.6" (ESPaDOnS) and 1.2" (SPIRou). \citet{thebault14} warn that radial velocity surveys aiming at exoplanet detection reject binary systems and therefore prevent from getting information about the planet formation in such systems. They mention a physical separation of about 100~AU below which the planet formation is affected. For the above mentioned limit in angular separation (2.0"), this corresponds to a distance from Earth to the multiple system of 50~pc.

A more complete statistics has been drawn from our catalog of multiple systems involving an M dwarf. Among 669 systems, 111 have a physical separation smaller than 100~AU (assuming they are all physical systems). Among those, 28 are close enough to have an angular separation larger than 2.0". This means that our observational constraints typically reject 75\% of the interesting sample where planet formation may be affected by the binarity.

Spectroscopic binaries are also rejected from most samples, especially SB2. In our sample of 440 M dwarfs we listed 55 SB2 already known or discovered by us, a rate of 12.5\%. About a third of them are also close visual binaries (angular separation smaller than 2.0"), allowing a good determination of their physical properties. 

In summary, about 80\% of interesting multiple systems for constraining the planet formation mechanism are lost to the size of the spectrograph fibers, linked to the atmospheric seeing.

\section{Summary and Conclusion}\label{sec:conclusion}

In this paper, we have been reporting on a sample of 440 M dwarfs observed with the ESPaDOnS spectro-polarimeter at CFHT. 114 of them correspond to observations conducted by our team in the framework of the CoolSnap collaboration. Two other papers \citep{moutou17} and Malo et al., in prep. report additional results from this program. Another 71 stars observed in polarimetric mode and 255 in spectroscopic mode (S+S) were extracted from the ESPaDOnS archive at CADC and cover the whole set of observations of M dwarfs conducted at CFHT between 2005 and 2015.

From this homogeneous set of observations, we measured spectral type using the TiO~5 index, effective temperatures and metallicities using the \textsc{\small{mcal}} method when the star is not active (H$\alpha$ index smaller than 0.25, see Section~\ref{sec:st-teff-feh}). We checked that our values generally agree with measurements obtained from similar or different methods in the literature.

As part of a larger project to identify multiple systems involving M dwarfs, we list all the stars in our sample belonging to such system, without limit on the separation. We also identify new spectroscopic binaries from our observations and summarize those already known from the literature.

We calibrate the measurement of the projected rotation velocity from the width of the LSD profile. This calibration is valid for other observations of late-type dwarfs observed with the ESPaDOnS spectro-polarimeter.

Finally, we estimate the radial-velocity content for each star of our sample, in order to select those which are expected to display the smallest radial velocity uncertainty possible with SPIRou. This work participates to the effort of selecting the targets for low-mass planet search using the new high-velocity precision near-infrared spectro-polarimeter SPIRou. In the first paper, \citet{moutou17} defined a merit function based on the star activity; in the present paper, we discarded close binaries and estimated the expected radial-velocity uncertainty; in the final paper of this series, Malo et al., in prep. uses the present measurements of $T_{\rm eff}$ and [Fe/H] to refine the planet-detection merit function used to define the initial sample, and combines it to the other merit function and selection criteria to finally select the best sample of targets for the new SPIRou instrument.

\section*{Acknowledgements}
The authors wish to recognize and acknowledge the very significant cultural role that the summit of Maunakea has always had within the indigenous Hawaiian community. We are most grateful to have the opportunity to conduct observations from this mountain. 

We are sincerely grateful to the anonymous referee, whose careful reading and suggested additions and corrections helped and clarified the paper.

PF gratefully acknowledges Simon Prunet's help with the python codes and the entire CFHT `Ohana for offering excellent conditions of work.

This research has made use of:
\begin{itemize}
    \item the SIMBAD database and the VizieR catalogue access tool, operated at CDS, Strasbourg, France. The original description of the SIMBAD database was published in A\&AS, 143, 9 (2000), and of the VizieR service in A\&AS, 143, 23 (2000).
    \item data products from the Two Micron All Sky Survey, which is a joint project of the University of Massachusetts and the Infrared Processing and Analysis Center/California Institute of Technology, funded by the National Aeronautics and Space Administration and the National Science Foundation.
    \item data products from the AAVSO Photometric All Sky Survey (APASS), funded by the Robert Martin Ayers Sciences Fund and the National Science Foundation.
    \item data from the European Space Agency (ESA) mission {\it Gaia} (\url{http://www.cosmos.esa.int/gaia}), processed by the {\it Gaia} Data Processing and Analysis Consortium (DPAC) (\url{http://www.cosmos.esa.int/web/gaia/dpac/consortium}). Funding for the DPAC has been provided by national institutions, in particular the institutions participating in the {\it Gaia} Multilateral Agreement.
    \item the facilities of the Canadian Astronomy Data Centre operated by the National Research Council of Canada with the support of the Canadian Space Agency.
    \item the Washington Double Star Catalog maintained at the U.S. Naval Observatory.
\end{itemize}

This research has made use of the Brazilian time at CFHT through the agreement between the Brazilian Ministry of Science Technology Innovation and Communications (MCTIC) and the CFHT. E.M. acknowledges the support of CNPq, under the Universal grant process number 443557/2014-4, and also the support of FAPEMIG, under project number 01/2014-23092.

X.D., T.F. and F.A. received funding from the French Programme National de Physique Stellaire (PNPS) and the Programme National de Plan\'{e}tologie of CNRS (INSU). This work has been partially supported by the Labex OSUG@2020. The computations of atmosphere models were performed in part on the Milky Way supercomputer, which is funded by the Deutsche Forschungsgemeinschaft (DFG) through the Collaborative Research Centre (SFB 881) "The Milky Way System" (sub-project Z2) and hosted at the University of Heidelberg Computing Centre, and at the P\^{o}le Scientifique de Mod\'{e}lisation Num\'{e}rique (PSMN) at the \'{E}cole Normale Sup\'{e}rieure (ENS) in Lyon, and at the Gesellschaft f\"{u}r Wissenschaftliche Datenverarbeitung G\"{o}ttingen in collaboration with the Institut f\"{u}r Astrophysik G\"{o}ttingen.

\addcontentsline{toc}{section}{Acknowledgements}
\addcontentsline{toc}{section}{References}
\addcontentsline{toc}{section}{Appendices}





\begin{thebibliography}{}
\makeatletter
\relax
\def\mn@urlcharsother{\let\do\@makeother \do\$\do\&\do\#\do\^\do\_\do\%\do\~}
\def\mn@doi{\begingroup\mn@urlcharsother \@ifnextchar [ {\mn@doi@}
  {\mn@doi@[]}}
\def\mn@doi@[#1]#2{\def\@tempa{#1}\ifx\@tempa\@empty \href
  {http://dx.doi.org/#2} {doi:#2}\else \href {http://dx.doi.org/#2} {#1}\fi
  \endgroup}
\def\mn@eprint#1#2{\mn@eprint@#1:#2::\@nil}
\def\mn@eprint@arXiv#1{\href {http://arxiv.org/abs/#1} {{\tt arXiv:#1}}}
\def\mn@eprint@dblp#1{\href {http://dblp.uni-trier.de/rec/bibtex/#1.xml}
  {dblp:#1}}
\def\mn@eprint@#1:#2:#3:#4\@nil{\def\@tempa {#1}\def\@tempb {#2}\def\@tempc
  {#3}\ifx \@tempc \@empty \let \@tempc \@tempb \let \@tempb \@tempa \fi \ifx
  \@tempb \@empty \def\@tempb {arXiv}\fi \@ifundefined
  {mn@eprint@\@tempb}{\@tempb:\@tempc}{\expandafter \expandafter \csname
  mn@eprint@\@tempb\endcsname \expandafter{\@tempc}}}

\bibitem[\protect\citeauthoryear{{Allard}, {Homeier}, {Freytag},
  {Schaffenberger}, {}  \& {Rajpurohit}}{{Allard} et~al.}{2013}]{allard13}
{Allard} F.,  {Homeier} D.,  {Freytag} B.,  {Schaffenberger} {} W.,
  {Rajpurohit} A.~S.,  2013, Memorie della Societa Astronomica Italiana
  Supplementi, \href {http://adsabs.harvard.edu/abs/2013MSAIS..24..128A} {24,
  128}

\bibitem[\protect\citeauthoryear{{Alonso-Floriano} et~al.,}{{Alonso-Floriano}
  et~al.}{2015}]{alonso15}
{Alonso-Floriano} F.~J.,  et~al., 2015, \mn@doi [\aap]
  {10.1051/0004-6361/201525803}, \href
  {http://adsabs.harvard.edu/abs/2015A%26A...577A.128A} {577, A128}

\bibitem[\protect\citeauthoryear{{Artigau} et~al.,}{{Artigau}
  et~al.}{2014}]{artigau14}
{Artigau} {\'E}.,  et~al., 2014, in Observatory Operations: Strategies,
  Processes, and Systems V. p. 914905 (\mn@eprint {arXiv} {1406.6927}),
  \mn@doi{10.1117/12.2056385}

\bibitem[\protect\citeauthoryear{{Baize}}{{Baize}}{1976}]{baize76}
{Baize} P.,  1976, \aaps, \href
  {http://adsabs.harvard.edu/abs/1976A%26AS...26..177B} {26, 177}

\bibitem[\protect\citeauthoryear{{Barnes}, {Jeffers}, {Haswell}, {Jones},
  {Shulyak}, {Pavlenko}  \& {Jenkins}}{{Barnes} et~al.}{2017}]{barnes17}
{Barnes} J.~R.,  {Jeffers} S.~V.,  {Haswell} C.~A.,  {Jones} H.~R.~A.,
  {Shulyak} D.,  {Pavlenko} Y.~V.,   {Jenkins} J.~S.,  2017, \mn@doi [\mnras]
  {10.1093/mnras/stx1482}, \href
  {http://adsabs.harvard.edu/abs/2017MNRAS.471..811B} {471, 811}

\bibitem[\protect\citeauthoryear{{Becklin} \& {Zuckerman}}{{Becklin} \&
  {Zuckerman}}{1988}]{becklin88}
{Becklin} E.~E.,  {Zuckerman} B.,  1988, \mn@doi [\nat] {10.1038/336656a0},
  \href {http://adsabs.harvard.edu/abs/1988Natur.336..656B} {336, 656}

\bibitem[\protect\citeauthoryear{{Benedict}, {McArthur}, {Franz}, {Wasserman}
  \& {Henry}}{{Benedict} et~al.}{2000}]{benedict00}
{Benedict} G.~F.,  {McArthur} B.~E.,  {Franz} O.~G.,  {Wasserman} L.~H.,
  {Henry} T.~J.,  2000, \mn@doi [\aj] {10.1086/301495}, \href
  {http://adsabs.harvard.edu/abs/2000AJ....120.1106B} {120, 1106}

\bibitem[\protect\citeauthoryear{{Bergfors} et~al.,}{{Bergfors}
  et~al.}{2010}]{bergfors10}
{Bergfors} C.,  et~al., 2010, \mn@doi [\aap] {10.1051/0004-6361/201014114},
  \href {http://adsabs.harvard.edu/abs/2010A%26A...520A..54B} {520, A54}

\bibitem[\protect\citeauthoryear{{Beuzit} et~al.,}{{Beuzit}
  et~al.}{2004}]{beuzit04}
{Beuzit} J.-L.,  et~al., 2004, \mn@doi [\aap] {10.1051/0004-6361:20048006},
  \href {http://adsabs.harvard.edu/abs/2004A%26A...425..997B} {425, 997}

\bibitem[\protect\citeauthoryear{{Bieger Smith}}{{Bieger
  Smith}}{1964}]{bieger64}
{Bieger Smith} G.~S.,  1964, \mn@doi [\aj] {10.1086/109356}, \href
  {http://adsabs.harvard.edu/abs/1964AJ.....69..804B} {69, 804}

\bibitem[\protect\citeauthoryear{{Blackwell}, {Petford}, {Arribas}, {Haddock}
  \& {Selby}}{{Blackwell} et~al.}{1990}]{blackwell90}
{Blackwell} D.~E.,  {Petford} A.~D.,  {Arribas} S.,  {Haddock} D.~J.,   {Selby}
  M.~J.,  1990, \aap, \href
  {http://adsabs.harvard.edu/abs/1990A%26A...232..396B} {232, 396}

\bibitem[\protect\citeauthoryear{{Boisse} et~al.,}{{Boisse}
  et~al.}{2010}]{boisse10}
{Boisse} I.,  et~al., 2010, \mn@doi [\aap] {10.1051/0004-6361/201014909}, \href
  {http://adsabs.harvard.edu/abs/2010A%26A...523A..88B} {523, A88}

\bibitem[\protect\citeauthoryear{{Bonfils}, {Delfosse}, {Udry}, {Santos},
  {Forveille}  \& {S{\'e}gransan}}{{Bonfils} et~al.}{2005}]{bonfils05}
{Bonfils} X.,  {Delfosse} X.,  {Udry} S.,  {Santos} N.~C.,  {Forveille} T.,
  {S{\'e}gransan} D.,  2005, \mn@doi [\aap] {10.1051/0004-6361:20053046}, \href
  {http://adsabs.harvard.edu/abs/2005A%26A...442..635B} {442, 635}

\bibitem[\protect\citeauthoryear{{Bonfils} et~al.,}{{Bonfils}
  et~al.}{2013}]{bonfils13}
{Bonfils} X.,  et~al., 2013, \mn@doi [\aap] {10.1051/0004-6361/201014704},
  \href {http://adsabs.harvard.edu/abs/2013A%26A...549A.109B} {549, A109}

\bibitem[\protect\citeauthoryear{{Bowler}, {Liu}, {Shkolnik}, {Dupuy}, {Cieza},
  {Kraus}  \& {Tamura}}{{Bowler} et~al.}{2012}]{bowler12a}
{Bowler} B.~P.,  {Liu} M.~C.,  {Shkolnik} E.~L.,  {Dupuy} T.~J.,  {Cieza}
  L.~A.,  {Kraus} A.~L.,   {Tamura} M.,  2012, \mn@doi [\apj]
  {10.1088/0004-637X/753/2/142}, \href
  {http://adsabs.harvard.edu/abs/2012ApJ...753..142B} {753, 142}

\bibitem[\protect\citeauthoryear{{Bowler}, {Liu}, {Shkolnik}  \&
  {Tamura}}{{Bowler} et~al.}{2015a}]{bowler15a}
{Bowler} B.~P.,  {Liu} M.~C.,  {Shkolnik} E.~L.,   {Tamura} M.,  2015a, \mn@doi
  [\apjs] {10.1088/0067-0049/216/1/7}, \href
  {http://adsabs.harvard.edu/abs/2015ApJS..216....7B} {216, 7}

\bibitem[\protect\citeauthoryear{{Bowler} et~al.,}{{Bowler}
  et~al.}{2015b}]{bowler15b}
{Bowler} B.~P.,  et~al., 2015b, \mn@doi [\apj] {10.1088/0004-637X/806/1/62},
  \href {http://adsabs.harvard.edu/abs/2015ApJ...806...62B} {806, 62}

\bibitem[\protect\citeauthoryear{{Boyajian} et~al.,}{{Boyajian}
  et~al.}{2012}]{boyajian12}
{Boyajian} T.~S.,  et~al., 2012, \mn@doi [\apj] {10.1088/0004-637X/757/2/112},
  \href {http://adsabs.harvard.edu/abs/2012ApJ...757..112B} {757, 112}

\bibitem[\protect\citeauthoryear{{Brandt} et~al.,}{{Brandt}
  et~al.}{2014}]{brandt14}
{Brandt} T.~D.,  et~al., 2014, \mn@doi [\apj] {10.1088/0004-637X/786/1/1},
  \href {http://adsabs.harvard.edu/abs/2014ApJ...786....1B} {786, 1}

\bibitem[\protect\citeauthoryear{{Browning}, {Basri}, {Marcy}, {West}  \&
  {Zhang}}{{Browning} et~al.}{2010}]{browning10}
{Browning} M.~K.,  {Basri} G.,  {Marcy} G.~W.,  {West} A.~A.,   {Zhang} J.,
  2010, \mn@doi [\aj] {10.1088/0004-6256/139/2/504}, \href
  {http://adsabs.harvard.edu/abs/2010AJ....139..504B} {139, 504}

\bibitem[\protect\citeauthoryear{{Casagrande}, {Flynn}  \&
  {Bessell}}{{Casagrande} et~al.}{2008}]{casagrande08}
{Casagrande} L.,  {Flynn} C.,   {Bessell} M.,  2008, \mn@doi [\mnras]
  {10.1111/j.1365-2966.2008.13573.x}, \href
  {http://adsabs.harvard.edu/abs/2008MNRAS.389..585C} {389, 585}

\bibitem[\protect\citeauthoryear{{Claudi} et~al.,}{{Claudi}
  et~al.}{2016}]{claudi16}
{Claudi} R.,  et~al., 2016, in Ground-based and Airborne Instrumentation for
  Astronomy VI. p. 99081A (\mn@eprint {arXiv} {1611.07603}),
  \mn@doi{10.1117/12.2231845}

\bibitem[\protect\citeauthoryear{{Cort{\'e}s-Contreras}
  et~al.,}{{Cort{\'e}s-Contreras} et~al.}{2017}]{cortes17}
{Cort{\'e}s-Contreras} M.,  et~al., 2017, \mn@doi [\aap]
  {10.1051/0004-6361/201629056}, \href
  {http://adsabs.harvard.edu/abs/2017A%26A...597A..47C} {597, A47}

\bibitem[\protect\citeauthoryear{{Daemgen}, {Siegler}, {Reid}  \&
  {Close}}{{Daemgen} et~al.}{2007}]{daemgen07}
{Daemgen} S.,  {Siegler} N.,  {Reid} I.~N.,   {Close} L.~M.,  2007, \mn@doi
  [\apj] {10.1086/509109}, \href
  {http://adsabs.harvard.edu/abs/2007ApJ...654..558D} {654, 558}

\bibitem[\protect\citeauthoryear{{Davison} et~al.,}{{Davison}
  et~al.}{2014}]{davison14}
{Davison} C.~L.,  et~al., 2014, \mn@doi [\aj] {10.1088/0004-6256/147/2/26},
  \href {http://adsabs.harvard.edu/abs/2014AJ....147...26D} {147, 26}

\bibitem[\protect\citeauthoryear{Deeg \& Belmonte}{Deeg \&
  Belmonte}{2018}]{deeg18}
Deeg H.,  Belmonte J.,  2018, Handbook of Exoplanets.
Handbook of Exoplanets, Springer International Publishing, \url
  {https://books.google.com/books?id=Gq1LnQAACAAJ}

\bibitem[\protect\citeauthoryear{{Delfosse}, {Forveille}, {Perrier}  \&
  {Mayor}}{{Delfosse} et~al.}{1998}]{delfosse98a}
{Delfosse} X.,  {Forveille} T.,  {Perrier} C.,   {Mayor} M.,  1998, \aap, \href
  {http://adsabs.harvard.edu/abs/1998A%26A...331..581D} {331, 581}

\bibitem[\protect\citeauthoryear{{Delfosse}, {Forveille}, {Mayor}, {Burnet}  \&
  {Perrier}}{{Delfosse} et~al.}{1999a}]{delfosse99a}
{Delfosse} X.,  {Forveille} T.,  {Mayor} M.,  {Burnet} M.,   {Perrier} C.,
  1999a, \aap, \href {http://adsabs.harvard.edu/abs/1999A%26A...341L..63D}
  {341, L63}

\bibitem[\protect\citeauthoryear{{Delfosse}, {Forveille}, {Beuzit}, {Udry},
  {Mayor}  \& {Perrier}}{{Delfosse} et~al.}{1999b}]{delfosse99b}
{Delfosse} X.,  {Forveille} T.,  {Beuzit} J.-L.,  {Udry} S.,  {Mayor} M.,
  {Perrier} C.,  1999b, \aap, \href
  {http://adsabs.harvard.edu/abs/1999A%26A...344..897D} {344, 897}

\bibitem[\protect\citeauthoryear{{Delfosse} et~al.,}{{Delfosse}
  et~al.}{2013}]{delfosse13}
{Delfosse} X.,  et~al., 2013, \mn@doi [\aap] {10.1051/0004-6361/201219013},
  \href {http://adsabs.harvard.edu/abs/2013A%26A...553A...8D} {553, A8}

\bibitem[\protect\citeauthoryear{{Dommanget} \& {Nys}}{{Dommanget} \&
  {Nys}}{2000a}]{dommanget00a}
{Dommanget} J.,  {Nys} O.,  2000a, \aap, \href
  {http://adsabs.harvard.edu/abs/2000A%26A...363..991D} {363, 991}

\bibitem[\protect\citeauthoryear{{Dommanget} \& {Nys}}{{Dommanget} \&
  {Nys}}{2000b}]{dommanget00b}
{Dommanget} J.,  {Nys} O.,  2000b, \aap, \href
  {http://adsabs.harvard.edu/abs/2000A%26A...364..927D} {364, 927}

\bibitem[\protect\citeauthoryear{{Donati}, {Semel}, {Carter}, {Rees}  \&
  {Collier Cameron}}{{Donati} et~al.}{1997}]{donati97}
{Donati} J.-F.,  {Semel} M.,  {Carter} B.~D.,  {Rees} D.~E.,   {Collier
  Cameron} A.,  1997, \mn@doi [\mnras] {10.1093/mnras/291.4.658}, \href
  {http://adsabs.harvard.edu/abs/1997MNRAS.291..658D} {291, 658}

\bibitem[\protect\citeauthoryear{{Donati}, {Forveille}, {Collier Cameron},
  {Barnes}, {Delfosse}, {Jardine}  \& {Valenti}}{{Donati}
  et~al.}{2006}]{donati06}
{Donati} J.-F.,  {Forveille} T.,  {Collier Cameron} A.,  {Barnes} J.~R.,
  {Delfosse} X.,  {Jardine} M.~M.,   {Valenti} J.~A.,  2006, \mn@doi [Science]
  {10.1126/science.1121102}, \href
  {http://adsabs.harvard.edu/abs/2006Sci...311..633D} {311, 633}

\bibitem[\protect\citeauthoryear{{Donati} et~al.,}{{Donati}
  et~al.}{2008}]{donati08}
{Donati} J.-F.,  et~al., 2008, \mn@doi [\mnras]
  {10.1111/j.1365-2966.2008.13799.x}, \href
  {http://adsabs.harvard.edu/abs/2008MNRAS.390..545D} {390, 545}

\bibitem[\protect\citeauthoryear{{Engle}, {Guinan}  \& {Mizusawa}}{{Engle}
  et~al.}{2009}]{engle09}
{Engle} S.~G.,  {Guinan} E.~F.,   {Mizusawa} T.,  2009, in {van Steenberg}
  M.~E.,  {Sonneborn} G.,  {Moos} H.~W.,   {Blair} W.~P.,  eds,  American
  Institute of Physics Conference Series Vol. 1135, American Institute of
  Physics Conference Series. pp 221--224 (\mn@eprint {arXiv} {0902.3444}),
  \mn@doi{10.1063/1.3154054}

\bibitem[\protect\citeauthoryear{{Epchtein} et~al.,}{{Epchtein}
  et~al.}{1999}]{epchtein99}
{Epchtein} N.,  et~al., 1999, \aap, \href
  {http://adsabs.harvard.edu/abs/1999A%26A...349..236E} {349, 236}

\bibitem[\protect\citeauthoryear{{Figueira} et~al.,}{{Figueira}
  et~al.}{2016}]{figueira16}
{Figueira} P.,  et~al., 2016, \mn@doi [\aap] {10.1051/0004-6361/201526900},
  \href {http://adsabs.harvard.edu/abs/2016A%26A...586A.101F} {586, A101}

\bibitem[\protect\citeauthoryear{{Frith} et~al.,}{{Frith}
  et~al.}{2013}]{frith13}
{Frith} J.,  et~al., 2013, \mn@doi [\mnras] {10.1093/mnras/stt1436}, \href
  {http://adsabs.harvard.edu/abs/2013MNRAS.435.2161F} {435, 2161}

\bibitem[\protect\citeauthoryear{{Gaidos} et~al.,}{{Gaidos}
  et~al.}{2014}]{gaidos14}
{Gaidos} E.,  et~al., 2014, \mn@doi [\mnras] {10.1093/mnras/stu1313}, \href
  {http://adsabs.harvard.edu/abs/2014MNRAS.443.2561G} {443, 2561}

\bibitem[\protect\citeauthoryear{{Gizis} \& {Reid}}{{Gizis} \&
  {Reid}}{1996}]{gizis96}
{Gizis} J.~E.,  {Reid} N.~I.,  1996, \mn@doi [\aj] {10.1086/117789}, \href
  {http://adsabs.harvard.edu/abs/1996AJ....111..365G} {111, 365}

\bibitem[\protect\citeauthoryear{{Goldin} \& {Makarov}}{{Goldin} \&
  {Makarov}}{2007}]{goldin07}
{Goldin} A.,  {Makarov} V.~V.,  2007, \mn@doi [\apjs] {10.1086/520513}, \href
  {http://adsabs.harvard.edu/abs/2007ApJS..173..137G} {173, 137}

\bibitem[\protect\citeauthoryear{{Goldman}, {Marsat}, {Henning}, {Clemens}  \&
  {Greiner}}{{Goldman} et~al.}{2010}]{goldman10}
{Goldman} B.,  {Marsat} S.,  {Henning} T.,  {Clemens} C.,   {Greiner} J.,
  2010, \mn@doi [\mnras] {10.1111/j.1365-2966.2010.16524.x}, \href
  {http://adsabs.harvard.edu/abs/2010MNRAS.405.1140G} {405, 1140}

\bibitem[\protect\citeauthoryear{{Gomes da Silva}, {Santos}, {Bonfils},
  {Delfosse}, {Forveille}  \& {Udry}}{{Gomes da Silva} et~al.}{2011}]{gomes11}
{Gomes da Silva} J.,  {Santos} N.~C.,  {Bonfils} X.,  {Delfosse} X.,
  {Forveille} T.,   {Udry} S.,  2011, \mn@doi [\aap]
  {10.1051/0004-6361/201116971}, \href
  {http://adsabs.harvard.edu/abs/2011A%26A...534A..30G} {534, A30}

\bibitem[\protect\citeauthoryear{{Gould} \& {Chanam{\'e}}}{{Gould} \&
  {Chanam{\'e}}}{2004}]{gould04}
{Gould} A.,  {Chanam{\'e}} J.,  2004, \mn@doi [\apjs] {10.1086/381147}, \href
  {http://adsabs.harvard.edu/abs/2004ApJS..150..455G} {150, 455}

\bibitem[\protect\citeauthoryear{{Harrington} \& {Dahn}}{{Harrington} \&
  {Dahn}}{1984}]{harrington84}
{Harrington} R.~S.,  {Dahn} C.~C.,  1984, \iaucirc, \href
  {http://adsabs.harvard.edu/abs/1984IAUC.3989....1H} {3989}

\bibitem[\protect\citeauthoryear{{Hartkopf}, {McAlister}, {Mason}, {Barry},
  {Turner}  \& {Fu}}{{Hartkopf} et~al.}{1994}]{hartkopf94}
{Hartkopf} W.~I.,  {McAlister} H.~A.,  {Mason} B.~D.,  {Barry} D.~J.,  {Turner}
  N.~H.,   {Fu} H.-H.,  1994, \mn@doi [\aj] {10.1086/117242}, \href
  {http://adsabs.harvard.edu/abs/1994AJ....108.2299H} {108, 2299}

\bibitem[\protect\citeauthoryear{{Hartman}, {Bakos}, {Noyes}, {Sip{\H o}cz},
  {Kov{\'a}cs}, {Mazeh}, {Shporer}  \& {P{\'a}l}}{{Hartman}
  et~al.}{2011}]{hartman11}
{Hartman} J.~D.,  {Bakos} G.~{\'A}.,  {Noyes} R.~W.,  {Sip{\H o}cz} B.,
  {Kov{\'a}cs} G.,  {Mazeh} T.,  {Shporer} A.,   {P{\'a}l} A.,  2011, \mn@doi
  [\aj] {10.1088/0004-6256/141/5/166}, \href
  {http://adsabs.harvard.edu/abs/2011AJ....141..166H} {141, 166}

\bibitem[\protect\citeauthoryear{{Hauschildt}, {Allard}, {Ferguson}, {Baron}
  \& {Alexander}}{{Hauschildt} et~al.}{1999}]{hauschildt99}
{Hauschildt} P.~H.,  {Allard} F.,  {Ferguson} J.,  {Baron} E.,   {Alexander}
  D.~R.,  1999, \mn@doi [\apj] {10.1086/307954}, \href
  {http://adsabs.harvard.edu/abs/1999ApJ...525..871H} {525, 871}

\bibitem[\protect\citeauthoryear{{Heintz}}{{Heintz}}{1994}]{heintz94}
{Heintz} W.~D.,  1994, \mn@doi [\aj] {10.1086/117247}, \href
  {http://adsabs.harvard.edu/abs/1994AJ....108.2338H} {108, 2338}

\bibitem[\protect\citeauthoryear{{Herbig} \& {Moorhead}}{{Herbig} \&
  {Moorhead}}{1965}]{herbig65}
{Herbig} G.~H.,  {Moorhead} J.~M.,  1965, \mn@doi [\apj] {10.1086/148150},
  \href {http://adsabs.harvard.edu/abs/1965ApJ...141..649H} {141, 649}

\bibitem[\protect\citeauthoryear{{Houdebine} \& {Mullan}}{{Houdebine} \&
  {Mullan}}{2015}]{houdebine15}
{Houdebine} E.~R.,  {Mullan} D.~J.,  2015, \mn@doi [\apj]
  {10.1088/0004-637X/801/2/106}, \href
  {http://adsabs.harvard.edu/abs/2015ApJ...801..106H} {801, 106}

\bibitem[\protect\citeauthoryear{{Hu{\'e}lamo} et~al.,}{{Hu{\'e}lamo}
  et~al.}{2008}]{huelamo08}
{Hu{\'e}lamo} N.,  et~al., 2008, \mn@doi [\aap] {10.1051/0004-6361:200810596},
  \href {http://adsabs.harvard.edu/abs/2008A%26A...489L...9H} {489, L9}

\bibitem[\protect\citeauthoryear{{Irwin}, {Berta}, {Burke}, {Charbonneau},
  {Nutzman}, {West}  \& {Falco}}{{Irwin} et~al.}{2011}]{irwin11}
{Irwin} J.,  {Berta} Z.~K.,  {Burke} C.~J.,  {Charbonneau} D.,  {Nutzman} P.,
  {West} A.~A.,   {Falco} E.~E.,  2011, \mn@doi [\apj]
  {10.1088/0004-637X/727/1/56}, \href
  {http://adsabs.harvard.edu/abs/2011ApJ...727...56I} {727, 56}

\bibitem[\protect\citeauthoryear{{Janson} et~al.,}{{Janson}
  et~al.}{2012}]{janson12}
{Janson} M.,  et~al., 2012, \mn@doi [\apj] {10.1088/0004-637X/754/1/44}, \href
  {http://adsabs.harvard.edu/abs/2012ApJ...754...44J} {754, 44}

\bibitem[\protect\citeauthoryear{{Janson}, {Bergfors}, {Brandner},
  {Kudryavtseva}, {Hormuth}, {Hippler}  \& {Henning}}{{Janson}
  et~al.}{2014}]{janson14a}
{Janson} M.,  {Bergfors} C.,  {Brandner} W.,  {Kudryavtseva} N.,  {Hormuth} F.,
   {Hippler} S.,   {Henning} T.,  2014, \mn@doi [\apj]
  {10.1088/0004-637X/789/2/102}, \href
  {http://adsabs.harvard.edu/abs/2014ApJ...789..102J} {789, 102}

\bibitem[\protect\citeauthoryear{{Jenkins}, {Ramsey}, {Jones}, {Pavlenko},
  {Gallardo}, {Barnes}  \& {Pinfield}}{{Jenkins} et~al.}{2009}]{jenkins09}
{Jenkins} J.~S.,  {Ramsey} L.~W.,  {Jones} H.~R.~A.,  {Pavlenko} Y.,
  {Gallardo} J.,  {Barnes} J.~R.,   {Pinfield} D.~J.,  2009, \mn@doi [\apj]
  {10.1088/0004-637X/704/2/975}, \href
  {http://adsabs.harvard.edu/abs/2009ApJ...704..975J} {704, 975}

\bibitem[\protect\citeauthoryear{{J{\'o}dar}, {P{\'e}rez-Garrido},
  {D{\'{\i}}az-S{\'a}nchez}, {Vill{\'o}}, {Rebolo}  \&
  {P{\'e}rez-Prieto}}{{J{\'o}dar} et~al.}{2013}]{jodar13}
{J{\'o}dar} E.,  {P{\'e}rez-Garrido} A.,  {D{\'{\i}}az-S{\'a}nchez} A.,
  {Vill{\'o}} I.,  {Rebolo} R.,   {P{\'e}rez-Prieto} J.~A.,  2013, \mn@doi
  [\mnras] {10.1093/mnras/sts382}, \href
  {http://adsabs.harvard.edu/abs/2013MNRAS.429..859J} {429, 859}

\bibitem[\protect\citeauthoryear{{Johnson} \& {Apps}}{{Johnson} \&
  {Apps}}{2009}]{johnson09}
{Johnson} J.~A.,  {Apps} K.,  2009, \mn@doi [\apj]
  {10.1088/0004-637X/699/2/933}, \href
  {http://adsabs.harvard.edu/abs/2009ApJ...699..933J} {699, 933}

\bibitem[\protect\citeauthoryear{{Joy} \& {Sanford}}{{Joy} \&
  {Sanford}}{1926}]{joy26}
{Joy} A.~H.,  {Sanford} R.~F.,  1926, \mn@doi [\apj] {10.1086/143009}, \href
  {http://adsabs.harvard.edu/abs/1926ApJ....64..250J} {64}

\bibitem[\protect\citeauthoryear{{Karata{\c s}}, {Bilir}, {Eker}  \&
  {Demircan}}{{Karata{\c s}} et~al.}{2004}]{karatas04}
{Karata{\c s}} Y.,  {Bilir} S.,  {Eker} Z.,   {Demircan} O.,  2004, \mn@doi
  [\mnras] {10.1111/j.1365-2966.2004.07588.x}, \href
  {http://adsabs.harvard.edu/abs/2004MNRAS.349.1069K} {349, 1069}

\bibitem[\protect\citeauthoryear{{Kiraga}}{{Kiraga}}{2012}]{kiraga12}
{Kiraga} M.,  2012, \actaa, \href
  {http://adsabs.harvard.edu/abs/2012AcA....62...67K} {62, 67}

\bibitem[\protect\citeauthoryear{{Kiraga} \& {St{\c e}pie{\'n}}}{{Kiraga} \&
  {St{\c e}pie{\'n}}}{2013}]{kiraga13}
{Kiraga} M.,  {St{\c e}pie{\'n}} K.,  2013, \actaa, \href
  {http://adsabs.harvard.edu/abs/2013AcA....63...53K} {63, 53}

\bibitem[\protect\citeauthoryear{{Kiraga} \& {Stepien}}{{Kiraga} \&
  {Stepien}}{2007}]{kiraga07}
{Kiraga} M.,  {Stepien} K.,  2007, \actaa, \href
  {http://adsabs.harvard.edu/abs/2007AcA....57..149K} {57, 149}

\bibitem[\protect\citeauthoryear{{Kirkpatrick}}{{Kirkpatrick}}{2000}]{kirkpatrick00}
{Kirkpatrick} J.~D.,  2000, in {Griffith} C.~A.,  {Marley} M.~S.,  eds,
  Astronomical Society of the Pacific Conference Series Vol. 212, From Giant
  Planets to Cool Stars. p.~20

\bibitem[\protect\citeauthoryear{{Kirkpatrick} et~al.,}{{Kirkpatrick}
  et~al.}{1999}]{kirkpatrick99}
{Kirkpatrick} J.~D.,  et~al., 1999, \mn@doi [\apj] {10.1086/307414}, \href
  {http://adsabs.harvard.edu/abs/1999ApJ...519..802K} {519, 802}

\bibitem[\protect\citeauthoryear{{Kirkpatrick} et~al.,}{{Kirkpatrick}
  et~al.}{2012}]{kirkpatrick12}
{Kirkpatrick} J.~D.,  et~al., 2012, \mn@doi [\apj]
  {10.1088/0004-637X/753/2/156}, \href
  {http://adsabs.harvard.edu/abs/2012ApJ...753..156K} {753, 156}

\bibitem[\protect\citeauthoryear{{Kurucz}}{{Kurucz}}{1993a}]{kurucz93a}
{Kurucz} R.,  1993a, ATLAS9 Stellar Atmosphere Programs and 2 km/s grid.~Kurucz
  CD-ROM No.~13.~ Cambridge, Mass.: Smithsonian Astrophysical Observatory,
  1993., \href {http://adsabs.harvard.edu/abs/1993KurCD..13.....K} {13}

\bibitem[\protect\citeauthoryear{{Kurucz}}{{Kurucz}}{1993b}]{kurucz93b}
{Kurucz} R.,  1993b, SYNTHE Spectrum Synthesis Programs and Line Data.~Kurucz
  CD-ROM No.~18.~Cambridge, Mass.: Smithsonian Astrophysical Observatory,
  1993., \href {http://adsabs.harvard.edu/abs/1993KurCD..18.....K} {18}

\bibitem[\protect\citeauthoryear{{Law}, {Hodgkin}  \& {Mackay}}{{Law}
  et~al.}{2006}]{law06}
{Law} N.~M.,  {Hodgkin} S.~T.,   {Mackay} C.~D.,  2006, \mn@doi [\mnras]
  {10.1111/j.1365-2966.2006.10265.x}, \href
  {http://adsabs.harvard.edu/abs/2006MNRAS.368.1917L} {368, 1917}

\bibitem[\protect\citeauthoryear{{Law}, {Hodgkin}  \& {Mackay}}{{Law}
  et~al.}{2008}]{law08}
{Law} N.~M.,  {Hodgkin} S.~T.,   {Mackay} C.~D.,  2008, \mn@doi [\mnras]
  {10.1111/j.1365-2966.2007.12675.x}, \href
  {http://adsabs.harvard.edu/abs/2008MNRAS.384..150L} {384, 150}

\bibitem[\protect\citeauthoryear{{L{\'e}pine} \& {Bongiorno}}{{L{\'e}pine} \&
  {Bongiorno}}{2007}]{lepine07b}
{L{\'e}pine} S.,  {Bongiorno} B.,  2007, \mn@doi [\aj] {10.1086/510333}, \href
  {http://adsabs.harvard.edu/abs/2007AJ....133..889L} {133, 889}

\bibitem[\protect\citeauthoryear{{L{\'e}pine} \& {Gaidos}}{{L{\'e}pine} \&
  {Gaidos}}{2011}]{lepine11}
{L{\'e}pine} S.,  {Gaidos} E.,  2011, \mn@doi [\aj]
  {10.1088/0004-6256/142/4/138}, \href
  {http://adsabs.harvard.edu/abs/2011AJ....142..138L} {142, 138}

\bibitem[\protect\citeauthoryear{{L{\'e}pine}, {Rich}  \& {Shara}}{{L{\'e}pine}
  et~al.}{2007}]{lepine07a}
{L{\'e}pine} S.,  {Rich} R.~M.,   {Shara} M.~M.,  2007, \mn@doi [\apj]
  {10.1086/521614}, \href {http://adsabs.harvard.edu/abs/2007ApJ...669.1235L}
  {669, 1235}

\bibitem[\protect\citeauthoryear{{Leung} \& {Schneider}}{{Leung} \&
  {Schneider}}{1978}]{leung78}
{Leung} K.-C.,  {Schneider} D.~P.,  1978, \mn@doi [\aj] {10.1086/112244}, \href
  {http://adsabs.harvard.edu/abs/1978AJ.....83..618L} {83, 618}

\bibitem[\protect\citeauthoryear{{Lindgren} \& {Heiter}}{{Lindgren} \&
  {Heiter}}{2017}]{lindgren17}
{Lindgren} S.,  {Heiter} U.,  2017, \mn@doi [\aap]
  {10.1051/0004-6361/201730715}, \href
  {http://adsabs.harvard.edu/abs/2017A%26A...604A..97L} {604, A97}

\bibitem[\protect\citeauthoryear{{Lindgren}, {Heiter}  \&
  {Seifahrt}}{{Lindgren} et~al.}{2016}]{lindgren16}
{Lindgren} S.,  {Heiter} U.,   {Seifahrt} A.,  2016, \mn@doi [\aap]
  {10.1051/0004-6361/201526602}, \href
  {http://adsabs.harvard.edu/abs/2016A%26A...586A.100L} {586, A100}

\bibitem[\protect\citeauthoryear{{Lippincott}}{{Lippincott}}{1977}]{lippincott77}
{Lippincott} S.~L.,  1977, \mn@doi [\aj] {10.1086/112147}, \href
  {http://adsabs.harvard.edu/abs/1977AJ.....82..925L} {82, 925}

\bibitem[\protect\citeauthoryear{{Mahadevan} et~al.,}{{Mahadevan}
  et~al.}{2012}]{mahadevan12}
{Mahadevan} S.,  et~al., 2012, in Ground-based and Airborne Instrumentation for
  Astronomy IV. p. 84461S (\mn@eprint {arXiv} {1209.1686}),
  \mn@doi{10.1117/12.926102}

\bibitem[\protect\citeauthoryear{{Makarov} \& {Kaplan}}{{Makarov} \&
  {Kaplan}}{2005}]{makarov05}
{Makarov} V.~V.,  {Kaplan} G.~H.,  2005, \mn@doi [\aj] {10.1086/429590}, \href
  {http://adsabs.harvard.edu/abs/2005AJ....129.2420M} {129, 2420}

\bibitem[\protect\citeauthoryear{{Maldonado} et~al.,}{{Maldonado}
  et~al.}{2015}]{maldonado15}
{Maldonado} J.,  et~al., 2015, \mn@doi [\aap] {10.1051/0004-6361/201525797},
  \href {http://adsabs.harvard.edu/abs/2015A%26A...577A.132M} {577, A132}

\bibitem[\protect\citeauthoryear{{Malo}, {Artigau}, {Doyon}, {Lafreni{\`e}re},
  {Albert}  \& {Gagn{\'e}}}{{Malo} et~al.}{2014a}]{malo14a}
{Malo} L.,  {Artigau} {\'E}.,  {Doyon} R.,  {Lafreni{\`e}re} D.,  {Albert} L.,
   {Gagn{\'e}} J.,  2014a, \mn@doi [\apj] {10.1088/0004-637X/788/1/81}, \href
  {http://adsabs.harvard.edu/abs/2014ApJ...788...81M} {788, 81}

\bibitem[\protect\citeauthoryear{{Malo}, {Doyon}, {Feiden}, {Albert},
  {Lafreni{\`e}re}, {Artigau}, {Gagn{\'e}}  \& {Riedel}}{{Malo}
  et~al.}{2014b}]{malo14b}
{Malo} L.,  {Doyon} R.,  {Feiden} G.~A.,  {Albert} L.,  {Lafreni{\`e}re} D.,
  {Artigau} {\'E}.,  {Gagn{\'e}} J.,   {Riedel} A.,  2014b, \mn@doi [\apj]
  {10.1088/0004-637X/792/1/37}, \href
  {http://adsabs.harvard.edu/abs/2014ApJ...792...37M} {792, 37}

\bibitem[\protect\citeauthoryear{{Mann}, {Brewer}, {Gaidos}, {L{\'e}pine}  \&
  {Hilton}}{{Mann} et~al.}{2013a}]{mann13b}
{Mann} A.~W.,  {Brewer} J.~M.,  {Gaidos} E.,  {L{\'e}pine} S.,   {Hilton}
  E.~J.,  2013a, \mn@doi [\aj] {10.1088/0004-6256/145/2/52}, \href
  {http://adsabs.harvard.edu/abs/2013AJ....145...52M} {145, 52}

\bibitem[\protect\citeauthoryear{{Mann}, {Gaidos}  \& {Ansdell}}{{Mann}
  et~al.}{2013b}]{mann13a}
{Mann} A.~W.,  {Gaidos} E.,   {Ansdell} M.,  2013b, \mn@doi [\apj]
  {10.1088/0004-637X/779/2/188}, \href
  {http://adsabs.harvard.edu/abs/2013ApJ...779..188M} {779, 188}

\bibitem[\protect\citeauthoryear{{Mann}, {Feiden}, {Gaidos}, {Boyajian}  \&
  {von Braun}}{{Mann} et~al.}{2015}]{mann15}
{Mann} A.~W.,  {Feiden} G.~A.,  {Gaidos} E.,  {Boyajian} T.,   {von Braun} K.,
  2015, \mn@doi [\apj] {10.1088/0004-637X/804/1/64}, \href
  {http://adsabs.harvard.edu/abs/2015ApJ...804...64M} {804, 64}

\bibitem[\protect\citeauthoryear{{Martin}, {Basri}, {Delfosse}  \&
  {Forveille}}{{Martin} et~al.}{1997}]{martin97}
{Martin} E.~L.,  {Basri} G.,  {Delfosse} X.,   {Forveille} T.,  1997, \aap,
  \href {http://adsabs.harvard.edu/abs/1997A%26A...327L..29M} {327, L29}

\bibitem[\protect\citeauthoryear{{Mart{\'{\i}}n}, {Delfosse}, {Basri},
  {Goldman}, {Forveille}  \& {Zapatero Osorio}}{{Mart{\'{\i}}n}
  et~al.}{1999}]{martin99}
{Mart{\'{\i}}n} E.~L.,  {Delfosse} X.,  {Basri} G.,  {Goldman} B.,  {Forveille}
  T.,   {Zapatero Osorio} M.~R.,  1999, \mn@doi [\aj] {10.1086/301107}, \href
  {http://adsabs.harvard.edu/abs/1999AJ....118.2466M} {118, 2466}

\bibitem[\protect\citeauthoryear{{Martinache}, {Lloyd}, {Ireland}, {Yamada}  \&
  {Tuthill}}{{Martinache} et~al.}{2007}]{martinache07}
{Martinache} F.,  {Lloyd} J.~P.,  {Ireland} M.~J.,  {Yamada} R.~S.,   {Tuthill}
  P.~G.,  2007, \mn@doi [\apj] {10.1086/513868}, \href
  {http://adsabs.harvard.edu/abs/2007ApJ...661..496M} {661, 496}

\bibitem[\protect\citeauthoryear{{Mason}, {Wycoff}, {Hartkopf}, {Douglass}  \&
  {Worley}}{{Mason} et~al.}{2001}]{mason01}
{Mason} B.~D.,  {Wycoff} G.~L.,  {Hartkopf} W.~I.,  {Douglass} G.~G.,
  {Worley} C.~E.,  2001, \mn@doi [\aj] {10.1086/323920}, \href
  {http://adsabs.harvard.edu/abs/2001AJ....122.3466M} {122, 3466}

\bibitem[\protect\citeauthoryear{{McCarthy}, {Zuckerman}  \&
  {Becklin}}{{McCarthy} et~al.}{2001}]{mccarthy01}
{McCarthy} C.,  {Zuckerman} B.,   {Becklin} E.~E.,  2001, \mn@doi [\aj]
  {10.1086/321076}, \href {http://adsabs.harvard.edu/abs/2001AJ....121.3259M}
  {121, 3259}

\bibitem[\protect\citeauthoryear{{Melo}, {Pasquini}  \& {De Medeiros}}{{Melo}
  et~al.}{2001}]{melo01}
{Melo} C.~H.~F.,  {Pasquini} L.,   {De Medeiros} J.~R.,  2001, \mn@doi [\aap]
  {10.1051/0004-6361:20010897}, \href
  {http://adsabs.harvard.edu/abs/2001A%26A...375..851M} {375, 851}

\bibitem[\protect\citeauthoryear{{Montagnier} et~al.,}{{Montagnier}
  et~al.}{2006}]{montagnier06}
{Montagnier} G.,  et~al., 2006, \mn@doi [\aap] {10.1051/0004-6361:20066120},
  \href {http://adsabs.harvard.edu/abs/2006A%26A...460L..19M} {460, L19}

\bibitem[\protect\citeauthoryear{{Morin} et~al.,}{{Morin}
  et~al.}{2008a}]{morin08a}
{Morin} J.,  et~al., 2008a, \mn@doi [\mnras]
  {10.1111/j.1365-2966.2007.12709.x}, \href
  {http://adsabs.harvard.edu/abs/2008MNRAS.384...77M} {384, 77}

\bibitem[\protect\citeauthoryear{{Morin} et~al.,}{{Morin}
  et~al.}{2008b}]{morin08b}
{Morin} J.,  et~al., 2008b, \mn@doi [\mnras]
  {10.1111/j.1365-2966.2008.13809.x}, \href
  {http://adsabs.harvard.edu/abs/2008MNRAS.390..567M} {390, 567}

\bibitem[\protect\citeauthoryear{{Morin}, {Donati}, {Petit}, {Delfosse},
  {Forveille}  \& {Jardine}}{{Morin} et~al.}{2010}]{morin10}
{Morin} J.,  {Donati} J.-F.,  {Petit} P.,  {Delfosse} X.,  {Forveille} T.,
  {Jardine} M.~M.,  2010, \mn@doi [\mnras] {10.1111/j.1365-2966.2010.17101.x},
  \href {http://adsabs.harvard.edu/abs/2010MNRAS.407.2269M} {407, 2269}

\bibitem[\protect\citeauthoryear{{Morin} et~al.,}{{Morin}
  et~al.}{2011}]{morin11}
{Morin} J.,  et~al., 2011, in {Prasad Choudhary} D.,  {Strassmeier} K.~G.,
  eds,  IAU Symposium Vol. 273, Physics of Sun and Star Spots. pp 181--187
  (\mn@eprint {arXiv} {1009.2589}), \mn@doi{10.1017/S1743921311015213}

\bibitem[\protect\citeauthoryear{{Moutou} et~al.,}{{Moutou}
  et~al.}{2007}]{moutou07}
{Moutou} C.,  et~al., 2007, \mn@doi [\aap] {10.1051/0004-6361:20077795}, \href
  {http://adsabs.harvard.edu/abs/2007A%26A...473..651M} {473, 651}

\bibitem[\protect\citeauthoryear{{Moutou} et~al.,}{{Moutou}
  et~al.}{2017}]{moutou17}
{Moutou} C.,  et~al., 2017, \mn@doi [\mnras] {10.1093/mnras/stx2306}, \href
  {http://adsabs.harvard.edu/abs/2017MNRAS.472.4563M} {472, 4563}

\bibitem[\protect\citeauthoryear{{Nakajima}, {Oppenheimer}, {Kulkarni},
  {Golimowski}, {Matthews}  \& {Durrance}}{{Nakajima}
  et~al.}{1995}]{nakajima95}
{Nakajima} T.,  {Oppenheimer} B.~R.,  {Kulkarni} S.~R.,  {Golimowski} D.~A.,
  {Matthews} K.,   {Durrance} S.~T.,  1995, \mn@doi [\nat] {10.1038/378463a0},
  \href {http://adsabs.harvard.edu/abs/1995Natur.378..463N} {378, 463}

\bibitem[\protect\citeauthoryear{{Nelson}, {Robertson}, {Payne}, {Pritchard},
  {Deck}, {Ford}, {Wright}  \& {Isaacson}}{{Nelson} et~al.}{2016}]{nelson16}
{Nelson} B.~E.,  {Robertson} P.~M.,  {Payne} M.~J.,  {Pritchard} S.~M.,  {Deck}
  K.~M.,  {Ford} E.~B.,  {Wright} J.~T.,   {Isaacson} H.~T.,  2016, \mn@doi
  [\mnras] {10.1093/mnras/stv2367}, \href
  {http://adsabs.harvard.edu/abs/2016MNRAS.455.2484N} {455, 2484}

\bibitem[\protect\citeauthoryear{{Neves} et~al.,}{{Neves}
  et~al.}{2012}]{neves12}
{Neves} V.,  et~al., 2012, \mn@doi [\aap] {10.1051/0004-6361/201118115}, \href
  {http://adsabs.harvard.edu/abs/2012A%26A...538A..25N} {538, A25}

\bibitem[\protect\citeauthoryear{{Neves}, {Bonfils}, {Santos}, {Delfosse},
  {Forveille}, {Allard}  \& {Udry}}{{Neves} et~al.}{2013}]{neves13}
{Neves} V.,  {Bonfils} X.,  {Santos} N.~C.,  {Delfosse} X.,  {Forveille} T.,
  {Allard} F.,   {Udry} S.,  2013, \mn@doi [\aap]
  {10.1051/0004-6361/201220574}, \href
  {http://adsabs.harvard.edu/abs/2013A%26A...551A..36N} {551, A36}

\bibitem[\protect\citeauthoryear{{Neves}, {Bonfils}, {Santos}, {Delfosse},
  {Forveille}, {Allard}  \& {Udry}}{{Neves} et~al.}{2014}]{neves14}
{Neves} V.,  {Bonfils} X.,  {Santos} N.~C.,  {Delfosse} X.,  {Forveille} T.,
  {Allard} F.,   {Udry} S.,  2014, \mn@doi [\aap]
  {10.1051/0004-6361/201424139}, \href
  {http://adsabs.harvard.edu/abs/2014A%26A...568A.121N} {568, A121}

\bibitem[\protect\citeauthoryear{{Newton}, {Charbonneau}, {Irwin},
  {Berta-Thompson}, {Rojas-Ayala}, {Covey}  \& {Lloyd}}{{Newton}
  et~al.}{2014}]{newton14}
{Newton} E.~R.,  {Charbonneau} D.,  {Irwin} J.,  {Berta-Thompson} Z.~K.,
  {Rojas-Ayala} B.,  {Covey} K.,   {Lloyd} J.~P.,  2014, \mn@doi [\aj]
  {10.1088/0004-6256/147/1/20}, \href
  {http://adsabs.harvard.edu/abs/2014AJ....147...20N} {147, 20}

\bibitem[\protect\citeauthoryear{{Newton}, {Irwin}, {Charbonneau},
  {Berta-Thompson}, {Dittmann}  \& {West}}{{Newton} et~al.}{2016}]{newton16b}
{Newton} E.~R.,  {Irwin} J.,  {Charbonneau} D.,  {Berta-Thompson} Z.~K.,
  {Dittmann} J.~A.,   {West} A.~A.,  2016, \mn@doi [\apj]
  {10.3847/0004-637X/821/2/93}, \href
  {http://adsabs.harvard.edu/abs/2016ApJ...821...93N} {821, 93}

\bibitem[\protect\citeauthoryear{{Newton}, {Irwin}, {Charbonneau}, {Berlind},
  {Calkins}  \& {Mink}}{{Newton} et~al.}{2017}]{newton17}
{Newton} E.~R.,  {Irwin} J.,  {Charbonneau} D.,  {Berlind} P.,  {Calkins}
  M.~L.,   {Mink} J.,  2017, \mn@doi [\apj] {10.3847/1538-4357/834/1/85}, \href
  {http://adsabs.harvard.edu/abs/2017ApJ...834...85N} {834, 85}

\bibitem[\protect\citeauthoryear{{Nidever}, {Marcy}, {Butler}, {Fischer}  \&
  {Vogt}}{{Nidever} et~al.}{2002}]{nidever02}
{Nidever} D.~L.,  {Marcy} G.~W.,  {Butler} R.~P.,  {Fischer} D.~A.,   {Vogt}
  S.~S.,  2002, \mn@doi [\apjs] {10.1086/340570}, \href
  {http://adsabs.harvard.edu/abs/2002ApJS..141..503N} {141, 503}

\bibitem[\protect\citeauthoryear{{Norton} et~al.,}{{Norton}
  et~al.}{2007}]{norton07}
{Norton} A.~J.,  et~al., 2007, \mn@doi [\aap] {10.1051/0004-6361:20077084},
  \href {http://adsabs.harvard.edu/abs/2007A%26A...467..785N} {467, 785}

\bibitem[\protect\citeauthoryear{{{\"O}nehag}, {Heiter}, {Gustafsson},
  {Piskunov}, {Plez}  \& {Reiners}}{{{\"O}nehag} et~al.}{2012}]{onehag12}
{{\"O}nehag} A.,  {Heiter} U.,  {Gustafsson} B.,  {Piskunov} N.,  {Plez} B.,
  {Reiners} A.,  2012, \mn@doi [\aap] {10.1051/0004-6361/201118101}, \href
  {http://adsabs.harvard.edu/abs/2012A%26A...542A..33O} {542, A33}

\bibitem[\protect\citeauthoryear{{Pettersen}, {Coleman}  \&
  {Evans}}{{Pettersen} et~al.}{1984}]{pettersen84}
{Pettersen} B.~R.,  {Coleman} L.~A.,   {Evans} D.~S.,  1984, \mn@doi [\apj]
  {10.1086/162194}, \href {http://adsabs.harvard.edu/abs/1984ApJ...282..214P}
  {282, 214}

\bibitem[\protect\citeauthoryear{{Pr{\v s}a} et~al.,}{{Pr{\v s}a}
  et~al.}{2016}]{prsa16}
{Pr{\v s}a} A.,  et~al., 2016, \mn@doi [\aj] {10.3847/0004-6256/152/2/41},
  \href {http://adsabs.harvard.edu/abs/2016AJ....152...41P} {152, 41}

\bibitem[\protect\citeauthoryear{{Quirrenbach} et~al.,}{{Quirrenbach}
  et~al.}{2014}]{quirrenbach14}
{Quirrenbach} A.,  et~al., 2014, in Ground-based and Airborne Instrumentation
  for Astronomy V. p. 91471F, \mn@doi{10.1117/12.2056453}

\bibitem[\protect\citeauthoryear{{Rajpurohit}, {Reyl{\'e}}, {Allard},
  {Homeier}, {Schultheis}, {Bessell}  \& {Robin}}{{Rajpurohit}
  et~al.}{2013}]{rajpurohit13}
{Rajpurohit} A.~S.,  {Reyl{\'e}} C.,  {Allard} F.,  {Homeier} D.,  {Schultheis}
  M.,  {Bessell} M.~S.,   {Robin} A.~C.,  2013, \mn@doi [\aap]
  {10.1051/0004-6361/201321346}, \href
  {http://adsabs.harvard.edu/abs/2013A%26A...556A..15R} {556, A15}

\bibitem[\protect\citeauthoryear{{Rajpurohit}, {Allard}, {Teixeira}, {Homeier},
  {Rajpurohit}  \& {Mousis}}{{Rajpurohit} et~al.}{2017}]{rajpurohit17}
{Rajpurohit} A.~S.,  {Allard} F.,  {Teixeira} G.~D.~C.,  {Homeier} D.,
  {Rajpurohit} S.,   {Mousis} O.,  2017, preprint, \href
  {http://adsabs.harvard.edu/abs/2017arXiv170806211R} {} (\mn@eprint {arXiv}
  {1708.06211})

\bibitem[\protect\citeauthoryear{{Rebolo}, {Zapatero Osorio}  \&
  {Mart{\'{\i}}n}}{{Rebolo} et~al.}{1995}]{rebolo95}
{Rebolo} R.,  {Zapatero Osorio} M.~R.,   {Mart{\'{\i}}n} E.~L.,  1995, \mn@doi
  [\nat] {10.1038/377129a0}, \href
  {http://adsabs.harvard.edu/abs/1995Natur.377..129R} {377, 129}

\bibitem[\protect\citeauthoryear{{Reid} \& {Gizis}}{{Reid} \&
  {Gizis}}{1997}]{reid97}
{Reid} I.~N.,  {Gizis} J.~E.,  1997, \mn@doi [\aj] {10.1086/118436}, \href
  {http://adsabs.harvard.edu/abs/1997AJ....113.2246R} {113, 2246}

\bibitem[\protect\citeauthoryear{{Reid}, {Hawley}  \& {Gizis}}{{Reid}
  et~al.}{1995}]{reid95}
{Reid} I.~N.,  {Hawley} S.~L.,   {Gizis} J.~E.,  1995, \mn@doi [\aj]
  {10.1086/117655}, \href {http://adsabs.harvard.edu/abs/1995AJ....110.1838R}
  {110, 1838}

\bibitem[\protect\citeauthoryear{{Reiners}}{{Reiners}}{2007}]{reiners07a}
{Reiners} A.,  2007, \mn@doi [\aap] {10.1051/0004-6361:20066991}, \href
  {http://adsabs.harvard.edu/abs/2007A%26A...467..259R} {467, 259}

\bibitem[\protect\citeauthoryear{{Reiners} \& {Basri}}{{Reiners} \&
  {Basri}}{2007}]{reiners07b}
{Reiners} A.,  {Basri} G.,  2007, \mn@doi [\apj] {10.1086/510304}, \href
  {http://adsabs.harvard.edu/abs/2007ApJ...656.1121R} {656, 1121}

\bibitem[\protect\citeauthoryear{{Reiners}, {Basri}  \& {Browning}}{{Reiners}
  et~al.}{2009}]{reiners09}
{Reiners} A.,  {Basri} G.,   {Browning} M.,  2009, \mn@doi [\apj]
  {10.1088/0004-637X/692/1/538}, \href
  {http://adsabs.harvard.edu/abs/2009ApJ...692..538R} {692, 538}

\bibitem[\protect\citeauthoryear{{Reiners}, {Joshi}  \& {Goldman}}{{Reiners}
  et~al.}{2012}]{reiners12}
{Reiners} A.,  {Joshi} N.,   {Goldman} B.,  2012, \mn@doi [\aj]
  {10.1088/0004-6256/143/4/93}, \href
  {http://adsabs.harvard.edu/abs/2012AJ....143...93R} {143, 93}

\bibitem[\protect\citeauthoryear{{Reiners} et~al.,}{{Reiners}
  et~al.}{2017}]{reiners18}
{Reiners} A.,  et~al., 2017, preprint, \href
  {http://adsabs.harvard.edu/abs/2017arXiv171106576R} {} (\mn@eprint {arXiv}
  {1711.06576})

\bibitem[\protect\citeauthoryear{{Reuyl}}{{Reuyl}}{1943}]{reuyl43}
{Reuyl} D.,  1943, \mn@doi [\apj] {10.1086/144511}, \href
  {http://adsabs.harvard.edu/abs/1943ApJ....97..186R} {97, 186}

\bibitem[\protect\citeauthoryear{{Robertson}, {Mahadevan}, {Endl}  \&
  {Roy}}{{Robertson} et~al.}{2014}]{robertson14}
{Robertson} P.,  {Mahadevan} S.,  {Endl} M.,   {Roy} A.,  2014, \mn@doi
  [Science] {10.1126/science.1253253}, \href
  {http://adsabs.harvard.edu/abs/2014Sci...345..440R} {345, 440}

\bibitem[\protect\citeauthoryear{{Rojas-Ayala}, {Covey}, {Muirhead}  \&
  {Lloyd}}{{Rojas-Ayala} et~al.}{2012}]{rojas-ayala12}
{Rojas-Ayala} B.,  {Covey} K.~R.,  {Muirhead} P.~S.,   {Lloyd} J.~P.,  2012,
  \mn@doi [\apj] {10.1088/0004-637X/748/2/93}, \href
  {http://adsabs.harvard.edu/abs/2012ApJ...748...93R} {748, 93}

\bibitem[\protect\citeauthoryear{{Schlaufman} \& {Laughlin}}{{Schlaufman} \&
  {Laughlin}}{2010}]{schlaufman10}
{Schlaufman} K.~C.,  {Laughlin} G.,  2010, \mn@doi [\aap]
  {10.1051/0004-6361/201015016}, \href
  {http://adsabs.harvard.edu/abs/2010A%26A...519A.105S} {519, A105}

\bibitem[\protect\citeauthoryear{{Shkolnik}, {Liu}, {Reid}, {Hebb}, {Cameron},
  {Torres}  \& {Wilson}}{{Shkolnik} et~al.}{2008}]{shkolnik08}
{Shkolnik} E.,  {Liu} M.~C.,  {Reid} I.~N.,  {Hebb} L.,  {Cameron} A.~C.,
  {Torres} C.~A.,   {Wilson} D.~M.,  2008, \mn@doi [\apj] {10.1086/589850},
  \href {http://adsabs.harvard.edu/abs/2008ApJ...682.1248S} {682, 1248}

\bibitem[\protect\citeauthoryear{{Shkolnik}, {Liu}  \& {Reid}}{{Shkolnik}
  et~al.}{2009}]{shkolnik09}
{Shkolnik} E.,  {Liu} M.~C.,   {Reid} I.~N.,  2009, \mn@doi [\apj]
  {10.1088/0004-637X/699/1/649}, \href
  {http://adsabs.harvard.edu/abs/2009ApJ...699..649S} {699, 649}

\bibitem[\protect\citeauthoryear{{Shkolnik}, {Hebb}, {Liu}, {Reid}  \& {Collier
  Cameron}}{{Shkolnik} et~al.}{2010}]{shkolnik10}
{Shkolnik} E.~L.,  {Hebb} L.,  {Liu} M.~C.,  {Reid} I.~N.,   {Collier Cameron}
  A.,  2010, \mn@doi [\apj] {10.1088/0004-637X/716/2/1522}, \href
  {http://adsabs.harvard.edu/abs/2010ApJ...716.1522S} {716, 1522}

\bibitem[\protect\citeauthoryear{{Shkolnik}, {Anglada-Escud{\'e}}, {Liu},
  {Bowler}, {Weinberger}, {Boss}, {Reid}  \& {Tamura}}{{Shkolnik}
  et~al.}{2012}]{shkolnik12}
{Shkolnik} E.~L.,  {Anglada-Escud{\'e}} G.,  {Liu} M.~C.,  {Bowler} B.~P.,
  {Weinberger} A.~J.,  {Boss} A.~P.,  {Reid} I.~N.,   {Tamura} M.,  2012,
  \mn@doi [\apj] {10.1088/0004-637X/758/1/56}, \href
  {http://adsabs.harvard.edu/abs/2012ApJ...758...56S} {758, 56}

\bibitem[\protect\citeauthoryear{{Skelly}, {Unruh}, {Collier Cameron},
  {Barnes}, {Donati}, {Lawson}  \& {Carter}}{{Skelly} et~al.}{2008}]{skelly08}
{Skelly} M.~B.,  {Unruh} Y.~C.,  {Collier Cameron} A.,  {Barnes} J.~R.,
  {Donati} J.-F.,  {Lawson} W.~A.,   {Carter} B.~D.,  2008, \mn@doi [\mnras]
  {10.1111/j.1365-2966.2008.12917.x}, \href
  {http://adsabs.harvard.edu/abs/2008MNRAS.385..708S} {385, 708}

\bibitem[\protect\citeauthoryear{{Skrutskie} et~al.,}{{Skrutskie}
  et~al.}{2006}]{skrutskie06}
{Skrutskie} M.~F.,  et~al., 2006, \mn@doi [\aj] {10.1086/498708}, \href
  {http://adsabs.harvard.edu/abs/2006AJ....131.1163S} {131, 1163}

\bibitem[\protect\citeauthoryear{{Strand}}{{Strand}}{1977}]{strand77}
{Strand} K.~A.,  1977, \mn@doi [\aj] {10.1086/112119}, \href
  {http://adsabs.harvard.edu/abs/1977AJ.....82..745S} {82, 745}

\bibitem[\protect\citeauthoryear{{Su{\'a}rez Mascare{\~n}o}, {Rebolo},
  {Gonz{\'a}lez Hern{\'a}ndez}  \& {Esposito}}{{Su{\'a}rez Mascare{\~n}o}
  et~al.}{2015}]{suarez15}
{Su{\'a}rez Mascare{\~n}o} A.,  {Rebolo} R.,  {Gonz{\'a}lez Hern{\'a}ndez}
  J.~I.,   {Esposito} M.,  2015, \mn@doi [\mnras] {10.1093/mnras/stv1441},
  \href {http://adsabs.harvard.edu/abs/2015MNRAS.452.2745S} {452, 2745}

\bibitem[\protect\citeauthoryear{{Su{\'a}rez Mascare{\~n}o}, {Rebolo}  \&
  {Gonz{\'a}lez Hern{\'a}ndez}}{{Su{\'a}rez Mascare{\~n}o}
  et~al.}{2016}]{suarez16}
{Su{\'a}rez Mascare{\~n}o} A.,  {Rebolo} R.,   {Gonz{\'a}lez Hern{\'a}ndez}
  J.~I.,  2016, \mn@doi [\aap] {10.1051/0004-6361/201628586}, \href
  {http://adsabs.harvard.edu/abs/2016A%26A...595A..12S} {595, A12}

\bibitem[\protect\citeauthoryear{{Su{\'a}rez Mascare{\~n}o}
  et~al.,}{{Su{\'a}rez Mascare{\~n}o} et~al.}{2017}]{suarez17}
{Su{\'a}rez Mascare{\~n}o} A.,  et~al., 2017, \mn@doi [\aap]
  {10.1051/0004-6361/201730957}, \href
  {http://adsabs.harvard.edu/abs/2017A%26A...605A..92S} {605, A92}

\bibitem[\protect\citeauthoryear{{Tamazian}, {Docobo}  \& {Balega}}{{Tamazian}
  et~al.}{2008}]{tamazian08}
{Tamazian} V.~S.,  {Docobo} J.~A.,   {Balega} Y.~Y.,  2008, in {Hubrig} S.,
  {Petr-Gotzens} M.,   {Tokovinin} A.,  eds, Multiple Stars Across the H-R
  Diagram. p.~71, \mn@doi{10.1007/978-3-540-74745-1_11}

\bibitem[\protect\citeauthoryear{{Terrien}, {Mahadevan}, {Bender}, {Deshpande},
  {Ramsey}  \& {Bochanski}}{{Terrien} et~al.}{2012}]{terrien12}
{Terrien} R.~C.,  {Mahadevan} S.,  {Bender} C.~F.,  {Deshpande} R.,  {Ramsey}
  L.~W.,   {Bochanski} J.~J.,  2012, \mn@doi [\apjl]
  {10.1088/2041-8205/747/2/L38}, \href
  {http://adsabs.harvard.edu/abs/2012ApJ...747L..38T} {747, L38}

\bibitem[\protect\citeauthoryear{{Terrien}, {Mahadevan}, {Deshpande}  \&
  {Bender}}{{Terrien} et~al.}{2015}]{terrien15b}
{Terrien} R.~C.,  {Mahadevan} S.,  {Deshpande} R.,   {Bender} C.~F.,  2015,
  \mn@doi [\apjs] {10.1088/0067-0049/220/1/16}, \href
  {http://adsabs.harvard.edu/abs/2015ApJS..220...16T} {220, 16}

\bibitem[\protect\citeauthoryear{{Thebault} \& {Haghighipour}}{{Thebault} \&
  {Haghighipour}}{2014}]{thebault14}
{Thebault} P.,  {Haghighipour} N.,  2014, preprint, \href
  {http://adsabs.harvard.edu/abs/2014arXiv1406.1357T} {} (\mn@eprint {arXiv}
  {1406.1357})

\bibitem[\protect\citeauthoryear{{Tokovinin}}{{Tokovinin}}{1997}]{tokovinin97}
{Tokovinin} A.~A.,  1997, \mn@doi [\aaps] {10.1051/aas:1997181}, \href
  {http://cdsads.u-strasbg.fr/abs/1997A%26AS..124...75T} {124}

\bibitem[\protect\citeauthoryear{{Tomkin} \& {Pettersen}}{{Tomkin} \&
  {Pettersen}}{1986}]{tomkin86}
{Tomkin} J.,  {Pettersen} B.~R.,  1986, \mn@doi [\aj] {10.1086/114278}, \href
  {http://adsabs.harvard.edu/abs/1986AJ.....92.1424T} {92, 1424}

\bibitem[\protect\citeauthoryear{{Torres}, {Stefanik}, {Latham}  \&
  {Mazeh}}{{Torres} et~al.}{1995}]{torres95}
{Torres} G.,  {Stefanik} R.~P.,  {Latham} D.~W.,   {Mazeh} T.,  1995, \mn@doi
  [\apj] {10.1086/176355}, \href
  {http://adsabs.harvard.edu/abs/1995ApJ...452..870T} {452, 870}

\bibitem[\protect\citeauthoryear{{Torres}, {Quast}, {da Silva}, {de La Reza},
  {Melo}  \& {Sterzik}}{{Torres} et~al.}{2006}]{torres06}
{Torres} C.~A.~O.,  {Quast} G.~R.,  {da Silva} L.,  {de La Reza} R.,  {Melo}
  C.~H.~F.,   {Sterzik} M.,  2006, \mn@doi [\aap] {10.1051/0004-6361:20065602},
  \href {http://adsabs.harvard.edu/abs/2006A%26A...460..695T} {460, 695}

\bibitem[\protect\citeauthoryear{{Vinter Hansen}}{{Vinter
  Hansen}}{1940}]{vinterhansen40}
{Vinter Hansen} J.~M.,  1940, \mn@doi [\pasp] {10.1086/125213}, \href
  {http://adsabs.harvard.edu/abs/1940PASP...52..329V} {52, 329}

\bibitem[\protect\citeauthoryear{{Ward-Duong} et~al.,}{{Ward-Duong}
  et~al.}{2015}]{ward15}
{Ward-Duong} K.,  et~al., 2015, \mn@doi [\mnras] {10.1093/mnras/stv384}, \href
  {http://adsabs.harvard.edu/abs/2015MNRAS.449.2618W} {449, 2618}

\bibitem[\protect\citeauthoryear{{Watson}}{{Watson}}{2006}]{watson06}
{Watson} C.~L.,  2006, Society for Astronomical Sciences Annual Symposium,
  \href {http://cdsads.u-strasbg.fr/abs/2006SASS...25...47W} {25, 47}

\bibitem[\protect\citeauthoryear{{West} et~al.,}{{West} et~al.}{2004}]{west04}
{West} A.~A.,  et~al., 2004, \mn@doi [\aj] {10.1086/421364}, \href
  {http://adsabs.harvard.edu/abs/2004AJ....128..426W} {128, 426}

\bibitem[\protect\citeauthoryear{{West}, {Weisenburger}, {Irwin},
  {Berta-Thompson}, {Charbonneau}, {Dittmann}  \& {Pineda}}{{West}
  et~al.}{2015}]{west15}
{West} A.~A.,  {Weisenburger} K.~L.,  {Irwin} J.,  {Berta-Thompson} Z.~K.,
  {Charbonneau} D.,  {Dittmann} J.,   {Pineda} J.~S.,  2015, \mn@doi [\apj]
  {10.1088/0004-637X/812/1/3}, \href
  {http://adsabs.harvard.edu/abs/2015ApJ...812....3W} {812, 3}

\bibitem[\protect\citeauthoryear{{Winters} et~al.,}{{Winters}
  et~al.}{2015}]{winters15}
{Winters} J.~G.,  et~al., 2015, \mn@doi [\aj] {10.1088/0004-6256/149/1/5},
  \href {http://adsabs.harvard.edu/abs/2015AJ....149....5W} {149, 5}

\bibitem[\protect\citeauthoryear{{Woolf} \& {Wallerstein}}{{Woolf} \&
  {Wallerstein}}{2005}]{woolf05}
{Woolf} V.~M.,  {Wallerstein} G.,  2005, \mn@doi [\mnras]
  {10.1111/j.1365-2966.2004.08515.x}, \href
  {http://adsabs.harvard.edu/abs/2005MNRAS.356..963W} {356, 963}

\bibitem[\protect\citeauthoryear{{Woolf} \& {Wallerstein}}{{Woolf} \&
  {Wallerstein}}{2006}]{woolf06}
{Woolf} V.~M.,  {Wallerstein} G.,  2006, \mn@doi [\pasp] {10.1086/498459},
  \href {http://adsabs.harvard.edu/abs/2006PASP..118..218W} {118, 218}

\bibitem[\protect\citeauthoryear{{van Gent}}{{van Gent}}{1926}]{vangent26}
{van Gent} H.,  1926, \bain, \href
  {http://adsabs.harvard.edu/abs/1926BAN.....3..121V} {3, 121}

\makeatother
\end{thebibliography}

\input{coolsnap1.bbl}



\section*{Appendices}

Long tables of the paper are given in theses appendices.

\subsection{Table of double-line spectroscopic binaries}

In Table~\ref{tab:spectro-bin2} are listed all the SB2 systems in our sample, detected in this work and from the literature.

\begin{table*}
	\centering
	\caption{Spectroscopic binaries detected in the observations of the CoolSnap sample, or listed in the literature and recovered from the ESPaDOnS archives (polarimetry and pure spectroscopy).}
	\label{tab:spectro-bin2}
	\begin{tabular}{llllcc}
		\hline
		2MASS name & Common name & SB type & Reference \\
		\hline
J00080642+4757025 & & SB2 & \citet{shkolnik10} \\
J00424820+3532554 & Gl~29.1A & SB2 & this work \\
J01351393-0712517 & & SB2 & \citet{malo14a} \\
J01434512-0602400 & & SB2 & this work \\
J01451820+4632077 & LHS~6032 & SB2 & \citet{shkolnik10} \\
J01591260+0331113 & GJ~1041B & SB2 & \citet{shkolnik10} \\
J02441245-1321387 & LP~711-62 & SB2 & this work \\
J03371407+6910498 & GJ~3236 & SB2 & \citet{shkolnik10} \\
J03373331+1751145 & GJ~3239 & SB2 & \citet{shkolnik10} \\
J04134585-0509049 & G~160-54 & SB3 & \citet{bowler15a} \\
J04244260-0647313 & & SB3 & \citet{shkolnik10} \\
J05031607+2123563 & HD~285190 & SB2 & this work \\
J06180730+7506032 & & SB3? & this work \\
J06573891+4951540 & & SB2 & this work \\
J07100180+3831457 & Gl 268 & SB2 & \citet{tomkin86} \\
J07282116+3345127 & & SB2 & \citet{shkolnik10,malo14a} \\
J07313848+4557173 & & SB2? & this work \\
J07343745+3152102 & Gl~278C & DESB2 & \citet{leung78} \\
J08313759+1923395 & GJ~2069A & DESB2 & \citet{delfosse99a} \\
J08585633+0828259 & GJ~3522 & SB2 & \citet{reid97,delfosse99b} \\
J09091563-1236184 & & SB3 & this work \\
J09201112-0110171 & G~161-13 & SB2 & this work \\
J09361593+3731456 & & SB2 & \citet{malo14a} \\
J10182870-3150029 & TWA~6 & SB2? & this work, but see Table~\ref{tab:special} \\
J10364812+5055041 & G~196-37 & SB2? & this work \\
J11220530-2446393 & TWA~4 & SB1+SB2 & \citet{karatas04} \\
J11250052+4319393 & LHS~2403 & SB2? & this work, but see Table~\ref{tab:special} \\
J11515681+0731262 & & SB2 & this work \\
J12165845+3109233 & GJ~3719 & SB2 & this work \\
J12290290+4143497 & GJ~3729 & SB2(3?) & \citet{shkolnik12} \\
J12490273+6606366 & Gl~487 & SB3 & \citet{delfosse99b} \\
J12521285+2908568 & LP321-163 & SB2? & this work \\
J14170294+3142472 & GJ~3839 & SB3 & \citet{delfosse99b}, Forveille p.c. \\
J14493338-2606205 & Gl~563.2A & SB2 & this work \\
J15235385+5609320 & & SB2 & this work \\
J16155939+3852102 & & SB2 & this work \\
J16170537+5516094 & Gl~616.2 & SB2 & \citet{shkolnik10} \\
J16411543+5344110 & & SB2? & this work \\
J16552880-0820103 & Gl~644 & SB2(3?) & \citet{pettersen84,delfosse99b} \\
J17035283+3211456 & NLTT~44114 & SB2? & this work \\
J17462934-0842362 & G~20-13 & SB2 & \citet{malo14a} \\
J18410977+2447143 & GJ~1230A & SB2 & \citet{gizis96,delfosse99b} \\
J18552740+0824090 & Gl~735 & SB2 & \citet{karatas04} \\
J18561590+5431479 & G~229-18 & SB2 & this work \\
J18580415-2953045 & TYC~6872-1011-1 & SB2? & this work \\
J19420065-2104051 & LP~869-19 & SB2 & \citet{malo14a} \\
J20103444+0632140 & NLTT~48838 & SB2 & \citet{shkolnik10} \\
J21000529+4004136 & Gl~815 & SB2 & \citet{karatas04} \\
J21293671+1738353 & Gl~829 & SB2 & \citet{delfosse99b} \\
J22143835-2141535 & BD-22~5866 & ESB2+SB2 & \citet{shkolnik08} \\
J22384559-2037160 & Gl~867A & SB2 & \citet{herbig65} \\
J23062378+1236269 & G~67-46 & SB2(3?) & \citet{shkolnik10} \\
J23172441+3812419 & GJ~4327 & SB2 & \citet{cortes17} \\
J23301341-2023271 & GJ~1284 & SB2 & \citet{torres06} \\
J23435944+6444291 & GJ~4359 & SB2 & \citet{shkolnik10} \\
J23483610-2739385 & GJ~4362 & SB2 & \citet{shkolnik10} \\
J23584342+4643452 & Gl~913 & SB2? & this work \\
		\hline
	\end{tabular}
\end{table*}

\subsection{Table of multiple systems involving M dwarfs in our sample}

In Table~\ref{tab:vis-bin}, we list 153 multiple systems from this compilation and involving at least one of the M dwarfs of our sample, detected by imagery, with the level of multiplicity and the component we measured in parentheses following the WDS notation; we also give the most recent projected separation and the corresponding position angle, or the semi-major axis when the orbit is known; in that case, the position angle is listed as "sma"; we list the physical status of the system (common proper motion, orbital monitoring), the year as given in the WDS, or the reference of discovery when more recent. For multiplicity larger than 2, we also list the separations and position angles for each pair composing the system (or the semi-major axes when "sma" is listed as position angle), with a classical notation used to define the targeted pair (AB, Aab, Bab, ...), following the WDS when possible. Twenty-three stars listed in Table~\ref{tab:spectro-bin2} and Table~\ref{tab:spectro-bin1}, which are only spectroscopic binaries are not repeated in this Table, but spectroscopic binaries belonging to visual systems of higher multiplicity are included.

\onecolumn
\begin{landscape}
\begin{longtable}{lllllrll}
	\caption{List of 153 multiple systems detected visually (adaptive optics, lucky imaging, coronagraphy) and involving M dwarfs from our sample.}
	\label{tab:vis-bin}\\
    \hline
    \noalign{\vskip 0.1cm}
    2MASS name & Common name & N (component) & Pair & $\rho$ or $a$ & $\theta$ & PM, orbit & Discovery \\
    & & & & arcsec & degree & \\
    \noalign{\vskip 0.1cm}
    \hline
    \noalign{\vskip 0.1cm}
    \endfirsthead
    \caption{continued.}\\
    \hline
    \noalign{\vskip 0.1cm}
    2MASS name & Common name & N (component) & Pair & $\rho$ or $a$ & $\theta$ & PM, orbit & Discovery \\
    & & & & arcsec & degree & \\
    \noalign{\vskip 0.1cm}
    \hline
    \noalign{\vskip 0.1cm}
    \endhead
    \noalign{\vskip 0.1cm}
    \hline
    \endfoot
    \endlastfoot
J00155808-1636578 & & 2 & & 0.1045 & 90 & & \citet{shkolnik12} \\
J00161455+1951385 & GJ~1006A & 3 (A) & AB & 25.2 & 58 & & 1936 \\
& & & AC & 9.6 & 336 & & 1969 \\
J00182256+4401222 & Gl~15A & 2 (A) & AB & 34.3 & 64 & CPM, OM & 1860 \\
J00182549+4401376 & Gl~15B & 2 (B) & AB & 34.3 & 64 & CPM, OM & 1860 \\
J00233468+2014282 & FK~Psc & 2 & & 1.6 & 143 & & Skiff p.c. to WDS \\
J00340843+2523498 & V493~And & 2 & & 1.5 & 103 & & Skiff p.c. to WDS \\
J00424820+3532554 & Gl~29.1A & 2 (A) & AB & 15.8 & 271 & & 1950 \\
& & & A & & & SB2 & this work \\
J00485822+4435091 & GJ~3058 & 2 (AB) & AB & 1.027 & 256 & CPM, OM & \citet{mccarthy01} \\
J01023895+6220422 & Gl~49 & 2 (A) & AB & 294.8 & 76 & & 1952 \\
J01031971+6221557 & Gl~51 & 2 (B) & AB & 294.8 & 76 & & 1952 \\
J01034013+4051288 & G~132-50 & 4 (A) & AB & 26.4 & 120 & CPM & 1960 \\
& & & Aab & 0.267 & 308 & & \citet{shkolnik12} \\
J01034210+4051158 & G~132-51 & 4 (BC) & BC & 2.477 & 97 & CPM & 1960 \\
J01112542+1526214 & GJ~3076 & 2 & & 0.327 & 241 & CPM, OM & \citet{beuzit04} \\
J01155017+4702023 & & 4 (AB) & AB-CD & 27.1 & 330 & CPM? & 1998 \\
& & & AB & 0.272 & 250 & & \citet{law08} \\
& & & CD & 0.271 & 268 & & \citet{janson12} \\
J01365516-0647379 & G~271-110 & 2 & & & & CPM with EX~Cet & \citet{shkolnik12} \\
J01373940+1835332 & TYC~1208-468-1 & 2 (A) & AB & 1.7 & 24 & & 1968 \\
J01390120-1757026 & Gl~65 & 2 (AB) & AB & 2.046 & sma & OM & 1935 \\
J01451820+4632077 & G~173-18 & 2 & & & & SB2, VB & \citet{shkolnik09,shkolnik10} \\
J01535076-1459503 & & 2 & AB & 2.879 & 292 & CPM & \citet{bergfors10} \\
J01591239+0331092 & GJ~1041A & 3 (A) & AB & 3.2 & 53 & & 1960 \\
J01591260+0331113 & GJ~1041B & 3 (Bab) & Bab & & & SB2 & \citet{shkolnik09} \\
J02110221-3540146 & HIP~10191 & 3 (A) & AB & 3.4 & 143 & & 1925 \\
& & & AC & 13.3 & 37 & & 1912 \\
J02132062+3648506 & & 2 & & 0.217 & 76 & CPM, OM & \citet{janson12} \\
J02155892-0929121 & & 4 (AabBC) & Aab & 0.042 & 308 & & \citet{bowler15b} \\
& & & AB & 0.576 & 290 & CPM, OM & \citet{bergfors10} \\
& & & AC & 3.43 & 299 & CPM & \citet{bergfors10} \\
J02272804+3058405 & BD+30~397B & 2 (B) & AB & 22.0 & 316 & & 1954 \\
J02272924+3058246 & AG~Tri & 2 (A) & AB & 22.0 & 316 & & 1954 \\
J03143273+5926160 & G~246-29 & 2 & & & & & this work \\
J03192872+6156045 & G~246-33 & 2 & & 0.384 & 241 & & \citet{janson14a} \\
J03323578+2843554 & & 3 & AB & 0.482 & 106 & CPM & \citet{janson12} \\
& & & BC & 0.098 & 282 & CPM & \citet{janson12} \\
J03373331+1751145 & GJ~3239 & 4 (Aab) & AB & 16.2 & 151 & & 1960 \\
& & & Aab & & & SB2 & \citet{shkolnik10} \\
& & & Bab & & & E?SB2 & \citet{shkolnik10} \\
J03591438+8020019 & & 2 & & 0.200 & 357 & & \citet{janson12} \\
J04134585-0509049 & G~160-54 & 4 (Aab) & AB & 3.332 & 108 & & \citet{bowler15a} \\
& & & Aab & 0.1667 & 123 & SB3 & \citet{bowler15a,bowler15b} \\
J04311147+5858375 & Gl~169.1A & 3 (Aab) & AB & 9.88 & 60 & CPM & 1908 \\
& & & Aab & 0.07 & sma & OM & \citet{strand77} \\
J05024924+7352143 & & 2 & & 0.301 & 82 & CPM & \citet{janson12} \\
J05031607+2123563 & HD~285190 & 4 (Aab) & AB & 166.3 & 241 & & 1960 \\
& & & Aab & & & SB2 & this work \\
& & & Bab & 0.302 & 168 & CPM & \citet{law08} \\
J05100427-2340407 & & 4 (Aab) & A-BC & 27.2 & 18 & & 1998 \\
& & & Aab & 0.522 &128 & CPM, OM & \citet{janson12} \\
& & & BC & 1.815 & 307 & CPM & \citet{janson12} \\
J05241914-1601153 & & 2 & & 0.613 & 68 & CPM, OM & \citet{bergfors10} \\
J06103462-2151521 & Gl~229 & 2 (A) & AB & 6.8 & 168 & & \citet{nakajima95} \\
J06211300+4414307 & GJ~3391 & 2 & & 1.319 & 204 & &  \citet{cortes17} \\
J07272450+0513329 & Gl~273 & 2 & & 0.17 & 327 & & \citet{ward15} \\
J07293108+3556003 & & 2 & & 0.198 & 262 & CPM, OM & \citet{janson12} \\
J07313848+4557173 & & 2 & & 0.206 & 353 & CPM, OM & \citet{janson12} \\
J07315735+3613477 & Gl~277B & 3 (B) & AB & 38.04 & 353 & CPM & 1930 \\
J07315773+3613102 & Gl~277A & 3 (A) & Aab & 1.53 & 194 & OM & \citet{beuzit04} \\
J07320291+1719103 & G~88-36 & 3 (Aa or Ab ?) & Aab & 5.1 & 116 & & Hipparcos \\
& & & AB & 11.2 & 281 & & 1960 \\
J07343745+3152102 & Gl~278C & 6 (C) & AB & 6.805 & sma & CPM, OM, DESB2 & 1778 \\
& & & Aab & & & SB1 & \citet{vinterhansen40} \\
& & & Bab & & & SB1 & \citet{vinterhansen40} \\
& & & AC & 70.1 & 163 & CPM & 1822 \\
& & & Cab & & & DESB2 & \citet{joy26,vangent26} \\
J07345632+1445544 & TYC 777-141-1 & 2 & & 1.00 & 293 & & \citet{cortes17} \\
J07384089-2113276 & LHS~1935 & 2 & & & & & \\
J07505369+4428181 & & 2 & & 2.031 & 142 & CPM & \citet{janson12} \\
J07583098+1530146 & GJ~3468 & 2 (A) & AB & 16.1 & 208 & & 1960 \\
J08081317+2106182 & GJ~3481 & 4 (A) & AB & 10.633 & 144 & CPM & 1893 \\
& & & Bab & & & SB2 & \citet{shkolnik10} \\
& & & BabC & 0.580 & 36 & & \citet{shkolnik10} \\
J08085639+3249118 & GJ~1108A & 4 (Aab) & AB & 13.9 & 240 & & 1950 \\
& & & Aab & 0.25 & & & \citet{brandt14} \\
& & & Bab & & & SB2 & \citet{shkolnik10} \\
J08103429-1348514 & Gl~297.2B & 3 (Bab) & AB & 97.3 & 236 & CPM & 1920 \\
& & & Bab & 0.913 & 283 & & \citet{jodar13} \\
J08310177+4012115 & & 2 & & 1.899 & 122 & CPM & \citet{mason01} \\
J08313744+1923494 & GJ~2069B & 5 (Bab) & Bab (or BD) & 0.957 & 191 & SB2 not confirmed & \citet{delfosse99b}, this work \\
J08313759+1923395 & GJ~2069A & 5 (AabE) & AB & 9.7 & 349 & CPM & 1936 \\
& & & Aab (or AC) & 0.0028 & sma & OM, DESB2 & \citet{delfosse99b} \\
& & & AE & 0.536 & 181 & & \citet{beuzit04} \\
J08524466+2230523 & NLTT~20426 & 2 & & 4.6 & & & \\
J08585633+0828259 & GJ~3522 & 3 (AC-B) & AC-B & 0.424 & sma & CPM, OM & \citet{delfosse99b} \\
& & & Aab (or AC) & & & SB2 & \citet{reid97} \\
J09142298+5241125 & Gl~338A & 3 (Aab) & AB & 16.725 & sma & CPM, OM & 1821 \\
& & & Aab & & & SB1 & \citet{cortes17} \\
J09142485+5241118 & Gl~338B & 3 (B) & AB & 16.725 & sma & CPM, OM & 1821 \\
J09423493+7002024 & Gl~360 & 2 (A) & AB & 89 & 77 & CPM & 1894 \\
J10141918+2104297 & GJ~2079 & 2 & & 0.095 & 320 & SB1 ? & \citet{makarov05} \\
J10193634+1952122 & Gl~388 & 4 (Cab) & AB & 4.7 & 127 & & 1820 \\
& & & AC & 336.0 & 288 & & 1851 \\
& & & Cab & 0.110 & sma & OM & \citet{reuyl43} \\
J10452148+3830422 & Gl~400 & 2 & & 1.791 & sma & OM & \citet{hartkopf94} \\
J11052903+4331357 & Gl~412A & 2 (A) & AB & 31.8 & 125 & CPM & 1950 \\
J11053133+4331170 & Gl~412B & 2 (B) & AB & 31.8 & 125 & CPM & 1950 \\
J11110245+3026415 & Gl~414B & 2 (B) & AB & 34.1 & 263 & CPM & 1844 \\
J11115176+3332111 & GJ~3647 & 2 & & 5.1 & & & \\
J11200526+6550470 & Gl~424 & 2 & & 0.132 & 334 & & \citet{tamazian08} \\
J11220530-2446393 & TWA~4 & 4 (AB) & AB & 1.030 & sma & & 1909 \\
& & & Aab & & & SB1 & \citet{torres95} \\
& & & Bab & 0.0233 & sma & OM, SB2 & \citet{torres95} \\
J11515681+0731262 & & 3 (AabB) & AB & 0.514 & 107 & & \citet{bowler15b} \\
& & & Aab & & & SB2 & \citet{bowler15b} \\
J12290290+4143497 & GJ~3729 & 2 & & 0.0503 & 256 & SB2 & \citet{shkolnik12} \\
J12490273+6606366 & Gl~487 & 3 (AabB) & AB & 0.297 & 15 & & \citet{delfosse99b} \\
& & & AabB & & & SB3 & \citet{delfosse99b} \\
J12574030+3513306 & Gl~490A & 4 (Aab) & AB & 16.0 & 227 & CPM & 1950 \\
& & & Aab & 0.10 & 240 & & \citet{shkolnik12} \\
J12573935+3513194 & Gl~490B & 4 (Bab) & Bab & 0.20 & 310 & & \citet{shkolnik12} \\
J13004666+1222325 & Gl~494 & 3 (AB) & AB & 0.051 & sma & OM & \citet{heintz94,beuzit04} \\
& & & AC & 102.1 & 220 & CPM & \citet{goldman10} \\
J13093495+2859065 & GJ~1167A & 2 & & 193.6 & 28 & CPM & 1965 \\
J13142039+1320011 & NLTT~33370 & 2 & & 0.2 & 50 & & \citet{law06} \\
J13282106-0221365 & Gl~512A & 2 (A) & AB & 8.5 & 52 & CPM & 1937 \\
J13314666+2916368 & GJ~3789 & 2 (AB) & AB & 0.190 & 85 & & \citet{beuzit04} \\
J13345147+3746195 & & 2 & & 0.082 & 198 & & \citet{daemgen07} \\
J13414631+5815197 & & 2 & & 0.699 & 251 & CPM, OM & \citet{janson12} \\
J14154197+5927274 & & 2 & & 5.064 & 231 & & \citet{cortes17} \\
J14170294+3142472 & GJ~3839 & 2 & & 0.439 & 219 & SB3 & \citet{delfosse99b}, Forveille p.c. \\
J14493338-2606205 & Gl~563.2A & 3 (Aab) & AB & 26.7 & 244 & & 1920 \\
& & & Aab & & & SB2 & this work \\
J14511044+3106406 & G~166-49 & 2 & & 2.353 & 48 & CPM, OM & \citet{janson12} \\
& & & Aab & & & SB2 & this work \\
J15123818+4543464 & GJ~3898 & 2 & & 0.481 & 220 & & \citet{mccarthy01} \\
J15235385+5609320 & & 2 & AB & 68 & 248 & & 1912 \\
& & & A & & & SB2 & this work \\
J15493833+3448555 & GJ~3919 & 2 & & 0.208 & 99 & & \citet{cortes17} \\
J15553178+3512028 & GJ~3928 & 2 & & 1.620 & 255 & & \citet{mccarthy01} \\
J15594729+4403595 & & 2 & & 5.67 & 284 & CPM? & 2000 \\
J16164537+6715224 & Gl~617B & 2 (B) & AB & 64.5 & 13 & & 1892 \\
J16170537+5516094 & Gl~616.2 & 2 & & 0.148 & sma & SB2, OM & \citet{shkolnik10} \\
J16240913+4821112 & Gl~623 & 3 (Aab) & AB & 176 & 288 & & 1911 \\
& & & Aab & 0.2397 & sma & OM, SB1 & \citet{martinache07} \\
J16352740+3500577 & GJ~3966 & 2 & & 0.0922 & 26 & & \citet{bowler15a} \\
J16411543+5344110 & & 2 & & 0.099 & 94 & CPM, OM, SB2? & \citet{janson12}, this work \\
J16552880-0820103 & Gl~644 & 5 (AB) & AB & 0.2256 & sma & OM & 1934 \\
& & & Bab & & & SB2 (3?) & \citet{pettersen84} \\
& & & AB-C & 72.2 & 313 & CPM & 1920 \\
J16553529-0823401 & Gl~644C & 5 (F) & AB-F & 230.6 & 155 & CPM & 1954 \\
J16575357+4722016 & Gl~649.1B & 3 (B) & AB & 2.79 & sma & OM & 1908 \\
& & & AC & 89.1 & sma & OM & 1823 \\
J16590962+2058160 & V1234~Her & 2 & & 0.689 & 139 & CPM, OM & \citet{janson12} \\
J17021204+5103284 & & 2 & & 0.816 & 63 & CPM & \citet{janson12} \\
J17035188+3211523 & LP~331-57B & 2 (B) & AB & 1.260 & 143 & CPM, OM & \citet{daemgen07} \\
J17035283+3211456 & LP~331-57A & 2 (A) & AB & 1.260 & 143 & CPM, OM, SB2? & \citet{daemgen07}, this work \\
J17155010+1900000 & GJ~3997 & 2 & & 1.841 & 267 & & \citet{jodar13} \\
J17195422+2630030 & Gl~669A & 3 (A) & AB & 16.7 & 269 & CPM & 1936 \\
& & & Bab & & & & \citet{shkolnik12} \\
J17294104-1748323 & BD-17~4821B & 2 (B) & AB & 8.5 & 194 & & 1830 \\
J17362594+6820220 & Gl~687 & 4 (Bab) & AB & 180.4 & 210 & & 1877 \\
& & & Aab & 0.30 & 352 & & 1984 \\
& & & Bab & 0.033 & sma & OM & \citet{lippincott77} \\
J17375330+1835295 & Gl~686 & 2 & & 0.040 & sma & OM & \citet{bieger64} \\
J17380077+3329457 & & 2 & & 1.029 & 158 & CPM? & \citet{janson12} \\
J17462507+2743014 & Gl~695BC & 5 (BC) & AD & 321.1 & 5 & & 1921 \\
& & & A-BC & 35.5 & 249 & CPM & 1781 \\
& & & Aab & 0.265 & sma & OM & 1998 \\
& & & BC & 1.36 & sma & OM & 1854 \\
J18130657+2601519 & GJ~4044 & 3 & & 1.45 & 226 & CPM & \citet{shkolnik12} \\
J18351833+4544379 & Gl~720A & 2 (A) & AB & 112.1 & 56 & & 1960 \\
J18410977+2447143 & GJ~1230A & 3 (Aab) & AB & 4.83 & 6 & & 1905 \\
& & & Aab & & & SB2 & \citet{gizis96} \\
J18424666+5937499 & Gl~725A & 3 (A) & AB & 13.88 & sma & CPM, OM & 1831 \\
J18424688+5937374 & Gl~725B & 3 (Bab) & Bab & 0.028 & sma & OM & \citet{baize76} \\
J18440971+7129178 & & 2 & AB & 2.30 & 97 & & 1963 \\
J18441019+7129175 & & 2 & AB & 2.30 & 97 & & 1963 \\
J18561590+5431479 & G~229-18 & 3 (Aab) & Aab & 0.4 & 306 & SB2 & 1991 \\
& & & AB & 118.8 & 170 & & 1905 \\
J19071320+2052372 & Gl~745B & 2 (A) & AB & 114.5 & 290 & CPM & 1897 \\
J19165762+0509021 & Gl~752B & 2 & & 75.8 & 152 & CPM & 1942 \\
J19445376-2337591 & LP~869-26 & 2 (AB) & AB & 0.60 & 341 & & \citet{montagnier06} \\
J19535443+4424541 & GJ~1245A & 3 (Aab) & Aab & 0.8267 & sma & OM & \citet{harrington84} \\
J19535508+4424550 & GJ~1245B & 3 (B) & AB & 6.454 & 70 & & 1954 \\
J20163382-0711456 & TYC~5174-242-1 & 2 & & 0.107 & 352 & CPM? & \citet{janson12} \\
J20294834+0941202 & Gl~791.2 & 2 (AB) & AB & 0.1037 & sma & OM & \citet{benedict00} \\
J20434114-2433534 & & 2 & & 1.48 & & & \citet{shkolnik12} \\
J20450949-3120266 & Gl~803 & 3 (A) & A-BC & & 213 & & 1920 \\
& & & BC & 3.18 & sma & OM & 1913 \\
J20531465-0221218 & LP~636-19 & 2 & AB & 0.086 & 321 & & \citet{janson12} \\
J21000529+4004136 & Gl~815 & 3 (AB) & AB & 0.685 & 39 & & 1934 \\
& & & Aab & & & SB2 & \citet{karatas04} \\
J21374019+0137137 & 2E~4498 & 2 & & 0.433 & 341 & & \citet{janson14a} \\
J21514831+1336154 & GJ~4228 & 2 & & 0.674 & 131 & & \citet{cortes17} \\
J22143835-2141535 & BD-22~5866 & 4 & & 0.104 & & ESB4 & \citet{shkolnik08} \\
J22171870-0848186 & Gl~852B & 3 (Bab) & AB & 7.954 & 213 & CPM & 1920 \\
& & & Bab & 0.970 & 317 & CPM, OM & \citet{beuzit04} \\
J22171899-0848122 & Gl~852A & 3 (A) & AB & 7.954 & 213 & CPM & 1920 \\
J22232904+3227334 & Gl~856 & 2 & AB & 1.61 & sma & CPM, OM & 1959 \\
J22384530-2036519 & Gl~867B & 4 (B) & AC-BD & 24.5 & 350 & CPM & 1830 \\
& & & Bab (or BD) & & & SB1 & \citet{davison14} \\
J22384559-2037160 & Gl~867A & 4 (A) & Aab (or AC) & & & SB2 & \citet{herbig65} \\
J22450004-3315258 & Gl~871.1B & 2 & AB & 35.8 & 133 & CPM & 1920 \\
J22465311-0707272 & UCAC4 415-145732 & 2 & & & & & \\
J22554384-3022392 & LP~933-24 & 2 (A) & AB & 5.8 & 163 & CPM & 1960 \\
J23062378+1236269 & G~67-46 & 3 (Aab) & AB & 37.3 & 36 & CPM & 1951 \\
& & & Aab & 0.426 & 317 & CPM, OM, SB2 (3?) & \citet{shkolnik10} \\
J23172441+3812419 & GJ~4327 & 3 (Bab) & AB & 18.1 & 253 & & 1929 \\
& & & Bab & & & SB2 & \citet{cortes17} \\
J23172807+1936469 & GJ~4326 & 2 & AB & 0.264 & sma & CPM, OM & \citet{beuzit04} \\
J23205766-0147373 & LP~642-48 & 2 & AB & 0.099 & 325 & & \citet{daemgen07} \\
J23292258+4127522 & GJ~4338B & 3 (Bab) & AB & 17.7 & 214 & & 1952 \\
& & & Bab & 0.257 & 209 & & \citet{shkolnik12} \\
J23292346+4128068 & GJ~4337A & 3 (A) & AB & 17.7 & 214 & & 1952 \\
J23315208+1956142 & Gl~896A & 2 (A) & AB & 7.6 & sma & CPM & 1941 \\
J23315244+1956138 & Gl~896B & 2 (B) & AB & 7.6 & sma & CPM & 1941 \\
J23495365+2427493 & & 2 & & 0.131 & 325 & CPM, OM & \citet{janson12} \\
J23513366+3127229 & & 2 (A) & AB & 2.386 & 92 & CPM & \citet{bowler12a} \\
J23574989+3837468 & GJ~4381 & 2 & AB & 0.50 & 247 & & \citet{mccarthy01} \\
J23581366-1724338 & LP~764-40 & 2 & AB & 1.989 & 356 & CPM, OM & \citet{daemgen07} \\
J23584342+4643452 & Gl~913 & 2 & & 0.0341 & sma & OM,SB2? & \citet{goldin07} \\
    \noalign{\vskip 0.1cm}
    \hline
\end{longtable}
\end{landscape}
\twocolumn

\subsection{Comparison of effective temperatures and metallicities between the present work and a reference \citep{mann15}}

Table~\ref{tab:calibrators_teff_feh} gives a comparison between our results for $T_{\rm eff}$ and [Fe/H] using the \textsc{\small{mcal}} method, with state of the art reference values taken from \citet{mann15}.

\onecolumn
\begin{longtable}{llclcr}
	\caption{List of 74 stars with measurements of $T_{\rm eff}$ and [Fe/H] in \citet{mann15} (called reference) compared to our measurements (called this work) when they exist (66 stars, SB1 and SB2 rejected). Stars with an * after the common name are the 29 used to re-calibrate the \textsc{\small{mcal}} method. Active stars have values in parentheses.}
	\label{tab:calibrators_teff_feh}\\
    \hline
    \noalign{\vskip 0.1cm}
    2MASS name & Common name & $T_{\rm eff}$ & $T_{\rm eff}$ & [Fe/H] & [Fe/H] \\
    & & (this work) & (reference) & (this work) & (reference) \\
    \noalign{\vskip 0.1cm}
    \hline
    \noalign{\vskip 0.1cm}
    \endfirsthead
    \caption{continued.}\\
    \hline
    \noalign{\vskip 0.1cm}
    2MASS name & Common name & $T_{\rm eff}$ & $T_{\rm eff}$ & [Fe/H] & [Fe/H] \\
    & & (this work) & (reference) & (this work) & (reference) \\
    \noalign{\vskip 0.1cm}
    \hline
    \noalign{\vskip 0.1cm}
    \endhead
    \noalign{\vskip 0.1cm}
    \hline
    \endfoot
    \endlastfoot
J00115302+2259047 & LP~348-40 & 3372 & 3359 & $+0.13$ & $+0.13$ \\
J00182256+4401222 & Gl~15A & 3562 & 3603 & $-0.33$ & $-0.30$ \\
J00182549+4401376 & Gl~15B & 3402 & 3218 & $-0.44$ & $-0.30$ \\
J01123052-1659570 & Gl~54.1* & (3344) &  3056 & ($-0.34$) & $-0.26$ \\
J01432015+0419172 & Gl~70 & 3482 & 3458 & $-0.10$ & $-0.13$ \\
J02122090+0334310 & Gl~87* & & 3638 & & $-0.36$ \\
J02190228+2352550 & GJ~3150 & (3058) & 3216 & ($-0.35$) & $-0.07$ \\
J02221463+4752481 & Gl~96 & 4001 & 3785 & $+0.34$ & $+0.14$ \\
J02333717+2455392 & Gl~102 & (3152) & 3199 & ($-0.31$) & 0.00 \\
J02361535+0652191 & Gl~105B* & & 3284 & & $-0.12$ \\
J02441537+2531249 & Gl~109 & 3423 & 3405 & $-0.10$ & $-0.10$ \\
J04374092+5253372 & Gl~172 & 3824 & 3929 & $+0.36$ & $-0.11$ \\
J04374188-1102198 & Gl~173 & 3747 & 3671 & $-0.02$ & $-0.04$ \\
J04425581+1857285 & Gl~176* & & 3680 & & $+0.14$ \\
J05015746-0656459 & LHS~1723 & 3519 & 3143 & $-0.38$ & $-0.06$ \\
J05032009-1722245 & LP~776-46 & 3398 & 3365 & $-0.21$ & $-0.12$ \\
J05312734-0340356 & Gl~205* & 3964 & 3801 & $+0.53$ & $+0.49$ \\
J05363099+1119401 & Gl~208 & 3937 & 3966 & $+0.52$ & $+0.05$ \\
J05420897+1229252 & Gl~213* & 3253 & 3250 & $-0.19$ & $-0.22$ \\
J06000351+0242236 & GJ~3379 & (2488) & 3214 & ($-0.14$) & $+0.07$ \\
J06011106+5935508 & GJ~3378 & 3241 & 3340 & $-0.06$ & $-0.09$ \\
J06521804-0511241 & Gl~250B* & & 3481 & & $+0.14$ \\
J06544902+3316058 & Gl~251 & 3415 & 3448 & $-0.03$ & $-0.02$ \\
J07272450+0513329 & Gl~273* & 3323 & 3317 & $-0.06$ & $-0.11$ \\
J07284541-0317524 & GJ~1097 & 3423 & 3448 & $-0.07$ & $-0.01$ \\
J07384089-2113276 & LHS~1935 & 3446 & 3358 & $-0.21$ & $-0.18$ \\
J08103429-1348514 & Gl~297.2B & 3912 & 3544 & $+0.15$ & 0.00 \\
J08160798+0118091 & GJ~2066* & 3571 & 3500 & $-0.10$ & $-0.12$ \\
J09142298+5241125 & Gl~338A & 3920 & 3920 & $+0.37$ & $-0.01$ \\
J10112218+4927153 & Gl~380 & 4172 & 4131 & $+0.83$ & $+0.24$ \\
J10121768-0344441 & Gl~382* & 3694 & 3623 & $+0.16$ & $+0.13$ \\
J10285555+0050275 & Gl~393* & 3576 & 3548 & $-0.13$ & $-0.18$ \\
J10505201+0648292 & Gl~402 & 3216 & 3238 & $-0.03$ & $+0.16$ \\
J11032023+3558117 & Gl~411 & 3561 & 3563 & $-0.44$ & $-0.38$ \\
J11052903+4331357 & Gl~412A & 3552 & 3619 & $-0.40$ & $-0.37$ \\
J11414471+4245072 & GJ~1148 & 3236 & 3304 & $+0.09$ & $+0.07$ \\
J11421096+2642251 & Gl~436* & 3500 & 3479 & $+0.01$ & $+0.01$ \\
J11474440+0048164 & Gl~447* & 3244 & 3192 & $-0.14$ & $-0.02$ \\
J11505787+4822395 & GJ 1151 & 3304 & 3118 & $-0.14$ & $+0.03$ \\
J12100559-1504156 & GJ~3707 & 3161 & 3385 & $+0.19$ & $+0.26$ \\
J12385241+1141461 & Gl~480 & 3384 & 3463 & $+0.22$ & $+0.26$ \\
J13282106-0221365 & Gl~512A & 3433 & 3498 & $+0.11$ & $+0.08$ \\
J13295979+1022376 & Gl~514* & 3747 & 3727 & $-0.01$ & $-0.09$ \\
J13454354+1453317 & Gl~526* & 3698 & 3649 & $-0.31$ & $-0.31$ \\
J14341683-1231106 & Gl~555* & 3211 & 3211 & $+0.11$ & $+0.17$ \\
J15192689-0743200 & Gl~581* & 3401 & 3395 & $-0.14$ & $-0.15$ \\
J16252459+5418148 & Gl~625 & 3557 & 3475 & $-0.40$ & $-0.35$ \\
J16301808-1239434 & Gl~628* & 3327 & 3272 & $-0.03$ & $-0.03$ \\
J16570570-0420559 & GJ 1207 & (1624) & 3229 & ($-0.14$) & $-0.09$ \\
J17302272+0532547 & Gl~678.1A* & & 3675 & & $-0.09$ \\
J17362594+6820220 & Gl~687 & 3424 & 3439 & $-0.03$ & $+0.05$ \\
J17375330+1835295 & Gl~686* & 3693 & 3657 & $-0.21$ & $-0.25$ \\
J17435595+4322441 & Gl~694 & 3557 & 3464 & $+0.05$ & 0.00 \\
J17574849+0441405 & Gl~699* & 3463 & 3228 & $-0.54$ & $-0.40$ \\
J17575096+4635182 & GJ~4040 & 3393 & 3470 & $+0.04$ & $+0.04$ \\
J18050755-0301523 & Gl~701* & & 3614 & & $-0.22$ \\
J18415908+3149498 & GJ~4070 & 3473 & 3400 & $-0.17$ & $-0.16$ \\
J18424666+5937499 & Gl~725A & 3470 & 3441 & $-0.32$ & $-0.23$ \\
J18424688+5937374 & Gl~725B & 3300 & 3345 & $-0.30$ & $-0.30$ \\
J19071320+2052372 & Gl~745B & 3495 & 3494 & $-0.44$ & $-0.35$ \\
J19165526+0510086 & Gl~752A* & & 3558 & & $+0.10$ \\
J20450403+4429562 & Gl~806 & 3748 & 3542 & $-0.14$ & $-0.15$ \\
J20523304-1658289 & LP~816-60 & 3196 & 3205 & $-0.05$ & $-0.02$ \\
J20564659-1026534 & Gl~811.1 & 3512 & 3473 & $+0.10$ & $+0.16$ \\
J21091740-1318080 & Gl~821 & 3633 & 3545 & $-0.65$ & $-0.45$ \\
J22021026+0124006 & Gl~846* & 3879 & 3848 & $+0.27$ & $+0.02$ \\
J22094029-0438267 & Gl~849* & 3490 & 3530 & $+0.22$ & $+0.37$ \\
J22531672-1415489 & Gl~876* & 3166 & 3247 & $+0.12$ & $+0.17$ \\
J23213752+1717284 & GJ~4333 & 3153 & 3324 & $+0.19$ & $+0.24$ \\
J22563497+1633130 & Gl~880* & 3887 & 3720 & $+0.27$ & $+0.21$ \\
J23055131-3551130 & Gl~887* & & 3688 & & $-0.06$ \\
J23415498+4410407 & Gl~905 & 3186 & 2930 & $-0.10$ & $+0.23$ \\
J23430628+3632132 & GJ~1289 & (3193) & 3173 & ($-0.08$) & $+0.05$ \\
J23491255+0224037 & Gl~908* & 3602 & 3646 & $-0.52$ & $-0.45$ \\
    \noalign{\vskip 0.1cm}
    \hline
\end{longtable}
\twocolumn

\subsection{Comparison of equatorial and projected rotation velocities}

Table~\ref{tab:prot_slow} gives the comparison of equatorial rotation velocities computed from photometric rotation periods and radii, with projected rotation velocities for the slow rotators.

\begin{table*}
	\centering
	\caption{List of 54 slow rotators with a time series measurement of the rotation period (in days), together with their adopted radius in \Rnom and derived $v_{\rm eq}$ in \kms, to be compared to our measure or upper limit of $v \sin i$ in \kms.}
	\label{tab:prot_slow}
	\begin{tabular}{llllcll}
		\hline
		2MASS name & Common name & $P_{\rm rot}$ & Ref. & Radius & $v_{\rm eq}$ & $v \sin i$ \\
		\hline
	J00161455+1951385 & GJ~1006A & 4.798 & \citet{newton16b} & 0.24 & 2.58 & $4.0\pm1.6$ \\
	J00240376+2626299 & & 29.84 & \citet{newton16b} & 0.21 & 0.36 & $2.6\pm1.0$ \\
	J01023895+6220422 & Gl~49 & 18.6 & \citet{donati08} & 0.46 & 1.26 & <2 \\
	J01123052-1659570 & Gl~54.1 & $69.2\pm0.1$ & \citet{suarez16} & 0.18 & 0.13 & $3.4\pm0.8$ \\
	J04274130+5935167 & GJ~3287 & 6.850 & \citet{newton16b} & 0.22 & 1.62 & $3.9\pm1.5$ \\
	J05015746-0656459 & LHS~1723 & 88.5 & \citet{kiraga12} & 0.21 & 0.12 & $3.8\pm1.3$ \\
	J05312734-0340356 & Gl~205 & $35.0\pm0.1$ & \citet{suarez15} & 0.62 & 0.89 & <2 \\
	J05335981-0221325 & & 7.25 & \citet{kiraga12} & 0.32 & 2.23 & $5.4\pm1.0$ \\
	J05363099+1119401 & Gl~208 & 12.04 & \citet{kiraga12} & 0.70 & 2.95 & $4.0\pm1.4$ \\
	J06103462-2151521 & Gl~229 & $27.3\pm0.1$ & \citet{suarez16} & 0.57 & 1.05 & <2 \\
	J07320291+1719103 & G~88-36 & 13.41 & \citet{hartman11} & 0.72 & 2.73 & $3.0\pm1.6$ \\
	J09360161-2139371 & Gl~357 & $74.30\pm1.70$ & \citet{suarez15} & 0.39 & 0.26 & $2.5\pm1.1$ \\
	J09562699+2239015 & LHS~2212 & 107.8 & \citet{newton16b} & 0.21 & 0.097 & <2 \\
	J10121768-0344441 & Gl~382 & 21.56 & \citet{kiraga12} & 0.45 & 1.17 & <2 \\
    J11023832+2158017 & Gl~410 & 14.0 & \citet{donati08} & 0.58 & 2.10 & $3.0\pm0.7$ \\
    J11032023+3558117 & Gl~411 & 48.00 & \citet{kiraga07} & 0.48 & 0.51 & <2 \\
    J11032125+1337571 & NLTT~26114 & 34.42 & \citet{newton16b} & 0.25 & 0.37 & $4.6\pm1.6$ \\
    J11115176+3332111 & GJ~3647 & 7.785 & \citet{newton16b} & 0.28 & 1.83 & $4.6\pm0.7$ \\
    J11200526+6550470 & Gl~424 & 149.7 & \citet{engle09} & 0.61 & 0.20 & <2 \\
    J11414471+4245072 & GJ~1148 & 73.498679 & \citet{hartman11} & 0.24 & 0.16 & <2 \\
    J11421096+2642251 & Gl~436 & $39.90\pm0.80$ & \citet{suarez15} & 0.35 & 0.45 & <2 \\
    J11474440+0048164 & Gl~447 & $165.1\pm0.8$ & \citet{suarez16} & 0.20 & 0.063 & $2.1\pm1.0$ \\
    J11505787+4822395 & GJ~1151 & 132 & \citet{irwin11} & 0.19 & 0.072 & $2.5\pm1.0$ \\
    J13101268+4745190 & LHS~2686 & 28.80 & \citet{newton16b} & 0.17 & 0.29 & $4.5\pm0.9$ \\
    J13295979+1022376 & Gl~514 & $28.0\pm2.9$ & \citet{suarez15} & 0.54 & 0.98 & $2.0\pm0.8$ \\
    J13454354+1453317 & Gl~526 & $52.3\pm1.7$ & \citet{suarez15} & 0.50 & 0.48 & <2 \\
    J14010324-0239180 & Gl~536 & $43.3\pm0.1$ & \citet{suarez16} & 0.53 & 0.62 & <2 \\
    J15192689-0743200 & Gl~581 & $130.00\pm2.00$ & \citet{robertson14} & 0.32 & 0.12 & <2 \\
    J15323737+4653048 & TYC~3483-856-1 & 10.585 & \citet{hartman11} & 0.49 & 2.35 & $3.4\pm1.6$ \\
    J15553178+3512028 & GJ~3928 & 3.542 & \citet{newton16b} & 0.19 & 2.71 & $6.9\pm0.8$ \\
    J15581883+3524236 & G~180-18 & 57.216476 & \citet{hartman11} & 0.29 & 0.26 & <2 \\
    J16252459+5418148 & Gl~625 & $77.8\pm5.5$ & \citet{suarez17} & 0.42 & 0.28 & $2.2\pm0.7$ \\
    J16301808-1239434 & Gl~628 & $119.3\pm0.5$ & \citet{suarez16} & 0.26 & 0.11 & <2 \\
    J16360563+0848491 & GJ~1204 & 6.331 & \citet{newton16b} & 0.22 & 1.80 & $3.0\pm0.7$ \\
    J17195422+2630030 & Gl~669A & 20.263417 & \citet{hartman11} & 0.27 & 0.68 & $3.2\pm0.7$ \\
    J17574849+0441405 & Gl~699 & 130 & \citet{kiraga07} & 0.24 & 0.095 & $3.1\pm1.2$ \\
    J17575096+4635182 & GJ~4040 & 31.643331 & \citet{hartman11} & 0.31 & 0.50 & $2.0\pm1.1$ \\
    J18073292-1557464 & GJ~1224 & <4.3 & \citet{morin10} & 0.18 & >2.0 & $4.3\pm0.7$ \\
    J18172513+4822024 & TYC~3529-1437-1 & 16.2578 & \citet{norton07} & 0.38 & 1.19 & $3.1\pm1.0$ \\
    J18424498+1354168 & GJ~4071 & 8.090 & \citet{newton16b} & 0.22 & 1.41 & $4.2\pm0.7$ \\
    J18441139+4814118 & & 21.522016 & \citet{hartman11} & 0.32 & 0.75 & $2.7\pm1.0$ \\
    J20414744+4938482 & & 104.50 & \citet{newton16b} & 0.22 & 0.11 & <2 \\
    J20523304-1658289 & LP~816-60 & $67.6\pm0.1$ & \citet{suarez16} & 0.22 & 0.17 & <2 \\
    J22004701+7949254 & NLTT~52801 & 75.41 & \citet{newton16b} & 0.30 & 0.20 & <2 \\
    J22094029-0438267 & Gl~849 & $39.2\pm6.3$ & \citet{suarez15} & 0.32 & 0.42 & <2 \\
    J22245593+5200190 & GJ~1268 & 81.77 & \citet{newton16b} & 0.18 & 0.11 & $3.6\pm1.0$ \\
    J22250174+3540079 & & 22.897888 & \citet{hartman11} & 0.39 & 0.86 & $2.3\pm1.0$ \\
    J22270871+7751579 & G~242-2 & 98.42 & \citet{newton16b} & 0.16 & 0.08 & $2.2\pm1.0$ \\
    J22523963+7504190 & NLTT~55174 & 107.3 & \citet{newton16b} & 0.18 & 0.087 & <2 \\
    J22531672-1415489 & Gl~876 & $95\pm1 $ & \citet{nelson16} & 0.25 & 0.13 & <2 \\
    J22563497+1633130 & Gl~880 & $37.5\pm0.1$ & \citet{suarez15} & 0.46 & 0.62 & <2 \\
    J23380819-1614100 & GJ~4352 & 61.66 & \citet{watson06} & 0.42 & 0.34 & $2.1\pm1.2$ \\
    J23415498+4410407 & Gl~905 & 99.58 & \citet{newton16b} & 0.14 & 0.07 & <2 \\
    J23545147+3831363 & & 4.755 & \citet{newton16b} & 0.25 & 2.64 & $5.4\pm1.3$ \\
		\hline
	\end{tabular}
\end{table*}

Table~\ref{tab:prot_resolved} gives the same comparison for the resolved rotators ($v_{\rm eq}>3$\kms). For LP~193-584, the rotation period from \citet{hartman11} is uncertain and therefore given in parentheses, as well as the affected value of $v_{\rm eq}$. For NLTT~3478, the very large difference between $v_{\rm eq}$ and $v \sin i$ would imply an improbable small value of the inclination. The photometric period should therefore be measured again.

\onecolumn
\begin{longtable}{llllccr}
	\caption{List of 93 resolved rotators with a time series measurement of the rotation period (in days), together with their adopted radius in \Rnom and derived $v_{\rm eq}$ in \kms, to be compared to our measure of $v \sin i$ in \kms.}
	\label{tab:prot_resolved}\\
    \hline
    \noalign{\vskip 0.1cm}
		2MASS name & Common name & $P_{\rm rot}$ & Reference & Radius & $v_{\rm eq}$ & $v \sin i$ \\
    \noalign{\vskip 0.1cm}
    \hline
    \noalign{\vskip 0.1cm}
    \endfirsthead
    \caption{continued.}\\
    \hline
    \noalign{\vskip 0.1cm}
		2MASS name & Common name & $P_{\rm rot}$ & Reference & Radius & $v_{\rm eq}$ & $v \sin i$ \\
    \noalign{\vskip 0.1cm}
    \hline
    \noalign{\vskip 0.1cm}
    \endhead
    \noalign{\vskip 0.1cm}
    \hline
    \endfoot
    \endlastfoot	
	J00233468+2014282 & FK~Psc & 7.9165 & \citet{norton07} & 0.77 & 4.91 & $3.2\pm0.7$ \\
	J00243478+3002295 & GJ~3033 & 1.0769 & \citet{west15} & 0.19 & 8.88 & $12.2\pm0.8$ \\
	J00340843+2523498 & V493~And & 3.1555 & \citet{norton07} & 0.83 & 13.3 & $11.3\pm1.9$ \\
	J00485822+4435091 & LP~193-584 & (1.305) & \citet{hartman11} & 0.30 & (11.5) & $15.6\pm1.4$ \\
	J01031971+6221557 & Gl~51 & 1.0237 & \citet{west15} & 0.19 & 9.34 & $12.5\pm0.7$ \\
	J01034013+4051288 & NLTT~3478 & 0.253982 & \citet{hartman11} & 0.86 & 172 & $5.4\pm1.1$ \\
	J01220441-3337036 & & 9.58 & \citet{kiraga12} & 0.76 & 4.04 & $4.1\pm1.1$ \\
	J01362619+4043443 & V539~And & 0.4357 & \citet{norton07} & 0.42 & 48.2 & $73.2\pm1.0$ \\
	J01373940+1835332 & TYC~1208-468-1 & 2.803 & \citet{kiraga12} & 0.56 & 10.0 & $16.3\pm1.4$ \\
	J01390120-1757026 & Gl~65A & $0.2430\pm0.0005$ & \citet{barnes17} & 0.14 & 29.8 & $29.5\pm0.7$ \\
	J01390120-1757026 & Gl~65B & $0.2268\pm0.0003$ & \citet{barnes17} & 0.15 & 33.2 & $37.9\pm1.4$ \\
	J01535076-1459503 & & 1.515 & \citet{kiraga12} & 0.28 & 9.22 & $11.6\pm1.7$ \\
	J02001277-0840516 & & 2.28 & \citet{kiraga12} & 0.36 & 8.08 & $12.2\pm2.1$ \\
	J02071032+6417114 & GJ~3134 & 1.177 & \citet{newton16b} & 0.20 & 8.68 & $11.4\pm1.0$ \\
	J02155892-0929121 & & 1.4374 & \citet{kiraga13} & 0.34 & 11.8 & $15.7\pm1.3$ \\
	J02272924+3058246 & AG~Tri & 13.6928 & \citet{norton07} & 0.98 & 3.61 & $4.9\pm2.0$ \\
	J02364412+2240265 & G~36-26 & 0.3697 & \citet{west15} & 0.15 & 21.1 & $11.2\pm1.4$ \\
	J03153783+3724143 & LP~247-13 & 1.2887 & \citet{hartman11} & 0.31 & 12.2 & $15.1\pm0.9$ \\
	J03472333-0158195 & G~80-21 & 3.881 & \citet{kiraga12} & 0.34 & 4.46 & $6.2\pm1.1$ \\
	J04302527+3951000 & Gl~170 & 0.7177 & \citet{west15} & 0.18 & 12.7 & $13.6\pm0.8$ \\
	J04353618-2527347 & LP~834-32 & 2.785 & \citet{kiraga12} & 0.26 & 4.70 & $7.1\pm1.0$ \\
	J04365738-1613065 & & 0.6105 & \citet{kiraga12} & 0.27 & 22.5 & $63.3\pm8.4$ \\
	J04435686+3723033 & V962~Per & 4.2878 & \citet{norton07} & 0.38 & 4.46 & $10.1\pm1.6$ \\
	J04571728-0621564 & & 0.7337 & \citet{kiraga12} & 0.63 & 43.2 & $11.0\pm1.8$ \\
	J04593483+0147007 & Gl~182 & 4.414 & \citet{kiraga12} & 0.63 & 7.19 & $8.7\pm1.6$ \\
	J05024924+7352143 & & 0.68204 & \citet{kiraga13} & 0.95 & 70.5 & $46.1\pm3.8$ \\
	J06000351+0242236 & GJ~3379 & 1.8088 & \citet{west15} & 0.22 & 6.29 & $5.9\pm1.4$ \\
	J06362522+4349473 & & 1.5945715 & \citet{hartman11} & 0.28 & 8.95 & $19.6\pm1.0$ \\
	J07310129+4600266 & & 1.33064 & \citet{hartman11} & 0.23 & 8.82 & $14.9\pm0.7$ \\
	J07444018+0333089 & Gl~285 & $2.7758\pm0.0006$ & \citet{morin08b} & 0.22 & 4.05 & $6.6\pm0.8$ \\
	J08085639+3249118 & GJ~1108A & 3.37045 & \citet{hartman11} & 0.65 & 9.80 & $9.5\pm1.6$ \\
	J08294949+2646348 & GJ~1111 & 0.459 & \citet{newton16b} & 0.19 & 20.7 & $11.4\pm0.7$ \\
	J09002359+2150054 & LHS~2090 & 0.439 & \citet{newton16b} & 0.12 & 14.2 & $15.0\pm1.0$ \\
	J09445422-1220544 & G~161-71 & 0.4417 & \citet{kiraga12} & 0.15 & 17.4 & $41.8\pm5.0$ \\
	J10141918+2104297 & GJ~2079 & 7.861 & \citet{kiraga12} & 0.60 & 3.87 & $5.5\pm1.6$ \\
	J10193634+1952122 & Gl~388 & $2.2399\pm0.0006$ & \citet{morin08b} & 0.29 & 6.48 & $4.1\pm0.7$ \\
    J10481258-1120082 & GJ~3622 & $1.5\pm0.2$ & \citet{morin10} & 0.13 & 4.42 & $3.3\pm0.7$ \\
    J10562886+0700527 & Gl~406 & <2.0 & \citet{morin10} & 0.12 & >3.1 & $2.9\pm0.8$ \\
    J11015191-3442170 & TW~Hya & $3.5683\pm0.0002$ & \citet{huelamo08} & 0.69 & 9.79 & $5.4\pm0.7$ \\
    J11053133+4331170 & Gl~412B & $0.78\pm0.02$ & \citet{morin10} & 0.13 & 8.56 & $8.0\pm0.7$ \\
    J11314655-4102473 & Gl~431 & 0.9328 & \citet{kiraga12} & 0.18 & 9.71 & $20.3\pm0.8$ \\
    J11324124-2651559 & TWA~8A & 4.638 & \citet{kiraga12} & 0.30 & 3.32 & $5.1\pm1.1$ \\
    J11432359+2518137 & GJ~3682 & 1.326 & \citet{newton16b} & 0.23 & 8.78 & $13.7\pm0.9$ \\
    J12141654+0037263 & GJ~1154 & 1.5835 & \citet{west15} & 0.16 & 4.95 & $6.1\pm0.7$ \\
    J12185939+1107338 & GJ~1156 & 0.491 & \citet{irwin11} & 0.14 & 14.9 & $15.6\pm0.8$ \\
    J12574030+3513306 & Gl~490A & 3.3664 & \citet{norton07} & 0.55 & 8.22 & $8.2\pm1.6$ \\
    J13003350+0541081 & Gl~493.1 & 0.600 & \citet{irwin11} & 0.18 & 14.9 & $15.6\pm0.8$ \\
    J13004666+1222325 & Gl~494 & 2.886 & \citet{kiraga12} & 0.47 & 8.22 & $9.6\pm0.9$ \\
    J13093495+2859065 & GJ~1167A & 0.215 & \citet{newton16b} & 0.20 & 47.3 & $51.3\pm1.5$ \\
    J13142039+1320011 & NLTT~33370 & 0.158 & \citet{newton16b} & 0.12 & 40.0 & $53.8\pm1.6$ \\
    J13314666+2916368 & GJ~3789 & 0.2683 & \citet{norton07} & 0.22 & 42.0 & $76.0\pm0.7$ \\
    J13345147+3746195 & & 3.0992 & \citet{hartman11} & 0.23 & 3.79 & $8.3\pm0.8$ \\
    J14142141-1521215 & GJ~3831 & 0.2982 & \citet{kiraga12} & 0.71 & 120 & $73.5\pm0.7$ \\
    J14200478+3903014 & GJ~3842 & 0.3693 & \citet{norton07} & 0.34 & 46.2 & $70.0\pm1.0$ \\
    J14321078+1600494 & GJ~3856 & 0.765 & \citet{newton16b} & 0.23 & 15.5 & $14.1\pm1.1$ \\
    J14372948+4128350 & LO~Boo & 2.09162 & \citet{hartman11} & 0.35 & 8.54 & $11.8\pm1.2$ \\
    J15040626+4858538 & & 1.02136 & \citet{hartman11} & 0.23 & 11.4 & $11.3\pm2.0$ \\
    J15123818+4543464 & GJ~3898 & 1.686 & \citet{newton16b} & 0.23 & 6.99 & $10.4\pm1.1$ \\
    J15215291+2058394 & GJ~9520 & 3.3829 & \citet{norton07} & 0.42 & 6.22 & $5.2\pm1.0$ \\
    J15565823+3738137 & & 0.30694 & \citet{hartman11} & 0.34 & 55.7 & $26.5\pm1.4$ \\
    J16352740+3500577 & GJ~3966 & 0.9166 & \citet{norton07} & 0.24 & 13.0 & $21.5\pm3.5$ \\
    J16400599+0042188 & GJ~3967 & 0.3114 & \citet{west15} & 0.20 & 32.7 & $31.0\pm0.8$ \\
    J16402068+6736046 & GJ~3971 & 0.3782 & \citet{west15} & 0.13 & 17.5 & $10.8\pm0.7$ \\
    J16553529-0823401 & Gl~644C & <1.0 & \citet{morin10} & 0.13 & >6.5 & $10.1\pm0.8$ \\
    J16570570-0420559 & GJ~1207 & 1.212 & \citet{kiraga12} & 0.39 & 16.4 & $11.5\pm1.5$ \\
    J16590962+2058160 & V1234~Her & 4.1037 & \citet{norton07} & 0.27 & 3.32 & $6.5\pm1.0$ \\
    J17365925+4859460 & V1279~Her & 2.613578 & \citet{hartman11} & 0.36 & 6.91 & $7.0\pm1.0$ \\
    J17380077+3329457 & & 12.184 & \citet{hartman11} & 0.80 & 3.30 & $7.2\pm2.7$ \\
    J18021660+6415445 & G~227-22 & 0.280 & \citet{newton16b} & 0.18 & 32.7 & $13.2\pm1.2$ \\
    J18130657+2601519 & GJ~4044 & 2.285 & \citet{newton16b} & 0.23 & 5.14 & $7.5\pm0.7$ \\
    J18315610+7730367 & LP~24-256 & 0.8607 & \citet{west15} & & & $15.8\pm0.7$ \\
    J19165762+0509021 & Gl~752B & <0.8 & \citet{morin10} & 0.13 & >8.2 & $5.3\pm0.9$ \\
    J19510930+4628598 & GJ~1243 & 0.59258 & \citet{hartman11} & 0.23 & 19.6 & $22.1\pm0.9$ \\
    J19535443+4424541 & GJ~1245AC & 0.263241 & \citet{hartman11} & 0.14 & 25.9 & $22.0\pm0.8$ \\
    J19535508+4424550 & GJ~1245B & $0.71\pm0.01$ & \citet{morin10} & 0.13 & 9.48 & $7.9\pm0.7$ \\
    J20294834+0941202 & Gl~791.2 & $0.3085\pm0.0005$ & \citet{barnes17} & 0.18 & 29.5 & $35.3\pm0.7$ \\
    J20450949-3120266 & Gl~803 & 4.852 & \citet{kiraga12} & 0.51 & 5.32 & $8.5\pm0.7$ \\
    J20465795-0259320 & & 3.644 & \citet{kiraga12} & 0.65 & 9.01 & $9.0\pm1.7$ \\
    J20560274-1710538 & TYC~6349-200-1 & 3.403 & \citet{kiraga12} & & & $12.4\pm0.8$ \\
    J21100535-1919573 & & 3.710 & \citet{kiraga13} & 0.38 & 5.14 & $8.3\pm0.9$ \\
    J21374019+0137137 & 2E~4498 & 0.213086 & \citet{kiraga12} & 0.22 & 51.5 & $49.9\pm0.9$ \\
    J22004158+2715135 & TYC~2211-1309-1 & 0.5235 & \citet{norton07} & 0.70 & 67.9 & $61.6\pm4.2$ \\
    J22011310+2818248 & GJ~4247 & $0.445654\pm0.000002$ & \citet{morin08a} & 0.23 & 26.2 & $36.9\pm0.7$ \\
    J22232904+3227334 & Gl~856 & 0.8539 & \citet{west15} & 0.32 & 18.8 & $16.2\pm0.6$ \\
    J22464980+4420030 & Gl~873 & $4.3715\pm0.0006$ & \citet{morin08b} & 0.26 & 3.03 & $5.9\pm0.7$ \\
    J22515348+3145153 & Gl~875.1 & 1.6404 & \citet{norton07} & 0.31 & 9.53 & $13.2\pm0.9$ \\
    J23060482+6355339 & GJ~9809 & 2.831 & \citet{kiraga13} & 0.57 & 10.1 & $7.0\pm1.4$ \\
    J23081954-1524354 & Gl~890 & 0.4311 & \citet{kiraga12} & 0.66 & 76.9 & $69.4\pm0.7$ \\
    J23315208+1956142 & Gl~896A & 1.0664 & \citet{norton07} & 0.29 & 13.6 & $14.5\pm0.8$ \\
    J23315244+1956138 & Gl~896B & $0.404\pm0.004$ & \citet{morin08b} & 0.17 & 21.0 & $25.4\pm0.9$ \\
    J23320018-3917368 & & 3.492 & \citet{kiraga12} & 0.28 & 4.06 & $6.0\pm1.1$ \\
    J23512227+2344207 & G~68-46 & 3.211 & \citet{newton16b} & 0.22 & 3.39 & $5.2\pm0.9$ \\
    J23581366-1724338 & LP~764-40 & 0.434093 & \citet{kiraga12} & 0.37 & 42.7 & $28.6\pm1.1$ \\
    \noalign{\vskip 0.1cm}
    \hline
\end{longtable}
\twocolumn

\subsection{Master table of properties for the stars in our sample}

A summary of the measurements for the whole sample of 440 stars is given in Table~\ref{tab:results}, an extract of which (0 < $RA$ < 2~h) is given here. There are actually 447 entries, as observations for some stars in the initial sample but later rejected are listed, and some close binaries appear with two different entries, one for each component. For each star are listed the number of observed spectra, including those finally rejected for the measurement of metallicity and effective temperature, the spectroscopic mode of observation (polarimetric or S+S), the spectral type from the TiO~5 index, the $V-K_{\rm s}$ color, the $H$ magnitude from the 2MASS PSC, the heliocentric radial velocity averaged over all spectra for the star (not given for SB2), the projected rotational velocity and its error (or <2\kms when unresolved), the H$\alpha$ index (above 0.25, $T_{\rm eff}$ and [Fe/H] cannot be reliably measured by the \textsc{\small{mcal}} method), our mean value of [Fe/H] and $T_{\rm eff}$ for the inactive stars (H$\alpha$ index <0.25), the predicted uncertainty of the radial velocity assuming a full correction of telluric lines (see Section~\ref{sec:discussion}), and a binarity flag (SB1 for a single-line spectroscopic binary, SB2 for a multiple-lines spectroscopic binary and VB for a visual binary with a projected separation smaller than 2.0\arcsec). The uncertainties on $T_{\rm eff}$ and [Fe/H] are computed from the individual internal uncertainties returned by the \textsc{\small{mcal}} method. They do not reflect systematic uncertainties associated with this method. Our measurements of $T_{\rm eff}$ and [Fe/H] are given the source code 1. When they are not available, we used the values listed in \citet{mann15} with a source code 2, or values without error derived using their Equation 7 and coefficients in Table 2 with a source code 3. The full table is available on-line. The first page is displayed here to illustrate the format.

\onecolumn
\begin{landscape}
\begin{table*}
	\centering
	\caption{Master list of data for the whole sample (440 dwarfs, 447 entries): number $N$ of measured spectra, number $n$ of rejected ones if any, instrument mode (P for polarimetry and/or S for S+S), spectral type (from TiO~5 index), $V-K_{\rm s}$ color, $H$ magnitude, heliocentric radial velocity, projected rotational velocity and error in \kms (<2 if not resolved), H$\alpha$ index, [Fe/H] and error, effective temperature and error in K, source code, radial velocity uncertainty, binarity flag (SB1, SB2, close VB).}
	\label{tab:results}
	\begin{tabular}{lllrrrrccrrccrccrcrl}
        \hline
		2MASS name & Common name & Alt name & $N$ & $n$ & P/S & ST & $V-K_{\rm s}$ & $H$ & $HRV$ & $v \sin i$ & err & H$\alpha$ index & [Fe/H] & err & $T_{\rm eff}$ & err & source & $\sigma (RV)$ & Binarity \\
		& & & & & & & mag & mag & \kms & \kms & \kms & & dex & dex & K & K & & \ms & \\
		\hline
J00080642+4757025 & & & 1 & & S & 4.3 & 5.102 & 8.000 & -76.06 & & & & & & 3129 & & 3 & & SB2 \\
J00115302+2259047 & LP 348-40 & & 1 & & S & 4.1 & 5.015 & 8.308 & -45.29 & <2 & & 0.05 & 0.126 & 0.036 & 3372 & 59 & 1 & 4.8 & \\
J00155808-1636578 & & & 1 & & S & 4.6 & 5.275 & 8.191 & 19.33 & 19.1 & 1.5 & 0.57 & & & 3077 & & 3 & 25.2 & VB \\
J00161455+1951385 & GJ 1006A & EZ Psc & 1 & & S & -1.0 & 5.058 & 7.322 &  4.72 & 4.0 & 1.6 & $-0.02$ & 1.540 & 0.052 & 3678 & 84 & 1 & 2.0 & \\
J00182256+4401222 & Gl 15A & GX And & 4 & & P & 1.4 & 4.116 & 4.476 & 11.73 & 2.6 & 0.7 & 0.04 & $-0.329$ & 0.005 & 3562 & 8 & 1 & 0.1 & \\
J00182549+4401376 & Gl 15B & GQ And & 3 & & P & 4.0 & 5.112 & 6.191 &  11.17 & 2.3 & 0.8 & 0.05 & $-0.437$ & 0.011 & 3402 & 18 & 1 & 0.7 & \\
J00210932+4456560 & & & 1 & & S & 3.8 & 4.972 & 8.967 & 3.16 & 14.3 & 1.4 & 0.58 & & & 3214 & & 3 & 57.3 & \\
J00233468+2014282 & FK Psc & & 3 & & S & $-0.3$ & 3.505 & 7.498 & $-2.19$ & 3.2 & 0.7 & 0.09-0.13 & 0.933 & 0.015 & 3757 & 25 & 1 & 2.3 & VB \\
J00240376+2626299 & & & 2 & & S & 4.6 & 5.436 & 9.592 & 20.31 & 2.6 & 1.0 & 0.19-0.20 & 0.003 & 0.041 & 3504 & 68 & 1 & 17.3 & \\
J00243478+3002295 & GJ 3033 & & 7 & & SP & 4.8 & 5.623 & 9.218 & 10.23 & 12.2 & 0.8 & 0.52-0.64 & & & 2995 & & 3 & 63.5 & \\
J00294322+0112384 & & & 2 & & P & 1.0 & 3.967 & 8.536 & $-31.57$ & <2 & & 0.03 & $-0.167$ & 0.010 & 3682 & 17 & 1 & 6.2 & \\
J00340843+2523498 & V493 And & & 2 & & S & $-0.5$ & 3.436 & 7.830 & $-9.63$ & 11.3 & 1.9 & 0.14 & & & 3866 & & 3 & 19.8 & VB \\
J00385879+3036583 & Gl 26 & Wolf 1056 & 2 & & P & 2.7 & 4.394 & 6.864 & $-0.37$ & <2 & & 0.04 & $-0.093$ & 0.010 & 3526 & 15 & 1 & 1.3 & \\
J00424820+3532554 & Gl 29.1A & FF And & 1 & & S & 1.4 & 4.106 & 6.506 & $-59.14$ & & & & & & 3508 & & 3 & & SB2 \\
J00433851+7545152 & & & 3 & & S & 4.4 & 5.218 & 9.674 & 2.85 & 4.8 & 0.8 & 0.39-0.42 & & & 3117 & & 3 & 15.5 & \\
J00485822+4435091 & LP 193-584 & GJ 3058 & 1 & & S & 3.7 & 4.851 & 8.485 & $-13.88$ & 15.6 & 1.4 & 0.48 & & & 3248 & & 3 & 36.7 & VB \\
J00570261+4505099 & G 172-30 & & 2 & & P & 3.4 & 4.731 & 7.459 & 6.62 & 2.1 & 1.1 & 0.05 & $-0.089$ & 0.010 & 3407 & 16 & 1 & 2.2 & \\
J00582789-2751251 & Gl 46 & & 2 & & P & 3.8 & 4.883 & 7.203 & 21.76 & 2.5 & 1.3 & 0.05 & $-0.018$ & 0.013 & 3361 & 20 & 1 & 1.7 & SB1 \\
J01012006+6121560 & Gl 47 & & 2 & & P & 2.3 & 4.323 & 6.710 & 7.58 & 2.7 & 1.2 & 0.04-0.05 & $-0.218$ & 0.008 & 3529 & 13 & 1 & 1.1 & \\
J01023213+7140475 & Gl 48 & & 2 & & P & 3.4 & 4.595 & 5.699 & 1.56 & <2 & & 0.03-0.04 & 0.044 & 0.010 & 3502 & 17 & 1 & 0.4 & \\
J01023895+6220422 & Gl 49 & HIP 4872 & 1 & & S & 1.6 & 4.194 & 5.582 & $-6.05$ & <2 & & 0.04 & 0.249 & 0.011 & 3750 & 18 & 1 & 0.4 & \\
J01031971+6221557 & Gl 51 & V388 Cas & 26 & 1 & P & 5.4 & 5.635 & 8.014 & $-5.76$ & 12.5 & 0.7 & 0.73-1.25 & & & 3003 & & 3 & 20.2 & \\
J01034013+4051288 & NLTT 3478 & G 132-50 & 1 & & S & 3.5 & 3.370 & 7.466 & $-10.82$ & 5.4 & 1.1 & 0.41 & & & 3908 & & 3 & 13.9 & VB \\
J01034210+4051158 & NLTT 3481 & G 132-51 & 1 & & S & 4.6 & 4.584 & 8.839 & $-11.06$ & 9.2 & 1.8 & 0.73 & & & 3301 & & 3 & 52.5 & \\
J01112542+1526214 & GJ 3076 & & 2 & & S & 5.6 & 6.222 & 8.512 & 3.18 & 18.1 & 2.2 & 0.70-0.71 & & & 2826 & & 3 & 27.0 & VB \\
J01123052-1659570 & Gl 54.1 & YZ Cet & 3 & & SP & 5.0 & 5.676 & 6.749 & 28.27 & 3.4 & 0.8 & 0.14-0.61 & & & 2953 & & 3 & 0.8 & \\
J01155017+4702023 & & & 5 & & S & 4.7 & 5.537 & 9.574 & 10.01 & 7.1 & 0.9 & 0.49-0.56 & & & 3044 & & 3 & 93.9 & VB \\
J01220441-3337036 & & & 2 & & S & 0.0 & 3.568 & 7.636 & 4.86 & 4.1 & 1.1 & 0.11 & 0.569 & 0.018 & 3764 & 29 & 1 & 2.6 & \\
J01351393-0712517 & & & 3 & & S & 4.5 & 5.348 & 8.387 & 7.48 & & & & & & 3081 & & 3 & & SB2 \\
J01362619+4043443 & V539 And & & 1 & & S & 1.6 & 4.364 & 9.073 & $-7.83$ & 73.2 & 1.0 & 0.23 & & & 3418 & & 3 & 67.9 & \\
J01365516-0647379 & G 271-110 & & 3 & & S & 4.3 & 5.196 & 9.137 & 12.77 & 11.5 & 0.8 & 0.70-0.71 & & & 3112 & & 3 & 64.0 & \\
J01373940+1835332 & TYC 1208-468-1 & & 2 & & S & $-1.0$ & 3.868 & 6.861 & 0.50 & 16.3 & 1.4 & 0.12 & & & 3571 & & 3 & 8.5 & VB \\
J01390120-1757026A & Gl 65A & BL Cet & 4 & & P & 5.9 & 6.320 & 6.520 & 19.82 & 29.5 & 0.7 & 0.39-0.67 & & & 2810 & & 3 & 4.1 & \\
J01390120-1757026B & Gl 65B & UV Cet & 8 & & P & 5.7 & 6.200 & 6.680 & 22.01 & 37.9 & 1.4 & 0.68-1.11 & & & 2842 & & 3 & 4.9 & \\
J01432015+0419172 & Gl 70 & & 4 & & P & 2.5 & 4.332 & 6.809 & $-25.94$ & <2 & & 0.04-0.05 & $-0.095$ & 0.007 & 3482 & 11 & 1 & 1.2 & \\
J01434512-0602400 & & & 1 & & S & 4.0 & 5.037 & 8.174 & $-13.72$ & & & & & & 3171 & & 3 & & SB2 \\
J01451820+4632077 & LHS 6032 & G 173-18 & 1 & & S & 2.6 & 4.339 & 7.423 & 36.79 & & & & & & 3415 & & 3 & & SB2 \\
J01515108+6426060 & GJ 3117A & & 2 & & P & 2.6 & 4.478 & 7.250 & $-12.79$ & 2.2 & 1.2 & 0.04 & 0.089 & 0.010 & 3620 & 17 & 1 & 1.9 & \\
J01535076-1459503 & & & 2 & & S & 3.8 & 4.940 & 7.297 & 10.34 & 11.6 & 1.7 & 0.56-0.57 & & & 3213 & & 3 & 11.8 & \\
J01591239+0331092 & GJ 1041A & NLTT 6637 & 2 & & S & 0.6 & 3.878 & 7.221 & $-7.98$ & 2.4 & 1.2 & 0.02 & 0.073 & 0.012 & 3821 & 19 & 1 & 1.7 & \\
J01591260+0331113 & NLTT 6638 & & 1 & & S & 3.2 & 5.351 & 7.384 & 4.21 & & & & & & 3087 & & 3 & & SB2 \\
J01592349+5831162 & Gl 82 & V596 Cas & 1 & & S & 4.7 & 5.194 & 7.224 & $-9.79$ & 13.4 & 2.0 & 0.55 & & & 3107 & & 3 & 10.4 & \\
		\hline
	\end{tabular}
\end{table*}
\end{landscape}
\twocolumn


\bsp	
\label{lastpage}
\end{document}